\documentclass[showpacs,nofootinbib,showkeys,eqsecnum,prd,aps,preprint]{revtex4-1}
\usepackage{amssymb}
\usepackage{amsmath}
\usepackage{amsfonts}
\usepackage{graphicx}
\usepackage{bm}
\usepackage[T1]{fontenc}
\usepackage{graphicx}

\begin{document}

\title{Exactly solvable cases in QED with $t$-electric potential steps}
\author{T. C. Adorno${}^{a}$}
\email{tg.adorno@gmail.com / tg.adorno@mail.tsu.ru}
\author{S.~P.~Gavrilov${}^{a,c}$}
\email{gavrilovsergeyp@yahoo.com, gavrilovsp@herzen.spb.ru}
\author{D.~M.~Gitman${}^{a,b,d}$}
\email{gitman@if.usp.br}
\date{\today }

\begin{abstract}
In this paper, we present in detail consistent QED (and scalar QED)
calculations of particle creation effects in external electromagnetic field
that correspond to three most important exactly solvable cases of $t$%
-electric potential steps: Sauter-like electric field, $T$-constant electric
field, and exponentially growing and decaying electric fields. In all these
cases, we succeeded to obtain new results, such as calculations in modified
configurations of the above mentioned steps and detailed considerations of
new limiting cases in already studied before steps. As was recently
discovered by us, the information derived from considerations of exactly
solvable cases allows one to make some general conclusions about quantum
effects in fields for which no closed form solutions of the Dirac (or
Klein-Gordon) equation are known. In the present article we briefly
represent such conclusions about an universal behavior of vacuum mean values
in slowly varying strong electric fields.

PACS numbers: 12.20.Ds,11.15.Tk,11.10.Kk
\end{abstract}

\keywords{Particle creation, Schwinger effect, time-dependent external
field, Dirac and Klein-Gordon equations.}

\affiliation{$^{a}$Tomsk State University, Russia\\
$^{b}$P.N. Lebedev Physical Institute, Russia\\
${}^{c}$Herzen State
Pedagogical University of Russia,
St.~Petersburg, Russia\\
$^{d}$Institute of Physics, University of
S\~{a}o Paulo, Brazil\\
}

\maketitle
\tableofcontents

\section{Introduction\label{S1}}

Quantum field theories (QFTs) with external backgrounds (external fields)
are to a certain extent the most appropriate models for calculating quantum
effects in strong fields of electromagnetic, gravitational, or other nature.
These calculations must be nonperturbative with respect to the interaction
with strong backgrounds. One of the most interesting effect of such kind
that attracts attention already for a long time is the particle creation
from the vacuum by strong external backgrounds \cite%
{Schwinger51,General1,General2,General3,General4,FGS}. In the framework of
the QFT, the particle creation is closely related to a violation of the
vacuum stability with time. Not all backgrounds violate the vacuum
stability. For example, electromagnetic backgrounds that violate the vacuum
stability have to be electriclike fields that are able to produce nonzero
work when interacting with charged particles. In such backgrounds any
process is accompanied by new created particles and, thus, turns out to a
many-particle process. That is why any consistent consideration of quantum
processes in the vacuum violating backgrounds has to be done in the
framework of a QFT; a consideration in the framework of the relativistic
quantum mechanics\ is restricted and may lead to paradoxes and even
incorrect results. It should be noted that at present, methods for the above
mentioned nonperturbative calculations are consistently formulated only in
QED (and scalar QED) with some specific types of external backgrounds.
Namely, these types are time-dependent external electric fields that are
switched on and off at the initial and the final time instants respectively,
see Refs. \cite{FGS,Gitman77a,Gitman77b,Gitman77c}, we refer to such kind of
external fields as the $t$-electric potential steps, and some
time-independent inhomogeneous external electric fields, which are called
conditionally, $x$-electric potential steps, see Refs. \cite{GavGit15}. It
should be stressed that the possibility of nonperturbative calculations in
the both above cases is based on the existence of exact solutions of the
Dirac (or Klein-Gordon) equation in the corresponding background fields. At
present, there are known only few such cases, which we call exactly solvable
cases in QED with $t$-electric or $x$-electric potential steps. Note that
solutions of the Dirac equation with $x$-electric potential steps are quite
similar to solutions obtained for $t$-electric steps where time $t$ is
replaced by the spatial coordinate $x$. Some of quantum effects in each of
these cases were considered in the literature using different approaches
(relativistic quantum mechanics, QED) with a certain levels of consistency
and details.

The case of homogeneous and constant electric field was examined by
Schwinger \cite{Schwinger51}, who obtained the probability for a vacuum to
remain a vacuum.{\large \ }Such field admit analytical solutions of the
Dirac and Klein-Gordon equations and was frequently used in various QFT
calculations; see Ref.~\cite{Dunn04} for a review. Different semiclassical
and numerical methods were applied to study Schwinger's effective action,
see Refs.~\cite{General1,General2,General3,General4} for a review. Nikishov
found the mean number of pairs created by the homogeneous and constant
electric field and established relation between the Schwinger's and Feynman'
representations of a causal propagator in this external field \cite%
{Nikis70a,Nikis79}. Subsequently he expanded and deepened his approach to
this problem in Refs.{\large \ }\cite%
{Nikis70b,NarNik76a,NarNik76b,Nikis04a,Nikis04b}.

In QED (and scalar QED) with $t$-electric potential steps, there exist a few
exactly solvable cases that have real physical importance. Those are
Sauter-like (or adiabatic or pulse) electric field, $T$-constant electric
field (a uniform electric field which effectively acts during a sufficiently
large but finite time interval $T$), an exponentially decaying electric
field, and their certain combinations. The particle creation effect in the
cases of the Sauter-like electric field was studied in Refs. \cite%
{NarNik70,GavG96a,DunHal98}, in the case of $T$-constant electric field in
Refs. \cite%
{GavG96a,GavG08,GavGT06,GavGS15,GG06-08a,GG06-08b,GavGitY12,DvoGavGit14,BagGiS75,HalL95}%
, and in the case of exponential electric fields in Refs. \cite%
{AdoGavGit14,AdoGavGit16}. One can see that quantum effects which can be
studied using the exactly solvable cases are important in astrophysics,
neutrino physics, cosmology, condense matter physics, and so on. In
particular, the particle creation effect due to the Sauter-like and $T$%
-constant electric fields is crucial for understanding the conductivity of
graphene or Weyl semimetals in the nonlinear regime as was reported, e.g.,
in Refs. \cite%
{GavGitY12,lewkowicz10a,lewkowicz10b,lewkowicz10c,Van+etal10,Zub12,KliMos13,Fil+McL15,VajDorMoe15}%
. Note that the cases of a constant and exponentially decaying electric
fields have many similarities with the case of the de Sitter background,
e.g., see Refs.~\cite{AndMot14,AkhmP15,StStHue16} and references therein.
One can also notice that the case of harmonically alternating electric field
is also exactly solvable \cite{d5a,d5b}. In this case the alternating
direction of the electric field can also contribute to the particle creation
effect, but it is not an example of $t$-electric potential step. Using
exactly solvable cases one can develop new approximation methods of
calculating quantum effects in QFT with unstable vacuum, see reviews \cite%
{General1,General2,General3,General4,FGS}. However, many already known
results are scattered over different publications and many new result were
not published at all.

In this article, we present in detail consistent QED (and scalar QED)
calculations of zero order\footnote{%
Processes that do not involve photons.} quantum effects in external
electromagnetic field that correspond to three most important exactly
solvable cases of $t$-electric potential steps: Sauter-like electric field, $%
T$-constant electric field, and exponentially growing and decaying electric
fields. In all these cases we succeeded to obtain new results, such as
calculations in modified configurations of the above mentioned $t$-electric
potential steps and a detailed consideration of new limiting cases
(asymptotics) in already studied before $t$-electric potential steps.
Considering all the cases, we tried to cite properly all previous relevant
works. In Sec. \ref{S2}, we briefly recall basic formulas for treatment of
zero-order processes in the framework of QED (and scalar QED) with $t$%
-electric potential steps. In\textbf{\ }Sets. \ref{S4}, \ref{S5} and \ref{S6}%
, we study quantum effects in the three most important exactly solvable
cases of $t$-electric potential steps, in their modified configurations, and
calculate carefully important limiting cases. As was recently discover in
our work \cite{GavGit17}, an information derived from considerations of
exactly solvable cases allows one to make some general conclusions about
quantum effects in slowly varying strong fields for which no closed form
solutions of the Dirac equation are known. In Sec. \ref{S7}, we briefly
represent such conclusions about an universal behavior of vacuum mean values
in slowly varying strong electric fields. Some asymptotic expansions are
placed in the Appendix \ref{Ap}. In the near future, we hope to present a
similar work about quantum effects in external electromagnetic field that
correspond to exactly solvable cases of $x$-electric potential steps.

\section{Vacuum instability description based on exact solutions \label{S2}}

Potentials $A^{\mu }\left( x\right) ,$ $x=(x^{\mu })=(x^{0}=t,\mathbf{r}),\;%
\mathbf{r}=(x^{i})$ of external electromagnetic fields\footnote{%
The Greek indexes span the Minkowisky space-time, $\mu =0,1,\dots ,D$, and
the Latin indexes span the Euclidean space, $i=1,\ldots ,D$. In what
follows, we use the system of units where $\hslash =c=1$.
\par
{}} corresponding to $t$-electric potential steps are defined as%
\begin{equation}
A^{0}=0\,,\ \ \mathbf{A}\left( t\right) =(A^{1}=A_{x}\left( t\right) \,,\ \
A^{l}=0\,,\ \ l=2,...,D),\ A_{x}\left( t\right) \overset{t\rightarrow \pm
\infty }{\longrightarrow }A_{x}\left( \pm \infty \right) ,  \label{2.6}
\end{equation}%
where $A_{x}\left( \pm \infty \right) $ are some constant quantities, and
the time derivative of the potential $A_{x}\left( t\right) $ does not change
its sign for any $t\in \mathbb{R}.$ For definiteness, we suppose that%
\begin{equation}
\dot{A}_{x}\left( t\right) \leq 0\Longrightarrow A_{x}\left( -\infty \right)
>A_{x}\left( +\infty \right) .  \label{2.7}
\end{equation}%
The magnetic field is always zero and electric fields are homogeneous and
have the form%
\begin{equation}
\mathbf{E}\left( t\right) =\left( E_{x}\left( t\right) ,0,...,0\right) ,\ \
E_{x}\left( t\right) =-\dot{A}_{x}\left( t\right) =E\left( t\right) \geq
0,\,E\left( t\right) \overset{\left\vert t\right\vert \rightarrow \infty }{%
\longrightarrow }0.\,  \label{2.4}
\end{equation}%
We stress that electric fields under consideration are switched off as $%
\left\vert t\right\vert \rightarrow \infty $ and do not have local minima.

As an example of a $t$-electric potential step, we refer\ to the so-called
Sauter-like (or adiabatic or pulse) electric field. This field and its
vector potential have the form%
\begin{equation}
E\left( t\right) =E\cosh ^{-2}\left( t/T_{\mathrm{S}}\right) \,,\ \
A_{x}\left( t\right) =-T_{\mathrm{S}}E\tanh \left( t/T_{\mathrm{S}}\right)
\,.  \label{2.8}
\end{equation}%
where the parameter $T_{\mathrm{S}}>0$ sets time scale.

The Dirac equation (in the Hamiltonian form) in $(d=D+1)$-dimensional
Minkowski space-time and with an external electromagnetic field of the form (%
\ref{2.6}) reads%
\begin{eqnarray}
&&i\partial _{t}\psi \left( x\right) =H\left( t\right) \psi \left( x\right)
\,,\ \ H\left( t\right) =\gamma ^{0}\left( \boldsymbol{\gamma }\mathbf{P}%
+m\right) ,  \notag \\
&&\,P_{x}=-i\partial _{x}-U\left( t\right) ,\ \ \mathbf{P}_{\bot }=-i%
\boldsymbol{\nabla }_{\perp },\ \ U\left( t\right) =qA_{x}\left( t\right) \,,
\label{2.9}
\end{eqnarray}%
where $H\left( t\right) $ is the one-particle Dirac Hamiltonian; $\psi (x)$
is a $2^{[d/2]}$-component spinor; $[d/2]$ stands for the integer part of $%
d/2$; $m\neq 0$ is the electron mass; the index $\perp $ stands for
components of the momentum operator that are perpendicular to the electric
field. Here, $\gamma ^{\mu }$ are the $\gamma $-matrices in $d$ dimensions
\cite{BraWey35},
\begin{equation*}
\lbrack \gamma ^{\mu },\gamma ^{\nu }]_{+}=2\eta ^{\mu \nu },\;\;\eta ^{\mu
\nu }=\mathrm{diag}(\underbrace{1,-1,-1,\ldots }_{d}).
\end{equation*}%
The number of spin degree of freedom is $J_{(d)}=2^{\left[ d/2\right] -1}$.

We choose the electron as the main particle with the charge $q=-e$, where $%
e>0$ is the absolute value of the electron charge, and we refer to $U\left(
t\right) $\ as the potential energy of an electron in the electric field.

Let us consider solutions of Dirac equation (\ref{2.9}) of the following the
form%
\begin{eqnarray}
&&\psi _{n}\left( x\right) =\exp \left( i\mathbf{pr}\right) \psi _{n}\left(
t\right) ,\;\ n=(\mathbf{p},\sigma ),  \notag \\
&&\psi _{n}\left( t\right) =\left\{ \gamma ^{0}i\partial _{t}-\gamma ^{1}
\left[ p_{x}-U\left( t\right) \right] -\boldsymbol{\gamma }\mathbf{p}_{\bot
}+m\right\} \phi _{n}(t)\,,  \label{2.10}
\end{eqnarray}%
where $\psi _{n}\left( t\right) $ and $\phi _{n}(t)$ are time-dependent
spinors. In fact, these are states with a definite momentum $\mathbf{p}%
=\left( p_{x},\mathbf{p}_{\bot }\right) $. We can separate spin variables by
the substitution
\begin{equation}
\phi _{n}(t)=\varphi _{n}\left( t\right) v_{\chi ,\sigma },\ \chi =\pm 1,\
\sigma =(\sigma _{1},\sigma _{2},\dots ,\sigma _{\lbrack d/2]-1}),\ \ \sigma
_{s}=\pm 1,  \label{2.12}
\end{equation}%
where $v_{\chi ,\sigma }$ is a set of constant orthonormalized spinors,
satisfying the following equations:%
\begin{equation}
\gamma ^{0}\gamma ^{1}v_{\chi ,\sigma }=\chi v_{\chi ,\sigma },\;\ v_{\chi
,\sigma }^{\dag }v_{\chi ^{\prime },\sigma ^{\prime }}=\delta _{\chi ,\chi
^{\prime }}\delta _{\sigma ,\sigma ^{\prime }}\ .  \label{e2a}
\end{equation}%
In the dimensions $d>3$, one can subject the spinors $v_{\chi }$ to some
supplementary conditions, which, for example, may be chosen as%
\begin{eqnarray}
&&i\gamma ^{2s}\gamma ^{2s+1}v_{\chi ,\sigma }=\sigma _{s}v_{\chi ,\sigma
},\ \mathrm{for}\ \mathrm{even\ }d,  \notag \\
&&i\gamma ^{2s+1}\gamma ^{2s+2}v_{\chi ,\sigma }=\sigma _{s}v_{\chi ,\sigma
},\ \mathrm{for\ odd}\ d.  \label{e2.5}
\end{eqnarray}%
Quantum numbers $\sigma _{s}$ describe the spin polarization [in the
dimensions $d=2,3$ there are no spin degrees of freedom that are described
by the quantum numbers $\sigma $], and, together with the additional index $%
\chi $, provide a convenient parametrization of the solutions. Then the
scalar functions $\varphi _{n}\left( t\right) $ have to obey the second
order differential equation
\begin{equation}
\left\{ \frac{d^{2}}{dt^{2}}+\left[ p_{x}-U\left( t\right) \right] ^{2}+\pi
_{\perp }^{2}-i\chi \dot{U}\left( t\right) \right\} \varphi _{n}\left(
t\right) =0\,,\;\pi _{\perp }=\sqrt{\mathbf{p}_{\perp }^{2}+m^{2}}.
\label{t3}
\end{equation}

The quantization of the Dirac field in the background under consideration is
based on the existence of solutions to the Dirac equation with special
asymptotics as $t\rightarrow \pm \infty $, see \cite{FGS,GavGT06} for
details. For instance, we let the electric field be switched on at $t_{%
\mathrm{in}}$ and switched off at $t_{\mathrm{out}}$, so that the
interaction between the Dirac field and the electric field vanishes at all
time instants outside the interval $t\in \left( t_{\mathrm{in}},t_{\mathrm{%
out}}\right) $, and the Dirac equation in the Hamiltonian form is given by%
\begin{eqnarray}
&&\left[ i\partial _{t}-H\left( t\right) \right] \ _{\zeta }\psi _{n}\left(
x\right) =0\,,\ \ t\in \left( -\infty ,t_{\mathrm{in}}\right]  \notag \\
&&\left[ i\partial _{t}-H\left( t\right) \right] \ ^{\zeta }\psi _{n}\left(
x\right) =0\,,\ \ t\in \left[ t_{\mathrm{out}},+\infty \right) \,.
\label{t3.1}
\end{eqnarray}%
where the additional quantum number $\zeta =\pm $ labels asymptotic states,
respectively. These asymptotic states are solutions of eigenvalue problems,%
\begin{eqnarray}
&&H\left( t\right) \ _{\zeta }\psi _{n}\left( x\right) =\ _{\zeta
}\varepsilon _{n}\ _{\zeta }\psi _{n}\left( x\right) \,,\ \ t\in \left(
-\infty ,t_{\mathrm{in}}\right] \,,\ _{\zeta }\varepsilon _{n}=\zeta
p_{0}\left( t_{\mathrm{in}}\right) \,,  \notag \\
&&H\left( t\right) \ ^{\zeta }\psi _{n}\left( x\right) =\ ^{\zeta
}\varepsilon _{n}\ ^{\zeta }\psi _{n}\left( x\right) \,,\ \ t\in \left[ t_{%
\mathrm{out}},+\infty \right) \,,\ ^{\zeta }\varepsilon _{n}=\zeta
p_{0}\left( t_{\mathrm{out}}\right) \,,  \notag \\
&&p_{0}\left( t\right) =\sqrt{\left[ p_{x}-U\left( t\right) \right] ^{2}+\pi
_{\perp }^{2}}\,,.  \label{t4b}
\end{eqnarray}%
In these asymptotic states, $\zeta =+$ correspond to free electrons and $%
\zeta =-$ corresponds to free positrons. We call the time interval $t\in
\left( -\infty ,t_{\mathrm{in}}\right] $ as the \textrm{in}-region, where
the \textrm{in}-set $\left\{ \ _{\zeta }\psi _{n}\left( x\right) \right\} $\
is defined. The time interval $t\in \left[ t_{\mathrm{out}},+\infty \right) $
is called the \textrm{out}-region, where the \textrm{out}-set $\left\{ \
^{\zeta }\psi _{n}\left( x\right) \right\} $ is defined. In these regions:%
\begin{equation}
\ _{\zeta }\varphi _{n}\left( t\right) =\ _{\zeta }\mathcal{N}e^{-i\ _{\zeta
}\varepsilon _{n}t}\,,\ \ t\in \left( -\infty ,t_{\mathrm{in}}\right] \,,\ \
^{\zeta }\varphi _{n}\left( t\right) =\ ^{\zeta }\mathcal{N}e^{-i\ ^{\zeta
}\varepsilon _{n}t}\,,\ \ t\in \left[ t_{\mathrm{out}},+\infty \right) \,,
\label{t4.1a}
\end{equation}%
where$\ _{\zeta }\mathcal{N},\ ^{\zeta }\mathcal{N},$ are normalization
constants, and there exists an energy gap between the electron and positron
states.

Then we construct two complete set of solutions to the Dirac equation,%
\begin{eqnarray}
\ _{\zeta }\psi _{n}\left( x\right) &=&\exp \left( i\mathbf{pr}\right)
\left\{ \gamma ^{0}i\partial _{t}-\gamma ^{1}\left[ p_{x}-U\left( t\right) %
\right] -\boldsymbol{\gamma }\mathbf{p}_{\bot }+m\right\} \;_{\zeta }\varphi
_{n}\left( t\right) v_{\chi ,\sigma }\,,  \notag \\
\ ^{\zeta }\psi _{n}\left( x\right) &=&\exp \left( i\mathbf{pr}\right)
\left\{ \gamma ^{0}i\partial _{t}-\gamma ^{1}\left[ p_{x}-U\left( t\right) %
\right] -\boldsymbol{\gamma }\mathbf{p}_{\bot }+m\right\} \;^{\zeta }\varphi
_{n}\left( t\right) v_{\chi ,\sigma }\,.  \label{t4.10}
\end{eqnarray}

We suppose additionally that the \textrm{in}- $\left\{ \ _{\zeta }\psi
_{n}\left( x\right) \right\} $ and \textrm{out}-sets $\left\{ \ ^{\zeta
}\psi _{n}\left( x\right) \right\} $ are complete and orthonormal with
respect to the standard definition of the inner product \cite{Schweber},%
\begin{equation}
\left( \psi ,\psi ^{\prime }\right) =\int \psi ^{\dagger }\left( x\right)
\psi ^{\prime }\left( x\right) d\mathbf{r}\,,\ \ d\mathbf{r}%
=dx^{1}...dx^{D}\,.  \label{t4.0}
\end{equation}

In calculating (\ref{t4.0}), we use the standard volume regularization, in
which the scattering problem is confined by a large spatial box of volume $%
V_{\left( d-1\right) }=L_{1}\times \cdots \times L_{D}$, with the Dirac
spinors subject to periodic boundary conditions at the spatial walls. This
inner product is time-independent. Since $\chi $ is not a physical quantum
number if $d>3$ (the spin operator $\gamma ^{0}\gamma ^{1}$ does not commute
with the Dirac Hamiltonian (\ref{2.9}) in case $m\neq 0$),\ one can select a
particular $\chi $ to calculate (\ref{t4.0}). For simplicity, we select the
same $\chi $ in the cases of $\psi (x)$ and $\psi ^{\prime }(x)$, so that
the inner product is simplified,%
\begin{eqnarray}
&&\left( \psi _{n},\psi _{n^{\prime }}\right) =V_{\left( d-1\right) }\delta
_{n,n^{\prime }}\mathcal{I},\ \ \mathcal{I}=\psi _{n}^{\dag }\left( x\right)
\psi _{n^{\prime }}\left( x\right) \,,  \notag \\
&&\mathcal{I}=\varphi _{n}^{\ast }\left( t\right) \left( i\overrightarrow{%
\partial }_{t}-i\overleftarrow{\partial }_{t}\right) \left\{ i%
\overrightarrow{\partial }_{t}-\chi \left[ p_{x}-U\left( t\right) \right]
\right\} \varphi _{n}\left( t\right) \,.  \label{t4.2}
\end{eqnarray}%
Then solutions (\ref{t4.10}) can be subject to the orthonormality conditions%
\begin{equation}
\left( \ _{\zeta }\psi _{n},\ _{\zeta ^{\prime }}\psi _{n^{\prime }}^{\prime
}\right) =\delta _{n,n^{\prime }}\delta _{\zeta ,\zeta ^{\prime }}\ ,\ \
\left( \ ^{\zeta }\psi _{n},\ ^{\zeta ^{\prime }}\psi _{n^{\prime }}^{\prime
}\right) =\delta _{n,n^{\prime }}\delta _{\zeta ,\zeta ^{\prime }}\ ,
\label{t4.3}
\end{equation}%
which, in particular, leads to the following expressions for the
normalization constants:%
\begin{eqnarray}
&&\ _{\zeta }\mathcal{N}\mathcal{=}_{\zeta }CY,\;\;\ ^{\zeta }\mathcal{N}=\
^{\zeta }CY,\;\;Y=V_{\left( d-1\right) }^{-1/2},  \notag \\
&&_{\zeta }C=\left[ 2p_{0}\left( t_{\mathrm{in}}\right) q_{\mathrm{in}%
}^{\zeta }\right] ^{-1/2}\,,\ \ \ ^{\zeta }C=\left[ 2p_{0}\left( t_{\mathrm{%
out}}\right) q_{\mathrm{out}}^{\zeta }\right] ^{-1/2}\,,  \notag \\
&&q_{\mathrm{in}/\mathrm{out}}^{\zeta }=p_{0}\left( t_{\mathrm{in}/\mathrm{%
out}}\right) -\chi \zeta \left[ p_{x}-U\left( t_{\mathrm{in}/\mathrm{out}%
}\right) \right] \,.  \label{t4.3.1}
\end{eqnarray}%
Completeness relations for the \textrm{in}- and \textrm{out}-sets are given
by%
\begin{equation}
\sum_{\zeta ,n}\ _{\zeta }\psi _{n}\left( t,\mathbf{r}\right) \ _{\zeta
}\psi _{n}^{\dagger }\left( t,\mathbf{r}^{\prime }\right) =\delta \left(
\mathbf{r}-\mathbf{r}^{\prime }\right) \mathbb{I}=\sum_{\zeta ,n}\ ^{\zeta
}\psi _{n}\left( t,\mathbf{r}\right) \ ^{\zeta }\psi _{n}^{\dagger }\left( t,%
\mathbf{r}^{\prime }\right) \,.  \label{t4.3.2}
\end{equation}

Due to property (\ref{t4.2}) inner products $\left( \ _{\zeta ^{\prime
}}\psi _{l},\ ^{\zeta }\psi _{n}\right) $ are diagonal in quantum numbers $n$
and $l,$%
\begin{equation}
\left( \ _{\zeta ^{\prime }}\psi _{l},\ ^{\zeta }\psi _{n}\right) =\delta
_{l,n}g\left( _{\zeta ^{\prime }}|^{\zeta }\right) ,\ \ g\left( ^{\zeta
^{\prime }}|_{\zeta }\right) =g\left( _{\zeta ^{\prime }}|^{\zeta }\right)
^{\ast }.  \label{t4.5}
\end{equation}%
The corresponding diagonal matrix elements $g$ relate \textrm{in}- and
\textrm{out}-solutions $\left\{ \ _{\zeta }\psi _{n}\left( x\right) \right\}
$ and $\left\{ \ ^{\zeta }\psi _{n}\left( x\right) \right\} $ for each $n$,%
\begin{eqnarray}
^{\zeta }\psi _{n}\left( x\right) &=&g\left( _{+}|^{\zeta }\right) \
_{+}\psi _{n}\left( x\right) +g\left( _{-}|^{\zeta }\right) \ _{-}\psi
_{n}\left( x\right) \,,  \notag \\
_{\zeta }\psi _{n}\left( x\right) &=&g\left( ^{+}|_{\zeta }\right) \
^{+}\psi _{n}\left( x\right) +g\left( ^{-}|_{\zeta }\right) \ ^{-}\psi
_{n}\left( x\right) \,.  \label{t4.4}
\end{eqnarray}

Substituting Eqs.~(\ref{t4.4}), into the orthonormality conditions, we
derive the unitarity relations%
\begin{equation}
\sum_{\varkappa }g\left( ^{\zeta }|_{\varkappa }\right) g\left( _{\varkappa
}|^{\zeta ^{\prime }}\right) =\sum_{\varkappa }g\left( _{\zeta }|^{\varkappa
}\right) g\left( ^{\varkappa }|_{\zeta ^{\prime }}\right) =\delta _{\zeta
,\zeta ^{\prime }}\,.  \label{3.16.1}
\end{equation}

Similar consideration is possible for a scalar fields that satisfies the
Klein--Gordon (KG) equation with $t$-electric potential step. A formal
transition to the case of scalar fields can be done by setting $\chi =0$\ in
Eq.~(\ref{t3}). The corresponding complete set of solutions to the KG
equation reads%
\begin{equation}
\phi _{n}\left( x\right) =\exp \left( i\mathbf{pr}\right) \varphi _{n}\left(
t\right) ,\;\ n=\mathbf{p}.  \label{KG1}
\end{equation}

One can also define complete \textrm{in}- $\left\{ \ _{\zeta }\phi
_{n}\right\} $ and \textrm{out}-sets $\left\{ \ ^{\zeta }\phi _{n}\right\} $
of solutions orthonormal with respect to the adequate inner product \cite%
{Schweber},%
\begin{equation}
\left( \phi ,\phi ^{\prime }\right) _{\mathrm{KG}}=i\int \phi ^{\ast }\left(
x\right) \left( \overrightarrow{\partial }_{t}-\overleftarrow{\partial }%
_{t}\right) \phi ^{\prime }\left( x\right) d\mathbf{r\,.}  \label{KG5}
\end{equation}%
Namely,%
\begin{equation}
\left( \ _{\zeta }\phi _{n},\ _{\zeta ^{\prime }}\phi _{n^{\prime }}^{\prime
}\right) _{\mathrm{KG}}=\zeta \delta _{n,n^{\prime }}\ \delta _{\zeta ,\zeta
^{\prime }}\ ,\ \ \left( \ ^{\zeta }\phi _{n},\ ^{\zeta ^{\prime }}\phi
_{n^{\prime }}^{\prime }\right) _{\mathrm{KG}}=\zeta \delta _{n,n^{\prime
}}\ \delta _{\zeta ,\zeta ^{\prime }}\ ,  \label{KG6}
\end{equation}%
with the normalization constants%
\begin{equation}
_{\zeta }C=\left[ 2p_{0}\left( t_{\mathrm{in}}\right) \right] ^{-1/2}\,,\ \
\ ^{\zeta }C=\left[ 2p_{0}\left( t_{\mathrm{out}}\right) \right] ^{-1/2}\,.
\label{KG8}
\end{equation}

Inner products $\ \left( \ _{\zeta ^{\prime }}\phi _{l},\ ^{\zeta }\phi
_{n}\right) _{\mathrm{KG}}$ are diagonal in quantum numbers $n$ and $l,$
\begin{equation}
\left( \ _{\zeta ^{\prime }}\phi _{l},\ ^{\zeta }\phi _{n}\right) _{\mathrm{%
KG}}=\delta _{l,n}g\left( _{\zeta ^{\prime }}|^{\zeta }\right) ,\ \ g\left(
^{\zeta ^{\prime }}|_{\zeta }\right) =g\left( _{\zeta ^{\prime }}|^{\zeta
}\right) ^{\ast }.  \label{kG8.3}
\end{equation}%
The corresponding diagonal matrix elements $g$ relate \textrm{in}- and
\textrm{out}-solutions,
\begin{eqnarray}
\ ^{\zeta }\phi _{n}\left( x\right) &=&g\left( _{+}|^{\zeta }\right) \
_{+}\phi _{n}\left( x\right) -g\left( _{-}|^{\zeta }\right) \ _{-}\phi
_{n}\left( x\right) \,,  \notag \\
\ _{\zeta }\phi _{n}\left( x\right) &=&g\left( ^{+}|_{\zeta }\right) \
^{+}\phi _{n}\left( x\right) -g\left( ^{-}|_{\zeta }\right) \ ^{-}\phi
_{n}\left( x\right) \,,  \label{KG8.2}
\end{eqnarray}%
and satisfy the unitarity relations%
\begin{equation}
\sum_{\varkappa }g\left( ^{\zeta }|_{\varkappa }\right) \varkappa g\left(
_{\varkappa }|^{\zeta ^{\prime }}\right) =\sum_{\varkappa }g\left( _{\zeta
}|^{\varkappa }\right) \varkappa g\left( ^{\varkappa }|_{\zeta ^{\prime
}}\right) =\zeta \delta _{\zeta ,\zeta ^{\prime }}\,.  \label{3.16.2}
\end{equation}

Decomposing the Dirac operator $\hat{\Psi}(x)$ in the complete sets of in-
and out-solutions \cite{FGS,GavGT06},%
\begin{equation}
\hat{\Psi}\left( x\right) =\sum_{n}\left[ a_{n}(\mathrm{in})\ _{+}\psi
_{n}(x)+b_{n}^{\dag }(\mathrm{in})\ _{\_}\psi _{n}(x)\right] =\sum_{n}\left[
a_{n}(\mathrm{out})\ ^{+}\psi _{n}(x)+b_{n}^{\dag }(\mathrm{out})\ ^{-}\psi
_{n}(x)\right] \,,  \label{3.1}
\end{equation}%
we introduce \textrm{in-} and \textrm{out-}creation and annihilation Fermi
operators. Their nonzero anticommutation relations are,%
\begin{equation}
\lbrack a_{n}(\mathrm{in}),a_{m}^{\dag }(\mathrm{in})]_{+}=[a_{n}(\mathrm{out%
}),a_{m}^{\dag }(\mathrm{out})]_{+}=[b_{n}(\mathrm{in}),b_{m}^{\dag }(%
\mathrm{in})]_{+}=[b_{n}(\mathrm{out}),b_{m}^{\dag }(\mathrm{out}%
)]_{+}=\delta _{nm}\,.  \label{3.4}
\end{equation}%
In these terms, the Heisenberg Hamiltonian is diagonalized at $t\leq t_{%
\mathrm{in}}$ and $t\geq t_{\mathrm{out}}\ ,$%
\begin{eqnarray}
&&\widehat{\mathbb{H}}(t)=\sum_{n}\left\{ \ _{+}\varepsilon _{n}a_{n}^{+}(%
\mathrm{in})a_{n}(\mathrm{in})+\left\vert \ _{-}\varepsilon _{n}\right\vert
b_{n}^{+}(\mathrm{in})b_{n}(\mathrm{in})\right\} \,,\ \ t\leq t_{\mathrm{in}%
}\ ,  \notag \\
&&\widehat{\mathbb{H}}(t)=\sum_{n}\left\{ \;^{+}\varepsilon _{n}a_{n}^{+}(%
\mathrm{out})a_{n}(\mathrm{out})+\left\vert \ ^{-}\varepsilon
_{n}\right\vert b_{n}^{+}(\mathrm{out})b_{n}(\mathrm{out})\right\} \,,\ \
t\geq t_{\mathrm{out}}\ ,  \label{3.5}
\end{eqnarray}%
where the diverging c-number parts have been omitted, as usual. The initial $%
|0,$\textrm{$in$}$\rangle $ and final $|0,$\textrm{$out$}$\rangle $ vacuum
vectors, as well as many-particle \textrm{in}\textbf{-} and \textrm{out}%
-states, are defined by%
\begin{eqnarray}
&&\ a_{n}(\mathrm{in})|0,\mathrm{in}\rangle =b_{n}(\mathrm{in})|0,\mathrm{in}%
\rangle =0,\ a_{n}(\mathrm{out})|0,\mathrm{out}\rangle =b_{n}(\mathrm{out}%
)|0,\mathrm{out}\rangle =0\,,  \notag \\
&&\ |\mathrm{in}\rangle =b_{n}^{+}(\mathrm{in})...a_{n}^{+}(\mathrm{in}%
)...|0,\mathrm{in}\rangle ,\ \ |\mathrm{out}\rangle =b_{n}^{+}(\mathrm{out}%
)...a_{n}^{+}(\mathrm{out})...|0,\mathrm{out}\rangle \,.  \label{3.6a}
\end{eqnarray}%
Using the charge operator one can see that $a_{n}^{\dag }$, $a_{n}$ are the
creation and annihilation operators of electrons, whereas $b_{n}^{\dag }$, $%
b_{n}$ are the creation and annihilation operators of positrons,
respectively.

Transition amplitudes in the Heisenberg representation have the form $M_{%
\mathrm{in}\rightarrow \mathrm{out}}=\langle \mathrm{out}|\mathrm{in}\rangle
\,.$ In particular, the vacuum-to-vacuum transition amplitude reads $%
c_{v}=\langle 0,\mathrm{out}|0,\mathrm{in}\rangle .$ Relative probability
amplitudes of particle scattering, pair creation and annihilation are:%
\begin{eqnarray}
&&w\left( +|+\right) _{n^{\prime }n}=c_{v}^{-1}\langle 0,\mathrm{out}%
\left\vert a_{n^{\prime }}\left( \mathrm{out}\right) a_{n}^{\dagger }(%
\mathrm{in})\right\vert 0,\mathrm{in}\rangle =\delta _{n,n^{\prime
}}w_{n}\left( +|+\right) ,  \notag \\
&&w\left( -|-\right) _{n^{\prime }n}=c_{v}^{-1}\langle 0,\mathrm{out}%
\left\vert b_{n^{\prime }}\left( \mathrm{out}\right) b_{n}^{\dagger }(%
\mathrm{in})\right\vert 0,\mathrm{in}\rangle =\delta _{n,n^{\prime
}}w_{n}\left( -|-\right) \,,  \notag \\
&&w\left( +-|0\right) _{n^{\prime }n}=c_{v}^{-1}\langle 0,\mathrm{out}%
\left\vert a_{n^{\prime }}\left( \mathrm{out}\right) b_{n}\left( \mathrm{out}%
\right) \right\vert 0,\mathrm{in}\rangle =\delta _{n,n^{\prime }}w_{n}\left(
+-|0\right) \,,  \notag \\
&&w\left( 0|-+\right) _{nn^{\prime }}=c_{v}^{-1}\langle 0,\mathrm{out}%
\left\vert b_{n}^{\dagger }(\mathrm{in})a_{n^{\prime }}^{\dagger }(\mathrm{in%
})\right\vert 0,\mathrm{in}\rangle \,=\delta _{n,n^{\prime }}w_{n}\left(
0|-+\right) .  \label{3.15}
\end{eqnarray}

The \textrm{in}\textbf{-} and \textrm{out}-operators are related by linear
canonical transformations,%
\begin{equation}
a_{n}(\mathrm{out})=g(^{+}|_{+})a_{n}(\mathrm{in})+g(^{+}|_{-})b_{n}^{\dag }(%
\mathrm{in})\,,\ \ b_{n}^{\dag }(\mathrm{out})=g(^{-}|_{+})a_{n}(\mathrm{in}%
)+g(^{-}|_{-})b_{n}^{\dag }(\mathrm{in})\,.  \notag
\end{equation}%
These relations allows one to calculate the differential mean numbers of
electrons $N_{n}^{a}\left( \mathrm{out}\right) $ and positrons $%
N_{n}^{b}\left( \mathrm{out}\right) $ created from the vacuum state as%
\begin{eqnarray}
&&N_{n}^{a}\left( \mathrm{out}\right) =\left\langle 0,\mathrm{in}\left\vert
a_{n}^{\dagger }(\mathrm{out})a_{n}(\mathrm{out})\right\vert 0,\mathrm{in}%
\right\rangle =\left\vert g\left( _{-}\left\vert ^{+}\right. \right)
\right\vert ^{2},  \notag \\
&&N_{n}^{b}\left( \mathrm{out}\right) =\left\langle 0,\mathrm{in}\left\vert
b_{n}^{\dagger }(\mathrm{out})b_{n}(\mathrm{out})\right\vert 0,\mathrm{in}%
\right\rangle =\left\vert g\left( _{+}\left\vert ^{-}\right. \right)
\right\vert ^{2},\ N_{n}^{\mathrm{cr}}=N_{n}^{b}\left( \mathrm{out}\right)
=N_{n}^{a}\left( \mathrm{out}\right) .  \label{3.23}
\end{eqnarray}%
By $N_{n}^{\mathrm{cr}}$\ we denote the differential numbers of created
pairs. The total number of pairs\ created from vacuum is given by the sum%
\begin{equation}
N=\sum_{n}N_{n}^{\mathrm{cr}}=\sum_{n}\left\vert g\left( _{-}\left\vert
^{+}\right. \right) \right\vert ^{2}.  \label{TN}
\end{equation}

Similar consideration is possible for a quantum scalar fields $\hat{\Phi}%
\left( x\right) $\ that satisfies the KG equation with $t$-electric
potential steps. Decomposing the quantum fields $\hat{\Phi}\left( x\right) $%
\ in the complete set of exact solutions $\left\{ _{\pm }\ \phi
_{n}(x)\right\} $\ and $\left\{ ^{\pm }\ \phi _{n}(x)\right\} $\ one \
introduces in- and out-creation and annihilation Bose operators and obtain
quite similar representations of relative probability amplitudes and the
differential numbers of created pairs via the corresponding diagonal matrix
elements $g$\ defined by Eq.~(\ref{kG8.3}) \cite{FGS,GavGT06}.

Both for fermions and bosons,{\large \ }relative probabilities (\ref{3.15}),
the vacuum-to-vacuum transition amplitude $c_{v}$, the probability for a
vacuum to remain a vacuum $P_{v}$ as well as the total number $N$ of pairs\
created from vacuum can be expressed via the distribution $N_{n}^{\mathrm{cr}%
}$,%
\begin{eqnarray}
&&\left\vert w_{n}\left( +-|0\right) \right\vert ^{2}=N_{n}^{\mathrm{cr}%
}\left( 1-\kappa N_{n}^{\mathrm{cr}}\right) ^{-1},\;\left\vert w_{n}\left(
-|-\right) \right\vert ^{2}=\left( 1-\kappa N_{n}^{\mathrm{cr}}\right) ^{-1},
\notag \\
&&P_{v}=|c_{v}|^{2}=\prod\limits_{n}\left( 1-\kappa N_{n}^{\mathrm{cr}%
}\right) ^{\kappa },\;\kappa =\left\{
\begin{array}{c}
+1\ \mathrm{for\ fermions} \\
-1\ \mathrm{for\ bosons}%
\end{array}%
\right. .  \label{vacprob}
\end{eqnarray}

The vacuum mean electric current, energy, and momentum are defined as
integrals over the spatial volume. Due to the translational invariance in
the uniform external field, all these mean values are proportional to the
space volume. Therefore, it is enough to calculate the vacuum mean values of
the current density vector $\langle j^{\mu }(t)\rangle $ and of the
energy-momentum tensor (EMT) $\langle T_{\mu \nu }(t)\rangle $, defined as
\begin{equation}
\langle j^{\mu }(t)\rangle =\langle 0,\mathrm{in}|j^{\mu }|0,\mathrm{in}%
\rangle ,\ \ \langle T_{\mu \nu }(t)\rangle =\langle 0,\mathrm{in}|T_{\mu
\nu }|0,\mathrm{in}\rangle \,.  \label{int1}
\end{equation}%
Here we stress the time dependence of mean values (\ref{int1}), which does
exist due to a time dependence of the external field. We recall for further
convenience the form of the operators of the current density and the EMT of
the quantum Dirac field,
\begin{align}
& j^{\mu }=\frac{q}{2}\left[ \overline{\hat{\Psi}}(x),\gamma ^{\mu }\hat{\Psi%
}\left( x\right) \right] \,,\quad T_{\mu \nu }=\frac{1}{2}(T_{\mu \nu
}^{can}+T_{\nu \mu }^{can})\,,  \notag \\
& T_{\mu \nu }^{can}=\frac{1}{4}\left\{ [\overline{\hat{\Psi}}(x),\gamma
_{\mu }P_{\nu }\hat{\Psi}\left( x\right) ]+[P_{\nu }^{\ast }\overline{\hat{%
\Psi}}(x),\gamma _{\mu }\hat{\Psi}\left( x\right) ]\right\} \,,  \notag \\
& P_{\mu }=i\partial _{\mu }-qA_{\mu }(x),\ \overline{\hat{\Psi}}(x)=\hat{%
\Psi}^{\dagger }\left( x\right) \gamma ^{0}.  \label{A1.0}
\end{align}

Note that the mean values (\ref{int1}) depend on the definition of the
initial vacuum, $|0,\mathrm{in}\rangle $ and on the evolution of the
electric field from the time $t_{\mathrm{in}}$ of switching it on up to the
current time instant $t$, but they do not depend on the further history of
the system. The renormalized vacuum mean values $\langle j^{\mu }(t)\rangle $
and $\langle T_{\mu \nu }(t)\rangle ,$ $t_{\mathrm{in}}$ $<t<$ $t_{\mathrm{%
out}}$ are sources in equations of motion for mean electromagnetic and
metric fields, respectively. In particular, complete description of the back
reaction is related to the calculation of these mean values for any $t$.

Mean values and probability amplitudes are calculated by the help of
different kind of propagators. The probability amplitudes are calculated
using Feynman diagrams with the causal (Feynman) propagator%
\begin{equation}
S^{c}(x,x^{\prime })=i\langle 0,\mathrm{out}|\hat{T}\hat{\Psi}\left(
x\right) \hat{\Psi}^{\dagger }\left( x^{\prime }\right) \gamma ^{0}|0,%
\mathrm{in}\rangle c_{v}^{-1}\,,  \label{A1.3}
\end{equation}%
where $\hat{T}$ denotes the chronological ordering operation. A perturbation
theory (with respect to radiative processes) uses the so-called \textrm{in-in%
} propagator $S_{\mathrm{in}}^{c}(x,x^{\prime })$ and $S^{p}(x,x^{\prime })$
propagator,
\begin{equation}
S_{\mathrm{in}}^{c}(x,x^{\prime })=i\langle 0,\mathrm{in}|\hat{T}\hat{\Psi}%
\left( x\right) \hat{\Psi}^{\dagger }\left( x^{\prime }\right) \gamma ^{0}|0,%
\mathrm{in}\rangle ,\ \ S^{p}(x,x^{\prime })=S_{\mathrm{in}}^{c}(x,x^{\prime
})-S^{c}(x,x^{\prime }).  \label{A1.1}
\end{equation}

All the above propagators can be expressed via the \textrm{in}- and \textrm{%
out}-solution as follows:%
\begin{eqnarray}
S^{c}\left( x,x^{\prime }\right) &=&i\left\{
\begin{array}{c}
\sum\limits_{n}\ ^{+}\psi _{n}(x)\omega _{n}(+|+)\ _{+}\bar{\psi}%
_{n}(x^{\prime }),\ \ t>t^{\prime } \\
-\sum\limits_{n}\ _{-}\psi _{n}(x)\omega _{n}(-|-)\ ^{-}\bar{\psi}%
_{n}(x^{\prime }),\ \ t<t^{\prime }%
\end{array}%
\right. \,,  \label{3.22} \\
S_{\mathrm{in}}^{c}\left( x,x^{\prime }\right) &=&i\left\{
\begin{array}{c}
\sum\limits_{n}\ _{+}\psi _{n}(x)\ _{+}\bar{\psi}_{n}(x^{\prime }),\ \
t>t^{\prime } \\
-\sum\limits_{n}\ _{-}\psi _{n}(x)\ _{-}\bar{\psi}_{n}(x^{\prime }),\ \
t<t^{\prime }%
\end{array}%
\right. \,,\ \ S^{p}(x,x^{\prime })=\ -i\sum_{n}\,_{-}{\psi }%
_{n}(x)w_{n}\left( 0|-+\right) \,{_{+}\bar{\psi}}_{n}(x^{\prime })\,.
\label{3.25}
\end{eqnarray}

The mean values of the operator (\ref{A1.0}) are expressed via the latter
propagators as%
\begin{align}
& \langle j^{\mu }(t)\rangle =\mathrm{Re}\,\langle j^{\mu }(t)\rangle ^{c}+%
\mathrm{Re}\,\langle j^{\mu }(t)\rangle ^{p}\,,\ \ \langle j^{\mu
}(t)\rangle ^{c,p}=iq\left. \mathrm{tr}\left[ \gamma ^{\mu
}S^{c,p}(x,x^{\prime })\right] \right\vert _{x=x^{\prime }}\,,  \notag \\
& \langle T_{\mu \nu }(t)\rangle =\mathrm{Re}\,\langle T_{\mu \nu
}(t)\rangle ^{c}+\mathrm{Re}\,\langle T_{\mu \nu }(t)\rangle ^{p}\,,\ \
\langle T_{\mu \nu }(t)\rangle ^{c,p}=i\left. \mathrm{tr}\left[ A_{\mu \nu
}S^{c,p}(x,x^{\prime })\right] \right\vert _{x=x^{\prime }}\,,  \notag \\
& A_{\mu \nu }=1/4\left[ \gamma _{\mu }\left( P_{\nu }+P_{\nu }^{\prime \ast
}\right) +\gamma _{\nu }\left( P_{\mu }+P_{\mu }^{\prime \ast }\right) %
\right] \,.  \label{A1.4}
\end{align}%
Here $\mathrm{tr}$ stands for the trace in the $\gamma $-matrices indices
and the limit $x\rightarrow x^{\prime }$ is understood as follows:
\begin{equation*}
\mathrm{tr}[R(x,x^{\prime })]_{x=x^{\prime }}=\frac{1}{2}\left[
\lim_{t\rightarrow t^{\prime }-0}\mathrm{tr}[R(x,x^{\prime
})]+\lim_{t\rightarrow t^{\prime }+0}\mathrm{tr}\left[ R(x,x^{\prime })%
\right] \right] _{\mathbf{x=x}^{\prime }},
\end{equation*}%
where $R(x,x^{\prime })$ is any two point matrix function.

The function $S^{p}(x,y)$ vanishes in the case of a stable vacuum. In this
case and only in this case $\langle j^{\mu }(t)\rangle =\mathrm{Re}\,\langle
j^{\mu }(t)\rangle ^{c}\ ,\ \ \ \langle T_{\mu \nu }(t)\rangle =\mathrm{Re}%
\,\langle T_{\mu \nu }(t)\rangle ^{c}.$

\section{Sauter-like electric field\label{S4}}

Here we consider quantum effects in a $t$-electric potential step which is
formed by the so-called Sauter-like electric field given by Eq. (\ref{2.8}).
The origin of the name of the field is the following: In his pioneer work
\cite{Sauter-pot} Sauter studied the Klein paradox considering the case of
an inhomogeneous field given by the $x$-electric potential step $-LE\tanh
\left( x/L\right) $,\ which is\ called at present the Sauter potential. The
homogeneous $t$-electric step which we are going to considered here has
similar form in $t$ coordinate. In a sense, solutions of the Dirac and KG
equations with the Sauter-like potential are formally similar to solutions
for the Sauter potential.

For the Sauter-like external field, the scalar functions $\varphi _{n}\left(
t\right) $ (see the previous section) satisfy equation (\ref{t3}) with $%
U\left( t\right) =T_{\mathrm{S}}eE\tanh \left( t/T_{\mathrm{S}}\right) $.
This field switches-on and -off adiabatically at $t_{\mathrm{in}}\rightarrow
-\infty $ and $t_{\mathrm{out}}\rightarrow +\infty $. In \textrm{in}- and
\textrm{out}-regions, Dirac spinors $\ _{\zeta }\psi _{n}\left( x\right) $
and $\ ^{\zeta }\psi _{n}\left( x\right) $ are solutions of the eigenvalue
problem (\ref{t4b}) and the plane-wave frequencies are%
\begin{equation}
\omega _{\pm }=p_{0}\left( \pm \infty \right) =\sqrt{\left( p_{x}\mp T_{%
\mathrm{S}}eE\right) ^{2}+\pi _{\perp }^{2}}\,.  \label{ta4}
\end{equation}

In the case under consideration, Eq. (\ref{t3}) is an equation for a
hypergeometric function \cite{BatE53}. Its solutions can be written as%
\begin{equation*}
\varphi _{n}\left( t\right) =y^{l}\left( 1-y\right) ^{m}f\left( y\right)
\,,\;\;y=\frac{1}{2}\left[ 1+\tanh \left( t/T_{\mathrm{S}}\right) \right] \,,
\end{equation*}%
where $l$ and $m$ are some constants, and the function $f\left( y\right) $
is a solution of the Gauss hypergeometric differential equation \cite{BatE53}%
. We will use complete sets of solutions $\ _{\zeta }\varphi _{n}\left(
t\right) $ and $\ ^{\zeta }\varphi _{n}\left( t\right) $,%
\begin{eqnarray}
&&_{\;\zeta }\varphi _{n}\left( t\right) =\;_{\zeta }\mathcal{N}\exp \left(
-i\zeta \omega _{-}t\right) \left[ 1+e^{2t/T_{\mathrm{S}}}\right] ^{\frac{%
iT_{\mathrm{S}}}{2}\left( \zeta \omega _{-}-\omega _{+}\right) }\ _{\zeta
}u\left( t\right) \,,  \notag \\
&&\ _{+}u\left( t\right) =F\left( a,b;c;y\right) ,\ _{-}u\left( t\right)
=F\left( a+1-c,b+1-c;2-c;y\right) \,;  \notag \\
&&\ ^{\zeta }\varphi _{n}\left( t\right) =\;^{\zeta }\mathcal{N}\exp \left(
-i\zeta \omega _{+}t\right) \left[ 1+e^{-2t/T_{\mathrm{S}}}\right] ^{\frac{%
iT_{\mathrm{S}}}{2}\left( \omega _{-}-\zeta \omega _{+}\right) }\ ^{\zeta
}u\left( t\right) \,,  \notag \\
&&\ ^{+}u\left( t\right) =F\left( a,b;a+b+1-c;1-y\right) \,,  \notag \\
&&\ ^{-}u\left( t\right) =F\left( c-a,c-b;c+1-a-b;1-y\right) \,,  \notag \\
&&a=\frac{iT_{\mathrm{S}}}{2}\left( \omega _{+}-\omega _{-}\right) +\frac{1}{%
2}+\left( \frac{1}{4}-\left( eET_{\mathrm{S}}^{2}\right) ^{2}-i\chi eET_{%
\mathrm{S}}^{2}\right) ^{1/2}\,,\ \ c=1-iT_{\mathrm{S}}\omega _{-}\,,  \notag
\\
&&b=\frac{iT_{\mathrm{S}}}{2}\left( \omega _{+}-\omega _{-}\right) +\frac{1}{%
2}-\left( \frac{1}{4}-\left( eET_{\mathrm{S}}^{2}\right) ^{2}-i\chi eET_{%
\mathrm{S}}^{2}\right) ^{1/2}\,,  \label{ta6.2.2}
\end{eqnarray}%
where $F\left( a,b;c;y\right) $ is the hypergeometric series in the variable
$y$ with the normalization $F\left( a,b;c;0\right) =1$ \cite{BatE53}. As was
already mentioned in Sec. \ref{S2}, the quantity $\chi $ can be chosen to be
either $\chi =+1$ or $\chi =-1$, and $\ _{\zeta }\mathcal{N}$ and $\ ^{\zeta
}\mathcal{N}$ are normalization factors given by Eq.~(\ref{t4.3.1}).

A formal transition to the Bose case can be done by setting $\chi =0$\ in
Eqs. (\ref{ta6.2.2}). In this case $n=\mathbf{p}$ and $\ _{\zeta }\mathcal{N}
$ and $\ ^{\zeta }\mathcal{N}$ are normalization factors given by Eq.~(\ref%
{KG8}).

In Fermi case, using Kummer's relations and Eqs.~(\ref{t4.5}), one can find
coefficients $g\left( _{+}\left\vert ^{-}\right. \right) ^{\ast }$ to be%
\begin{equation}
g\left( _{+}\left\vert ^{-}\right. \right) ^{\ast }=\frac{\ _{+}C\ \Gamma
\left( c\right) \Gamma \left( a+b-c\right) }{\ ^{-}C\ \Gamma \left( a\right)
\Gamma \left( b\right) }\,,  \label{ta7}
\end{equation}%
where $\ _{+}C$ and $\ ^{-}C$ are constants given by Eq.~(\ref{t4.3.1}) and $%
\Gamma (a)$ is the Euler gamma function. Then, using Eq.~(\ref{3.23}), we
obtain the mean number of created pairs,%
\begin{equation}
N_{n}^{\mathrm{cr}}=\frac{\sinh \left\{ \pi T_{\mathrm{S}}\left[ eET_{%
\mathrm{S}}+\frac{1}{2}\left( \omega _{+}-\omega _{-}\right) \right]
\right\} \sinh \left\{ \pi T_{\mathrm{S}}\left[ eET_{\mathrm{S}}-\frac{1}{2}%
\left( \omega _{+}-\omega _{-}\right) \right] \right\} }{\sinh \left( \pi T_{%
\mathrm{S}}\omega _{+}\right) \sinh \left( \pi T_{\mathrm{S}}\omega
_{-}\right) }\,.  \label{ta9}
\end{equation}%
In 3+1 QED the corresponding formula was found first in \cite{NarNik70}.

In the similar manner, in the Bose case, we obtain coefficients $g\left(
_{+}\left\vert ^{-}\right. \right) ^{\ast }$, where $\ _{+}C$ and $\ ^{-}C$
are given by Eqs.~(\ref{KG8}) and parameters $a,$ $b,$ and $c$ are given by
Eq.~(\ref{ta6.2.2}) at $\chi =0$. Here, the mean number for created pairs is%
\begin{equation}
N_{n}^{\mathrm{cr}}=\frac{\cosh ^{2}\left[ \pi \sqrt{\left( T_{\mathrm{S}%
}^{2}eE\right) ^{2}-\frac{1}{4}}\right] +\sinh ^{2}\left[ \frac{\pi T_{%
\mathrm{S}}}{2}\left( \omega _{+}-\omega _{-}\right) \right] }{\sinh \left(
\pi T_{\mathrm{S}}\omega _{+}\right) \sinh \left( \pi T_{\mathrm{S}}\omega
_{-}\right) }\,.  \label{ta10}
\end{equation}

It should be noted that mean numbers (\ref{ta9}) and (\ref{ta10}) are even
functions of all the momentum $\mathbf{p}$. In particular it can be seen
that $p_{x}\rightarrow -p_{x}$ leads to $\omega _{+}\rightarrow \omega _{-}$%
. The most important parameter in the case under consideration is $T_{%
\mathrm{S}}$. Its value determines the effective time of electric field
action.

We begin our analysis by considering $T_{\mathrm{S}}$ small and constant
values for the asymptotic

potentials,  $U\left( +\infty \right) =-U\left( -\infty \right) =\mathbb{U}%
/2=eET_{\mathrm{S}}$. In this case we deal with a very short pulse field.
The corresponding potential imitates sufficiently well a $t$-electric
rectangular potential step, and coincides with the latter as $T_{\mathrm{S}%
}\rightarrow 0$. Thus, the Sauter-like potential can be considered as a
regularization of the rectangular step. We assume that sufficiently small $%
T_{\mathrm{S}}$ for given $\omega _{\pm }$ satisfies the inequalities%
\begin{equation}
\mathbb{U}T_{\mathrm{S}}\ll 1,\;\;\max \left\{ T_{\mathrm{S}}\omega _{+},T_{%
\mathrm{S}}\omega _{-}\right\} \ll 1\,.  \label{ta10.1}
\end{equation}%
In such a case mean numbers of created pairs are
\begin{eqnarray}
N_{n}^{\mathrm{cr}} &=&\frac{\mathbb{U}^{2}-\left( \omega _{+}-\omega
_{-}\right) ^{2}}{4\omega _{+}\,\omega _{-}}\,\mathrm{\ in\ Fermi\ case,}
\label{ta10.2} \\
N_{n}^{\mathrm{cr}} &=&\frac{T_{\mathrm{S}}^{2}\mathbb{U}^{4}/4+\left(
\omega _{+}-\omega _{-}\right) ^{2}}{4\omega _{+}\,\omega _{-}}\,\mathrm{\
in\ Bose\ case.}  \label{ta10.3}
\end{eqnarray}%
The number of created fermions in Eq.~(\ref{ta10.2}) does not depend on $T_{%
\mathrm{S}}$. However, in contrast with the Fermi case, the limit $T_{%
\mathrm{S}}\rightarrow 0$ in Eq.~(\ref{ta10.3}) is possible only when the
difference $\left( \omega _{+}-\omega _{-}\right) ^{2}$ is not very small,
namely, when%
\begin{equation}
T_{\mathrm{S}}^{2}\mathbb{U}^{4}/4\ll \left( \omega _{+}-\omega _{-}\right)
^{2}\,.  \label{ta10.3b}
\end{equation}%
Only under the latter condition one can neglect an $T_{\mathrm{S}}$%
-depending term in Eq.~(\ref{ta10.3}) to obtain%
\begin{equation}
N_{n}^{\mathrm{cr}}=\frac{\left( \omega _{+}-\omega _{-}\right) ^{2}}{%
4\omega _{+}\,\omega _{-}}\,.  \label{ta10.4}
\end{equation}%
Unlike the Fermi case, where $N_{n}^{\mathrm{cr}}\leq 1$, \ in the Bose
case, the mean number of created particles is unlimited in two ranges of the
longitudinal kinetic momenta, namely when either $\omega _{+}/\omega
_{-}\rightarrow \infty $ or $\omega _{-}/\omega _{+}\rightarrow \infty $,
\begin{equation}
N_{n}^{\mathrm{cr}}\approx \frac{1}{4}\max \left\{ \omega _{+}/\omega
_{-},\omega _{-}/\omega _{+}\right\} \,.  \label{ta10.5}
\end{equation}

One can see that in the Fermi and Bose cases $N_{n}^{\mathrm{cr}}\rightarrow
0$ as $\pi _{\perp }\rightarrow \infty $. On a sufficiently high step and
small transversal momentum, $\pi _{\perp }/\mathbb{U}\ll 1$, one finds that
the maximum mean number of bosons is only limited by the potential
difference $\mathbb{U}$,
\begin{equation*}
\max N_{n}^{\mathrm{cr}}\approx \frac{\mathbb{U}}{4\pi _{\perp }}\,.
\end{equation*}

The maximum mean numbers of fermions $N_{n}^{\mathrm{cr}}\rightarrow 1$ are
in the range of small $\pi _{\perp }$ and $\left\vert p_{x}\right\vert $,
when longitudinal kinetic momenta are large, $\left( p_{x}\mp \mathbb{U}%
/2\right) \sim \mathbb{U}/2$.

The Sauter-like potential is suitable for imitating a slowly alternating
electric field. To this end the parameter $T_{\mathrm{S}}$ is taken to be
sufficiently large. Let us consider just this case, supposing that%
\begin{equation}
T_{\mathrm{S}}\gg \max \left( 1/\sqrt{eE},m/eE\right) \,.  \label{asy1}
\end{equation}

For both the Fermi and Bose cases, one can check that the mean numbers (\ref%
{ta9}) and (\ref{ta10}) are negligibly small,
\begin{equation}
N_{n}^{\mathrm{cr}}\ll e^{-\pi m^{2}/eE}\,,  \label{asy1.2}
\end{equation}%
for any given $p_{\bot }$ and for small kinetic momenta
\begin{equation*}
\left\vert p_{x}\pm eET_{\mathrm{S}}\right\vert =\sqrt{eE}K_{\mathrm{S}}\ll
eET_{\mathrm{S}}\,,\ \ K_{\mathrm{S}}\gg \max \left( 1,m/\sqrt{eE}\right) \,,
\end{equation*}%
where $K_{\mathrm{S}}$ is any given number.

For the range of large longitudinal kinetic momenta,%
\begin{equation}
\left\vert p_{x}\pm eET_{\mathrm{S}}\right\vert >\sqrt{eE}K_{\mathrm{S}%
}\Longleftrightarrow \left\vert p_{x}\right\vert <eET_{\mathrm{S}}-\sqrt{eE}%
K_{\mathrm{S}},\   \label{asy2.1}
\end{equation}%
and any given $p_{\bot }$, mean numbers (\ref{ta9}) and (\ref{ta10}) have
approximately the following form
\begin{equation}
N_{n}^{\mathrm{cr}}\approx N_{n}^{\mathrm{as}}=e^{-\pi \tau }\,,\ \ \tau =T_{%
\mathrm{S}}(\omega _{+}+\omega _{-}-2eET_{\mathrm{S}})\,.  \label{asy2}
\end{equation}%
The function $\tau $ has a minimum at $p_{x}=0$,%
\begin{equation}
\tau _{0}=\left. \tau \right\vert _{p_{x}=0}=T_{\mathrm{S}}\left[ 2\sqrt{\pi
_{\perp }^{2}+(eET_{\mathrm{S}})^{2}}-2eET_{\mathrm{S}}\right] \,,
\label{asy3}
\end{equation}%
and is growing monotonically as $\left\vert p_{x}\right\vert $ and $p_{\perp
}$ grow. One can see that mean numbers $N_{n}^{\mathrm{as}}$\ are
exponentially small in the range of large transversal momenta{\large , }$\pi
_{\perp }\gtrsim \sqrt{eE}K_{\mathrm{S}}${\large . }Therefore the following
range of $\pi _{\perp }$\ is of interest,%
\begin{equation}
\pi _{\perp }\ll \sqrt{eE}K_{\mathrm{S}}.  \label{asy2.2}
\end{equation}%
{\large \ }In this range, the following approximation holds true%
\begin{equation}
\tau \approx \frac{eET_{\mathrm{S}}^{2}\pi _{\perp }^{2}}{(eET_{\mathrm{S}%
})^{2}-p_{x}^{2}},\ \ \tau _{0}\approx \lambda =\frac{\pi _{\bot }^{2}}{eE}.
\label{asy3b}
\end{equation}%
The function $\tau $\ takes its maximum value%
\begin{equation*}
\tau _{\max }=\left. \tau \right\vert _{\left\vert p_{x}\right\vert =eET_{%
\mathrm{S}}-\sqrt{eE}K_{\mathrm{S}}}\approx \frac{\sqrt{eE}T_{\mathrm{S}%
}\lambda }{2K_{\mathrm{S}}}
\end{equation*}%
as $\left\vert p_{x}\right\vert $\ tends to its maximum.{\large \ }For $%
m\neq 0,$\ we see that $\tau _{\max }\rightarrow \infty $\ as $\sqrt{eE}T_{%
\mathrm{S}}\rightarrow \infty $. In the wide range of transversal momenta, $%
\pi _{\perp }\ll eET_{\mathrm{S}}$, the mean numbers $N_{n}^{\mathrm{as}}$
do not depend practically on the parameter $T_{\mathrm{S}}$ and coincide
with differential numbers of created particles in a constant electric field
\cite{Nikis70a,Nikis79}%
\begin{equation}
N_{n}^{\mathrm{as}}\approx N_{n}^{0}=e^{-\pi \lambda }\,.  \label{asy4}
\end{equation}

The total number of pairs\ created from a vacuum (defined by Eq.~(\ref{TN}))%
{\large \ }by an uniform electric field{\large ,} is proportional to the
space volume $V_{\left( d-1\right) }$ as $N^{\mathrm{cr}}=V_{\left(
d-1\right) }n^{\mathrm{cr}}$ and the corresponding number density $n^{%
\mathrm{cr}}$ has the form
\begin{equation}
n^{\mathrm{cr}}=\frac{1}{(2\pi )^{d-1}}\sum_{\sigma }\int d\mathbf{p}N_{n}^{%
\mathrm{cr}}\,.  \label{asy5}
\end{equation}%
In deriving Eq. (\ref{asy5}) the sum over all momenta $\mathbf{p}$\ was
transformed into an integral. Then the integral in the right hand side of
Eq. (\ref{asy5}) can be approximated by an integral over a subrange $\Omega $%
\ that gives the dominant contribution with respect to the total increment
to the number density of created particles,%
\begin{equation}
\Omega :n^{\mathrm{cr}}\approx \tilde{n}^{\mathrm{cr}}=\frac{1}{(2\pi )^{d-1}%
}\sum_{\sigma }\int_{\mathbf{p\in }\Omega }d\mathbf{p}N_{n}^{\mathrm{cr}}\,.
\label{asy6}
\end{equation}

Let us consider the number density of pairs created from the vacuum by the
Sauter-like potential with a large parameter $T_{\mathrm{S}}$. This quantity
can be calculated using Eq.~(\ref{asy6}) with differential numbers $N_{n}^{%
\mathrm{cr}}$ approximated by Eqs.~(\ref{asy2}) and (\ref{asy3b}). In this
case, the leading term, $\tilde{n}^{\mathrm{cr}}$, is formed over the range
given by Eqs.~(\ref{asy2.1}) and (\ref{asy2.2}), that is, this range is
chosen as a realization of the subrange $\Omega $\ in Eq.~(\ref{asy6}). In
this approximation, the numbers $N_{n}^{\mathrm{as}}$ are the same for
fermions and bosons and do not depend on the spin polarization parameters $%
\sigma _{s}$. Thus, in the Fermi case, probabilities and mean numbers summed
over all $\sigma _{s}$ obtain the factor $J_{(d)}=2^{\left[ d/2\right] -1}$.
We obtain that%
\begin{equation}
\tilde{n}^{\mathrm{cr}}=\frac{J_{(d)}}{(2\pi )^{d-1}}\int_{\mathbf{p\in }%
\Omega }d\mathbf{p}N_{n}^{\mathrm{cr}}\,.  \label{asy6b}
\end{equation}%
In the case of scalar bosons, $J_{(d)}=1$.

Taking into account Eqs.~(\ref{asy2}) and (\ref{asy3b}), we approximate
integral (\ref{asy6b}) as\textrm{\ }%
\begin{equation}
\tilde{n}^{\mathrm{cr}}\approx \frac{J_{(d)}}{(2\pi )^{d-1}}\int d\mathbf{p}%
_{\bot }I_{p_{\bot }}\,,\ \ I_{p_{\bot }}=2\int_{0}^{eET_{\mathrm{S}}-\sqrt{%
eE}K_{\mathrm{S}}}dp_{x}e^{-\pi \tau }\,.  \label{asy6c}
\end{equation}%
It is convenient to introduce a variable $t$, defined as $\tau =\lambda
t+\tau _{0}$. Taken into account Eq.~(\ref{asy3b}) we can find a relation
between $t$\ and $p_{x}$\ and see that%
\begin{equation}
dp_{x}=\frac{1}{2}eET_{\mathrm{S}}t^{-1/2}(t+1)^{-3/2}dt.  \label{asy7}
\end{equation}%
Neglecting the contribution from $\tau >\tau _{\max }$\ and using the
variable $t,$\ one can represent the quantity $I_{p_{\bot }}$\ as follows%
\begin{equation}
I_{p_{\bot }}\approx eET_{\mathrm{S}}\int_{0}^{\infty
}dtt^{-1/2}(t+1)^{-3/2}e^{-\pi \lambda \left( t+1\right) }.  \label{asy8}
\end{equation}%
In particular, using Eq.~(\ref{asy8}), one can find the number density of
created pairs with a given $p_{\bot }$\ for large and small $\lambda $\ in
the following form%
\begin{equation}
I_{p_{\bot }}\approx \frac{eET_{\mathrm{S}}}{\sqrt{\lambda }}e^{-\pi \lambda
}\,\;\mathrm{if}\;\lambda \gg 1,\;\;I_{p_{\bot }}\approx 2eET_{\mathrm{S}%
}\,\;\mathrm{if}\;\lambda \ll 1.  \label{asy9}
\end{equation}%
Finally, substituting Eq.~(\ref{asy8}) into integral (\ref{asy6c}) and
performing the integration over $p_{\bot }$, we obtain{\large \ }%
\begin{equation}
\tilde{n}^{\mathrm{cr}}=\frac{J_{(d)}T_{\mathrm{S}}\delta }{(2\pi )^{d-1}}%
\left( eE\right) ^{d/2}\exp \left( -\pi \frac{m^{2}}{eE}\right) \,,
\label{asy10}
\end{equation}%
where{\large \ }%
\begin{equation}
\delta =\int_{0}^{\infty }dtt^{-1/2}(t+1)^{-\left( d+1\right) /2}\exp \left(
-t\pi \frac{m^{2}}{eE}\right) =\sqrt{\pi }\Psi \left( \frac{1}{2},\frac{2-d}{%
2};\pi \frac{m^{2}}{eE}\right) \,.  \label{asy10b}
\end{equation}%
Here $\Psi \left( a,b;x\right) $ is the confluent hypergeometric function
\cite{BatE53}. This result was first obtained in Ref. \cite{GavG96a}%
\footnote{%
Unlike Eq.~(\ref{asy7}) the relation between $t$ and $p_{x}$ in Refs. \cite%
{GavG96a} is given for small $t$ only. This approximation is good enough for
$\lambda >1$. However, the final form of $\delta =\sqrt{\pi }\Psi \left(
\frac{1}{2},\frac{2-d}{2};\pi \frac{m^{2}}{eE}\right) $ is given correctly
for arbitrary $m^{2}/eE$.}.\textrm{\ }We see that the number density $\tilde{%
n}^{\mathrm{cr}}$, given by Eq.~(\ref{asy10}),\ is proportional to the total
increment of the longitudinal kinetic momentum, $\Delta U_{\mathrm{S}%
}=e\left\vert A_{x}\left( +\infty \right) -A_{x}\left( -\infty \right)
\right\vert =2eET_{\mathrm{S}}$.

From this result one can find the vacuum-to-vacuum probability $P_{v}$,
defined by Eq.~(\ref{vacprob}). Using the identity $\ln \left( 1\pm x\right)
=\pm x+\ldots ,$ and performing an integration following the considerations
above, one gets the following approximation%
\begin{eqnarray}
&&P_{v}\approx \exp \left( -\mu ^{\mathrm{S}}V_{\left( d-1\right) }\tilde{n}%
^{\mathrm{cr}}\right) \,,\ \ \mu ^{\mathrm{S}}=\sum_{l=0}^{\infty }\frac{%
(-1)^{(1-\kappa )l/2}\epsilon _{l+1}^{\mathrm{S}}}{(l+1)^{d/2}}\exp \left(
-l\pi \frac{m^{2}}{eE}\right) \,,  \notag \\
&&\epsilon _{l}^{\mathrm{S}}=\delta ^{-1}\sqrt{\pi }\Psi \left( \frac{1}{2},%
\frac{2-d}{2};l\pi \frac{m^{2}}{eE}\right) \,.  \label{asy11}
\end{eqnarray}%
In 3+1 QED the same result was found in a different way \cite{DunHal98}.

If the Sauter-like field is weak, $m^{2}/eE\gg $ $1$, one can use asymptotic
expression for the $\Psi $-function \cite{BatE53},%
\begin{equation}
\Psi \left( 1/2,\left( 2-d\right) /2;l\pi m^{2}/eE\right) =\left( eE/l\pi
m^{2}\right) ^{1/2}+O\left( \left[ eE/m^{2}\right] ^{3/2}\right) \,.
\label{asy12a}
\end{equation}%
Then $\delta \approx \sqrt{eE}/m,\;\epsilon _{l}^{\mathrm{S}}\approx l^{-%
\frac{1}{2}}$ and $\mu ^{\mathrm{S}}\approx 1$. In the case of a very strong
field, $m^{2}/eE\ll 1$, one obtains from  Ref.~\cite{BatE53} that the
leading term for the $\Psi $-function does not depend on the parameter $%
m^{2}/eE$,%
\begin{equation}
\Psi \left( 1/2,\left( 2-d\right) /2;\pi m^{2}/eE\right) \approx \Gamma
\left( d/2\right) /\Gamma \left( d/2+1/2\right) .  \label{asy12}
\end{equation}%
Then, for example, $\delta \approx \pi /2$ if $d=3$ and $\delta \approx 4/3$
if $d=4$. For the very strong field, $l\pi m^{2}/eE\ll 1$, the leading
contribution of $\epsilon _{l}^{\mathrm{S}}$ has a quite simple form and
does not depend on the dimension, $\epsilon _{l}^{\mathrm{S}}\approx 1$. In
this case%
\begin{equation}
\mu ^{\mathrm{S}}\approx \sum_{l=0}^{\infty }\frac{(-1)^{(1-\kappa )l/2}}{%
(l+1)^{d/2}}\,.  \label{asy13}
\end{equation}

\section{$T$-constant electric field\label{S5}}

In this section we present a detailed study of the particle creation problem
from the vacuum by a $T$-constant field. This field corresponds to a
regularized version of the constant field $E\left( t\right) =E$, in which
the electric field remains switched on for all the time $t\in \left( -\infty
,+\infty \right) $. This regularization was first considered in Ref.~\cite%
{BagGiS75} and then developed in Ref.~\cite{GavG96a}. In the present section
we explore additional peculiarities concerning particle creation,
supplementing the previous considerations with new details. The $T$-constant
electric field is constant within the time interval $T$ and is zero outside
of it,%
\begin{equation}
E\left( t\right) =\left\{
\begin{array}{l}
0\,,\ \ t\in \mathrm{I} \\
E\,,\ \ t\in \mathrm{II} \\
0\,,\ \ t\in \mathrm{III}%
\end{array}%
\right. \Longrightarrow A_{x}\left( t\right) =\left\{
\begin{array}{l}
-Et_{\mathrm{in}}\,,\ \ t\in \mathrm{I} \\
-Et\,,\ \ t\in \mathrm{II} \\
-Et_{\mathrm{out}}\,,\ \ t\in \mathrm{III}%
\end{array}%
\right. \,,  \label{t7}
\end{equation}%
where $\mathrm{I}$ denotes the in-region $t\in \left( -\infty ,t_{\mathrm{in}%
}\right] $, $\mathrm{II}$ is the intermediate region where the electric
field is non zero $t\in \left( t_{\mathrm{in}},t_{\mathrm{out}}\right) $ and
$\mathrm{III}$ is the out-region $t\in \left[ t_{\mathrm{out}},+\infty
\right) $ and $t_{\mathrm{out}}$, $t_{\mathrm{in}}$ are constants, $t_{%
\mathrm{out}}-t_{\mathrm{in}}=T$. We choose $t_{\mathrm{out}}=-t_{\mathrm{in}%
}=T/2$. It the\textrm{\ in}-region $\mathrm{I}$ and in \textrm{out}-region
\textrm{III}, Dirac spinors are solutions of the eigenvalue problem (\ref%
{t4b}).

For $t\in \mathrm{II}$, $U\left( t\right) =eEt$, equation (\ref{t3}) can be
written in the form%
\begin{equation}
\left[ \frac{\mathrm{d}^{2}}{\mathrm{d}\xi ^{2}}+\xi ^{2}-\mathrm{i}\chi
+\lambda \right] \varphi _{n}\left( t\right) =0,  \label{t9}
\end{equation}%
where
\begin{equation}
\xi =\frac{eEt-p_{x}}{\sqrt{eE}},\ \ \lambda =\frac{\pi _{\bot }^{2}}{eE}.
\label{t7.2}
\end{equation}%
The general solution of Eq.~ (\ref{t9}) is completely determined by an
appropriate pair of the linearly independent Weber parabolic cylinder
functions (WPCFs) \cite{BatE53}: either $D_{\rho }[(1-\mathrm{i})\xi ]$ and $%
D_{-1-\rho }[(1+\mathrm{i})\xi ],$ or $D_{\rho }[-(1-\mathrm{i})\xi ]$ and $%
D_{-1-\rho }[-(1+\mathrm{i})\xi ]$, where $\rho =\mathrm{i}\lambda /2-\left(
1-\chi \right) /2$. Then taking into account Eq.~(\ref{t4.4}), the functions
$\ _{-}\varphi _{n}\left( t\right) $ and $\ ^{+}\varphi _{n}\left( t\right) $
can be presented in the form%
\begin{eqnarray}
\ _{-}\varphi _{n}\left( t\right) &=&Y%
\begin{cases}
\ _{-}C\exp \left[ ip_{0}\left( t_{\mathrm{in}}\right) \left( t-t_{\mathrm{in%
}}\right) \right] , & t\in \mathrm{I} \\
\ _{-}C\left\{ a_{1}D_{\rho }[-(1-i)\xi ]+a_{2}D_{-1-\rho }[-(1+i)\xi
]\right\} , & t\in \mathrm{II} \\
g\left( ^{+}\left\vert _{-}\right. \right) \ ^{+}C\exp \left[ -ip_{0}\left(
t_{\mathrm{out}}\right) \left( t-t_{\mathrm{out}}\right) \right] +\kappa
g\left( ^{-}\left\vert _{-}\right. \right) \ ^{-}C\exp \left[ ip_{0}\left(
t_{\mathrm{out}}\right) \left( t-t_{\mathrm{out}}\right) \right] , & t\in
\mathrm{III}%
\end{cases}%
;  \notag \\
\ ^{+}\varphi _{n}\left( t\right) &=&Y%
\begin{cases}
g(_{+}|^{+})\ _{+}C\exp \left[ -ip_{0}\left( t_{\mathrm{in}}\right) \left(
t-t_{\mathrm{in}}\right) \right] +\kappa g(_{-}|^{+})\ _{-}C\exp \left[
ip_{0}\left( t_{\mathrm{in}}\right) \left( t-t_{\mathrm{in}}\right) \right] ,
& t\in \mathrm{I} \\
\ ^{+}C\left\{ a_{1}^{\prime }D_{\rho }[(1-i)\xi ]+a_{2}^{\prime }D_{-1-\rho
}[(1+i)\xi ]\right\} , & t\in \mathrm{II} \\
\ ^{+}C\exp \left[ -ip_{0}\left( t_{\mathrm{out}}\right) \left( t-t_{\mathrm{%
out}}\right) \right] , & t\in \mathrm{III}%
\end{cases}
\label{t8}
\end{eqnarray}%
on the whole axis $t$. Here $\kappa =1$ and the normalization constants are
given by Eqs.~(\ref{t4.3.1}). The functions $\ _{-}\varphi _{n}\left(
t\right) $ and $\ ^{+}\varphi _{n}\left( t\right) $ and their derivatives
satisfy the following gluing conditions:
\begin{equation}
\ _{-}^{+}\varphi _{n}(t_{\mathrm{in,out}}-0)=\ _{-}^{+}\varphi _{n}(t_{%
\mathrm{in,out}}+0),\quad \left. \partial _{t}\ _{-}^{+}\varphi
_{n}(t)\right\vert _{t=t_{\mathrm{in,out}}-0}=\left. \partial _{t}\
_{-}^{+}\varphi _{n}(t)\right\vert _{t=t_{\mathrm{in,out}}+0}.  \label{t8b}
\end{equation}

Using Eq.~(\ref{t8b}) and the Wronskian determinant of WPCFs \cite{BatE53},%
\begin{equation*}
D_{\rho }\left( z\right) \frac{d}{dz}D_{-\rho -1}\left( iz\right) -D_{-\rho
-1}\left( iz\right) \frac{d}{dz}D_{\rho }\left( z\right) =\exp \left[ -\frac{%
i\pi }{2}\left( \rho +1\right) \right] ,
\end{equation*}%
we find the coefficients $a_{j}$ and $a_{j}^{\prime },$
\begin{eqnarray}
a_{j} &=&\frac{(-1)^{j}}{\sqrt{2}}\exp \left[ \frac{i\pi }{2}\left( \rho +%
\frac{1}{2}\right) \right] \sqrt{\xi _{1}^{2}+\lambda }f_{j}^{(+)}(\xi _{1}),
\notag \\
a_{j}^{\prime } &=&\frac{(-1)^{j}}{\sqrt{2}}\exp \left[ \frac{i\pi }{2}%
\left( \rho +\frac{1}{2}\right) \right] \sqrt{\xi _{2}^{2}+\lambda }%
f_{j}^{(-)}(\xi _{2}),\quad j=1,2,  \label{t10.1}
\end{eqnarray}%
where%
\begin{eqnarray}
\xi _{1,2} &=&\left. \xi \right\vert _{t=t_{\mathrm{in,out}}}=\frac{\mp
eET/2-p_{x}}{\sqrt{eE}};  \label{t11.1} \\
f_{1}^{(\pm )}(\xi ) &=&\left( 1\pm \frac{i}{\sqrt{\xi ^{2}+\lambda }}\frac{d%
}{d\xi }\right) D_{-\rho -1}\left[ \mp (1+i)\xi \right] ,  \notag \\
f_{2}^{(\pm )}(\xi ) &=&\left( 1\pm \frac{i}{\sqrt{\xi ^{2}+\lambda }}\frac{d%
}{d\xi }\right) D_{\rho }\left[ \mp (1-i)\xi \right] .  \label{t10.2}
\end{eqnarray}%
Note that the following relations hold: $p_{0}\left( t_{\mathrm{in}}\right) /%
\sqrt{eE}=\sqrt{\xi _{1}^{2}+\lambda }$ and $p_{0}\left( t_{\mathrm{out}%
}\right) /\sqrt{eE}=\sqrt{\xi _{2}^{2}+\lambda }$. Using Eqs.~(\ref{t10.1})
one can determine the coefficients $g\left( {}_{\pm }|{}^{+}\right) $ and $%
g\left( {}^{\pm }|{}_{-}\right) $. It should be noted that we need to know
explicitly only the coefficients $g\left( {}_{-}|{}^{+}\right) $ and $%
g\left( {}^{+}|{}_{-}\right) $, which are%
\begin{eqnarray}
&&g\left( {}^{+}|{}_{-}\right) =AB\exp \left[ \left( \rho +1/2\right) i\pi /2%
\right] ,\ \ g\left( {}_{-}|{}^{+}\right) =A^{\prime }B^{\prime }\exp \left[
\left( \rho +1/2\right) i\pi /2\right] ,  \notag \\
&&A=\left[ \frac{\sqrt{\xi _{2}^{2}+\lambda }\sqrt{\xi _{1}^{2}+\lambda }%
\left( \sqrt{\xi _{2}^{2}+\lambda }+\chi \xi _{2}\right) }{8\left( \sqrt{\xi
_{1}^{2}+\lambda }-\chi \xi _{1}\right) }\right] ^{1/2},\ \
B=f_{2}^{(+)}(\xi _{1})f_{1}^{(+)}(\xi _{2})-f_{1}^{(+)}(\xi
_{1})f_{2}^{(+)}(\xi _{2}),  \notag \\
&&A^{\prime }=\left[ \frac{\sqrt{\xi _{2}^{2}+\lambda }\sqrt{\xi
_{1}^{2}+\lambda }\left( \sqrt{\xi _{1}^{2}+\lambda }-\chi \xi _{1}\right) }{%
8\left( \sqrt{\xi _{2}^{2}+\lambda }+\chi \xi _{2}\right) }\right] ^{1/2},\
\ B^{\prime }=f_{1}^{(-)}(\xi _{1})f_{2}^{(-)}(\xi _{2})-f_{2}^{(-)}(\xi
_{1})f_{1}^{(-)}(\xi _{2}).  \label{t10}
\end{eqnarray}%
One can see that $\left. \xi _{2}\right\vert _{p_{x}\rightarrow -p_{x}}=-\xi
_{1}$ then coefficients (\ref{t10}) obey the relations
\begin{equation}
\left. g\left( {}^{+}|{}_{-}\right) \right\vert _{p_{x}\rightarrow
-p_{x}}=g\left( {}_{-}|{}^{+}\right) .  \label{t10.3}
\end{equation}%
From these relations, we see that $\left\vert g\left( {}_{-}|{}^{+}\right)
\right\vert $ is an even function of momenta $\mathbf{p}$ and does not
depend on a spin polarization.

Taking into account Eq.~(\ref{KG8.2}), a formal transition to the
Klein-Gordon case can be done by setting $\chi =0$ and $\kappa =-1$\ in
Eqs.~(\ref{t8}). In this case $n=\mathbf{p}$ and the normalization factors
are given by Eq.~(\ref{KG8}). In the Klein-Gordon case, the coefficients $g$
are
\begin{eqnarray}
&&g\left( {}^{+}|{}_{-}\right) =\exp \left( -\lambda \pi /4\right) A_{%
\mathrm{sc}}\left. B\right\vert _{\chi =0},\ \ g\left( {}_{-}|{}^{+}\right)
=-\exp \left( -\lambda \pi /4\right) A_{\mathrm{sc}}\left. B^{\prime
}\right\vert _{\chi =0}\ ,  \notag \\
&&A_{\mathrm{sc}}=\left( \frac{1}{8}\sqrt{\xi _{2}^{2}+\lambda }\sqrt{\xi
_{1}^{2}+\lambda }\right) ^{1/2},\ \   \label{kk6}
\end{eqnarray}%
where $B$ and $B^{\prime }$ are given by Eqs.~(\ref{t10}). We stress that
this results are new.

The differential mean numbers of created pairs have the form $N_{n}^{\mathrm{%
cr}}=\left\vert g\left( _{-}\left\vert ^{+}\right. \right) \right\vert ^{2},$
see Eq.~(\ref{3.23}), where $g\left( _{-}\left\vert ^{+}\right. \right) $ is
given by Eqs.~(\ref{t10}) for Dirac particles and by Eqs.~(\ref{kk6}) for
Klein-Gordon particles. They depend only on the values $\xi _{1,2}$ for a
given $\lambda $. The $T$-constant field is a regularization for a constant
uniform electric field and it is suitable for imitating a slowly varying
field. That is why the $T$-constant field with a sufficiently large time
interval $T$,
\begin{equation}
\sqrt{eE}T\gg \max \left( 1,m^{2}/eE\right) ,  \label{T-large}
\end{equation}%
is of interest. In what follows, we suppose that these conditions hold true
and additionally assume that%
\begin{equation}
\sqrt{\lambda }<K_{\bot },  \label{p-fin}
\end{equation}%
where $K_{\bot }$ is any given number satisfying the condition$\;\sqrt{eE}%
T/2\gg K_{\bot }^{2}\gg \max \left\{ 1,m^{2}/eE\right\} .$

Let us analyze how the numbers $N_{n}^{\mathrm{cr}}$ depend on the
parameters $\xi _{1,2}$ and $\lambda $. Let for fermions $\chi =1$ and $\rho
=\mathrm{i}\lambda /2=\nu $. Since $N_{n}^{\mathrm{cr}}$ are even functions
of $p_{x}$, we can consider only the case of $p_{x}\leq 0$. In this case $%
\xi _{2}\geq \sqrt{eE}T/2$ is large, $\xi _{2}\gg \max \left\{ 1,\lambda
\right\} $, and the asymptotic expansions of WPCFs with respect to $\xi _{2}$
are valid. As to the parameter $\xi _{1}$, the whole interval $-\sqrt{eE}%
L/2\leq \xi _{1}\leq +\infty $ can be divided in three ranges:%
\begin{equation}
(a)\;-\sqrt{eE}T/2\leq \xi _{1}\leq -K,\;\;(b)\;-K<\xi _{1}<K,\;\;(c)\;\xi
_{1}\geq K,  \label{range}
\end{equation}%
where $K$ is any given number satisfying the condition $\sqrt{eE}T/2\gg K\gg
K_{\bot }^{2}$. Using the asymptotic expansions of WPCFs with respect to $%
\xi _{1,2}$ \cite{BatE53}, we get the following expansions of the
coefficients $f_{1,2}^{(-)}(\xi )$, given by Eqs. (\ref{t10.2}),
\begin{eqnarray}
f_{1}^{(-)}(\xi ) &=&e^{-\mathrm{i}\xi ^{2}/2}\left( \sqrt{2}e^{\mathrm{i}%
\pi /4}\xi \right) ^{-\nu -1}\left[ \frac{i}{\xi ^{2}}+O\left( \xi
^{-4}\right) \right] ,  \notag \\
f_{2}^{(-)}(\xi ) &=&e^{\mathrm{i}\xi ^{2}/2}\left( \sqrt{2}e^{-\mathrm{i}%
\pi /4}\xi \right) ^{\nu }2\left[ 1+i\frac{\nu \left( 1-\nu \right) }{4\xi
^{2}}+O\left( \xi ^{-4}\right) \right] \quad \mathrm{if}\quad \xi \geq K;
\notag \\
f_{1}^{(-)}(\xi ) &=&-e^{-\mathrm{i}\xi ^{2}/2}\left( \sqrt{2}e^{\mathrm{i}%
\pi /4}\left\vert \xi \right\vert \right) ^{-\nu -1}e^{i\pi \nu }\left[
2+i\left( \frac{3}{2}\nu +\nu ^{2}\right) \xi ^{-2}+O\left( \xi ^{-4}\right) %
\right]  \notag \\
&&+e^{\mathrm{i}\xi ^{2}/2}\left( \sqrt{2}e^{-\mathrm{i}\pi /4}\left\vert
\xi \right\vert \right) ^{\nu }e^{i\pi \nu /2}\frac{\sqrt{2\pi }}{2\Gamma
\left( \nu \right) \xi ^{4}},  \notag \\
f_{2}^{(-)}(\xi ) &=&ie^{-\mathrm{i}\xi ^{2}/2}\left( \sqrt{2}e^{\mathrm{i}%
\pi /4}\left\vert \xi \right\vert \right) ^{-\nu -1}e^{i\pi \nu /2}\frac{%
\sqrt{2\pi }}{\Gamma \left( -\nu \right) }\left[ 2+O\left( \xi ^{-2}\right) %
\right] \quad \mathrm{if}\quad \xi <0,\;\left\vert \xi \right\vert \geq K.
\label{f-exp}
\end{eqnarray}

One can use Eq.~(\ref{f-exp}) with respect to $\xi _{1}$ and $\xi _{2}$ for
the cases (a) and (c). In the case (c), we find that the quantity $N_{n}^{%
\mathrm{cr}}$ is very small,
\begin{equation}
N_{n}^{\mathrm{cr}}\sim \max \left\{ \left\vert \xi _{1}\right\vert
^{-6},\left\vert \xi _{2}\right\vert ^{-6}\right\} \;\;\mathrm{if}\;\;\min
\left\{ \left\vert \xi _{1}\right\vert ,\left\vert \xi _{2}\right\vert
\right\} \geq K.  \label{t11.2}
\end{equation}%
In the case (a), we obtain
\begin{eqnarray}
N_{n}^{\mathrm{cr}} &=&e^{-\pi \lambda }\left[ 1+\left( 1-e^{-\pi \lambda
}\right) ^{1/2}\frac{\sqrt{\lambda }}{2}\left( \frac{\sin \phi _{1}}{%
\left\vert \xi _{1}\right\vert ^{3}}+\frac{\sin \phi _{2}}{\left\vert \xi
_{2}\right\vert ^{3}}\right) +O\left( \left\vert \xi _{1}\right\vert
^{-4}\right) +O\left( \left\vert \xi _{2}\right\vert ^{-4}\right) \right] ,
\notag \\
\phi _{1,2} &=&\left( \xi _{1,2}\right) ^{2}+\lambda \ln \left( \sqrt{2}%
\left\vert \xi _{1,2}\right\vert \right) -\arg \Gamma \left( i\lambda
/2\right) -\pi /4.  \label{t12}
\end{eqnarray}%
Consequently, the quantity (\ref{t12}) is almost constant over the wide
range of longitudinal momentum $p_{x}$ for any given $\lambda $ satisfying
Eq.~(\ref{p-fin}). Note that the next-to-leading oscillating terms in Eq.~(%
\ref{t12}) presented here{\large \ }improve an approximation obtained before
in Ref.~\cite{GavG96a}.{\large \ }When $\sqrt{eE}T\rightarrow \infty $, one
obtains the result in a constant uniform electric field, given by Eq.~(\ref%
{asy4}), setting $\left\vert \xi _{1,2}\right\vert \rightarrow \infty $ in
Eq.~(\ref{t12}).

In the intermediate range (b), using the only asymptotics with respect to $%
\xi _{2}$ given by Eq.~(\ref{f-exp}) and the exact form of $f_{1}^{(-)}(\xi
_{1})$ given by Eq.~(\ref{t10.2}), we find that
\begin{equation}
N_{n}^{\mathrm{cr}}=\frac{1}{4}e^{-\pi \lambda /4}\sqrt{\xi _{1}^{2}+\lambda
}\left( \sqrt{\xi _{1}^{2}+\lambda }-\xi _{1}\right) \left\vert
f_{1}^{(-)}(\xi _{1})\right\vert ^{2}.  \label{t13}
\end{equation}%
One can make some conclusions about the contribution of this region to the
integral over the longitudinal momentum in Eq.~(\ref{asy5}). Taking into
account that $N_{n}^{\mathrm{cr}}$ is always less than unity for fermions,
one can get a rough estimation of the integral%
\begin{equation*}
\int_{\left\vert \xi _{1}\right\vert <K}N_{n}^{\mathrm{cr}}dp_{x}<2eEK
\end{equation*}%
and conclude that it is not essential in comparison with the integral over
the longitudinal momentum in the range (a) at $T\rightarrow \infty $. A more
accurate estimations can be made numerically. Thus, in the case of strong
field, $\lambda \lesssim 1$, one can see that the contribution from the
intermediate region (b) to the integral is much less than that given by a
rough estimate. In particular, one can see that the value $K=3$ is
sufficiently large for the problem in question.

For bosons $\chi =0$ and $\rho =\mathrm{i}\lambda /2-1/2$. We have to
consider the same three ranges (\ref{range}), wherein only the range (a) is
essential. In this range, using the asymptotic expansions of WPCFs we obtain
that the differential mean number of scalar particles created is%
\begin{eqnarray}
N_{n}^{\mathrm{cr}} &\simeq &e^{-\pi \lambda }\left\{ 1+\frac{1}{2}\left(
1+e^{\pi \lambda }\right) ^{1/2}\left( \frac{\sin \vartheta _{1}}{\left\vert
\xi _{1}\right\vert ^{2}}+\frac{\sin \vartheta _{2}}{\xi _{2}^{2}}\right)
\right\} \,,  \notag \\
\vartheta _{1,2} &=&\left( \xi _{1,2}\right) ^{2}+\lambda \log \left( \sqrt{2%
}\left\vert \xi _{1,2}\right\vert \right) +\arg \left[ \Gamma \left( \frac{%
1+i\lambda }{2}\right) +\frac{3\pi }{8}\right] .  \label{kk14}
\end{eqnarray}%
Note that the next-to-leading term approximation (\ref{kk14}) is presented
here for the first time.{\large \ }When $\sqrt{eE}T\rightarrow \infty $, one
obtains the same limit form (\ref{asy4}) in a constant uniform electric
field, setting $\left\vert \xi _{1,2}\right\vert \rightarrow \infty $ in
Eq.~(\ref{kk14}).

Let us consider the number density $n^{\mathrm{cr}}$ of pairs created by the
$T$-constant field of large time duration $T$. In general, the number
density $n^{\mathrm{cr}}$ of pairs created by uniform field is given by
integral (\ref{asy5}) both for fermions and bosons. The parameter $K$ plays
the role of a sharp cutoff in integral (\ref{asy5}). Therefore, the main
contribution to this integral is due to an subrange $D$ that is defined by
Eq.~(\ref{p-fin}) and the range (a) given by Eq.~(\ref{range}) for $%
p_{x}\leq 0$. Taking into account that $n^{\mathrm{cr}}$ is an even function
of $p_{x}$, we find the complete subrange $D$ as%
\begin{eqnarray}
&&D:\sqrt{\lambda }<K_{\bot },\;\;\left\vert p_{x}\right\vert /\sqrt{eE}<%
\sqrt{eE}T/2-K,  \notag \\
&&\sqrt{eE}T/2\gg K\gg K_{\bot }^{2}\gg \max \left\{ 1,m^{2}/eE\right\} .
\label{D}
\end{eqnarray}%
In this subrange $N_{n}^{\mathrm{cr}}\approx e^{-\pi \lambda }$ both for
fermions and bosons. Then one can find the total number of created particles
with given transversal momentum and spin polarization but with all possible
values of longitudinal momentum:
\begin{equation}
N_{\mathbf{p_{\perp }},\sigma }=\frac{L}{2\pi }\int dp_{x}N_{n}^{\mathrm{cr}%
}=\Delta _{\mathrm{long}}e^{-\pi \lambda },\;\;\Delta _{\mathrm{long}}=\frac{%
\sqrt{eE}L}{2\pi }\left[ \sqrt{eE}T+O(K)\right] ,  \label{t17}
\end{equation}%
where $L$ is the regularization length in the direction of the field. The
factor $\Delta _{\mathrm{long}}$ can be interpreted as a number of quantum
states with a longitudinal momentum, in which the particles can be created.
If $\sqrt{eE}T$ is big enough, the dependence on $K$ and $K_{\bot }$ can be
ignored, that is, the form of $N_{n}^{\mathrm{cr}}$ is unchanged in the
inner subrange $D$. Thus, the definition of the subrange $D$ (\ref{D}) can
be also treated as the stabilization condition for $N_{n}^{\mathrm{cr}}$,\
that is, this range is chosen as a realization of the subrange $\Omega $\ in
Eq.~(\ref{asy6}).{\large \ }In this approximation, the numbers $N_{n}^{%
\mathrm{cr}}$\ for fermions and bosons are equal, such that one can
represent the number density $\tilde{n}^{\mathrm{cr}}$\ in the form Eq.~(\ref%
{asy6b}).

Using Eq.~(\ref{t17}) we approximate integral (\ref{asy6b}) as \
\begin{equation}
\tilde{n}^{\mathrm{cr}}=\frac{J_{(d)}}{(2\pi )^{d-1}}\int_{D}d\mathbf{p}%
e^{-\pi \lambda }\approx \frac{J_{(d)}}{(2\pi )^{d-1}}\int_{\sqrt{\lambda }%
<K_{\bot }}d\mathbf{p}_{\bot }I_{p_{\bot }},\ \ I_{p_{\bot }}=eETe^{-\pi
\lambda }.  \label{t18.1}
\end{equation}%
Performing the integration over $\mathbf{p}_{\bot }$, and neglecting
exponentially small contribution from the range $\sqrt{\lambda }>K_{\bot }$,
we finally obtain \cite{GavG96a}
\begin{equation}
\tilde{n}^{\mathrm{cr}}=r^{\mathrm{cr}}\left[ T+\frac{O(K)}{\sqrt{eE}}\right]
,\;\;r^{\mathrm{cr}}=\frac{J_{(d)}\left( eE\right) ^{d/2}}{(2\pi )^{d-1}}%
\exp \left\{ -\pi \frac{m^{2}}{eE}\right\} .  \label{t18.2}
\end{equation}%
We see that the number density $\tilde{n}^{\mathrm{cr}}$, given by Eq.~(\ref%
{t18.2}),\ is proportional to the total increment of the longitudinal
kinetic momentum,{\large \ }$\Delta U_{\mathrm{T}}=e\left\vert A_{x}\left(
+\infty \right) -A_{x}\left( -\infty \right) \right\vert =eET${\large .}
Note that this density is a function of the time duration of the field. The
same result was obtained in 3+1 dimensions, using the functional Schr\"{o}%
dinger picture \cite{HalL95}. The quantity $r^{\mathrm{cr}}=d\tilde{n}^{%
\mathrm{cr}}/dT$ is often called the pairs production rate. It is constant
if $\sqrt{eE}T$ is big enough. It is useful to compare this result with one
obtained for the model of pair creation by the $L$-constant field \cite%
{GavGit15L}. In fact, the $L$-constant field is a constant uniform electric
field confined between two capacitor plates separated by a finite distance $%
L $. This field can create pairs with transversal momenta that satisfy the
inequality{\large \ }$2\pi _{\bot }\leq eEL$. In this case the total number
of created pairs, $N^{\mathrm{cr}}$, is a function of the field length $L$.
The $T$-constant and $L$-constant fields are physically distinct. However,
if $\sqrt{eE}L$ is big enough, the process of pair creation can be
characterized by the same density $r^{\mathrm{cr}}$. Thus, only in the
asymptotic case when $T\rightarrow \infty $ and $L\rightarrow \infty ,$ one
can consider these fields as regularizations of a constant uniform electric
field given by two distinct gauge conditions on the electromagnetic
potentials $A^{\mu }\left( x\right) $.

Using Eq.~(\ref{vacprob}) and performing the integration following the
considerations above, we get the vacuum-to-vacuum probability $P_{v}$ in the
form
\begin{equation}
P_{v}=\exp \left( -\mu ^{\mathrm{T}}V_{\left( d-1\right) }\tilde{n}^{\mathrm{%
cr}}\right) ,\;\;\mu ^{\mathrm{T}}=\sum_{l=0}^{\infty }\frac{(-1)^{(1-\kappa
)l/2}}{(l+1)^{d/2}}\exp \left( -l\pi \frac{m^{2}}{eE}\right) \;,  \label{t19}
\end{equation}%
where $\tilde{n}^{\mathrm{cr}}$ \ is given by Eq.~(\ref{t18.2}). The formula
above coincide, for $d=4$, with the well known Schwinger's result \cite%
{Schwinger51} obtained for a constant electric field $T\rightarrow \infty $.

\section{Peak electric field\label{S6}}

In this section we present a complete discussion concerning particle
creation from the vacuum by a third set of exactly solvable $t$-electric
potential steps, namely, the peak electric field and the exponentially
decaying electric field, both considered previously by us in Refs. \cite%
{AdoGavGit14,AdoGavGit16}. Here we supplement the former studies with new
details and discussions particular to these fields.

\subsection{General form of peak electric field\label{Ss6.1}}

The peak electric field $E\left( t\right) $ is composed of two parts, one of
them is increasing exponentially on the time-interval $\mathrm{I}=\left(
-\infty ,0\right] $, and reaches its maximal magnitude $E>0$ at the end of
the interval $t=0,$ the second part decreases exponentially on the
time-interval $\mathrm{II}=\left( 0,+\infty \right) $ having at $t=0$ the
same magnitude $E.$ The vector potential $A_{x}\left( t\right) $ and the
field $E_{x}\left( t\right) $ are
\begin{equation}
A_{x}\left( t\right) =E\left\{
\begin{array}{l}
k_{1}^{-1}\left( -e^{k_{1}t}+1\right) ,\ \ t\in \mathrm{I}\, \\
k_{2}^{-1}\left( e^{-k_{2}t}-1\right) \,,\ \ t\in \mathrm{II}%
\end{array}%
\right. ,\ E\left( t\right) =E\left\{
\begin{array}{l}
e^{k_{1}t}\,,\ \ t\in \mathrm{I}\, \\
e^{-k_{2}t}\,,\ \ t\in \mathrm{II}%
\end{array}%
\right. ,  \label{ns4.0}
\end{equation}%
and $A_{y}=A_{z}=E_{y}=E_{z}=0$. Here $k_{1}$ and $k_{2}$ are positive
constants. The field $E\left( t\right) $ is continuous at $t=0$, but its
time-derivative is not in the general case,%
\begin{equation}
\lim_{t\rightarrow -0}E\left( t\right) =\lim_{t\rightarrow +0}E\left(
t\right) =E,\ \ \forall t:E\left( t\right) \leq E,\ \lim_{t\rightarrow -0}%
\dot{E}\left( t\right) =k_{1}E\neq \lim_{t\rightarrow +0}\dot{E}\left(
t\right) =-k_{2}E\,.  \label{s3.3}
\end{equation}

Note that the so-called exponentially decreasing electric field with the
potential%
\begin{equation}
A_{x}^{\mathrm{ed}}\left( t\right) =E\left\{
\begin{array}{l}
0\,,\ \ t\in \mathrm{I}\, \\
k_{2}^{-1}\left( e^{-k_{2}t}-1\right) \,,\ \ t\in \mathrm{II}%
\end{array}%
\right. \,,  \label{com1}
\end{equation}%
can be considered as a particular case of the peak field, when the latter
switches on abruptly at $t=0$, i.e., when $k_{1}$ is sufficiently large, $%
k_{1}\rightarrow \infty $. Similarly can be treated the exponentially
increasing electric field.

Exact solutions of the Dirac equation with the exponentially decreasing and
the peak electric fields have been obtained by us previously in Refs. \cite%
{AdoGavGit14,AdoGavGit16}. Following the same way, we introduce new
variables $\eta _{j}$,%
\begin{equation}
\eta _{1}=ih_{1}e^{k_{1}t}\,,\ \ \eta _{2}=ih_{2}e^{-k_{2}t}\,,\ \
h_{j}=2eEk_{j}^{-2},\ \ j=1,2\,,  \label{i.0}
\end{equation}%
in place of $t$ and represent the scalar functions $\varphi _{n}\left(
t\right) $ as%
\begin{equation}
\varphi _{n}^{j}\left( t\right) =e^{-\eta _{j}/2}\eta _{j}^{\nu _{j}}\tilde{%
\varphi}^{j}\left( \eta _{j}\right) ,\ \nu _{j}=\frac{i\omega _{j}}{k_{j}},\
\ \omega _{j}=\sqrt{\pi _{j}^{2}+\pi _{\perp }^{2}},\ \ \pi
_{j}=p_{x}-\left( -1\right) ^{j}\frac{eE}{k_{j}}\,,  \label{i.2}
\end{equation}%
where the subscript $j$ distinguishes quantities associated to the
time-intervals $\mathrm{I}$ and $\mathrm{II}$. The functions $\tilde{\varphi}%
^{j}\left( \eta _{j}\right) $ satisfy confluent hypergeometric equations
\cite{BatE53},%
\begin{eqnarray}
&&\left[ \eta _{j}\frac{d^{2}}{d\eta _{j}^{2}}+\left( c_{j}-\eta _{j}\right)
\frac{d}{d\eta _{j}}-a_{j}\right] \tilde{\varphi}^{j}\left( \eta _{j}\right)
=0\,,  \notag \\
&&c_{j}=1+2\nu _{j}\,,\ \ a_{j}=\frac{1}{2}\left( 1+\chi \right) +\left(
-1\right) ^{j}\frac{i\pi _{j}}{k_{j}}+\nu _{j}\,.  \label{i.3}
\end{eqnarray}%
In accordance with Eq.~(\ref{e2a}), the quantity $\chi $ can be chosen to be
either $\chi =+1$ or $\chi =-1$.

A fundamental set of solutions for the latter equation consists of two
linearly independent confluent hypergeometric functions $\Phi \left(
a_{j},c_{j};\eta _{j}\right) \ $\ and$\ \ \eta _{j}^{1-c_{j}}e^{\eta
_{j}}\Phi \left( 1-a_{j},2-c_{j};-\eta _{j}\right) ,$ where%
\begin{equation}
\Phi \left( a,c;\eta \right) =1+\frac{a}{c}\frac{\eta }{1!}+\frac{a\left(
a+1\right) }{c\left( c+1\right) }\frac{\eta ^{2}}{2!}+\ldots \,.  \label{chf}
\end{equation}%
Thus, the general solution of Eq.~(\ref{t3}) in the intervals $\mathrm{I}$
and $\mathrm{II}$ can be written as the following linear superposition:%
\begin{align}
& \varphi _{n}^{j}\left( t\right) =b_{2}^{j}y_{1}^{j}\left( \eta _{j}\right)
+b_{1}^{j}y_{2}^{j}\left( \eta _{j}\right) \,,  \notag \\
& y_{1}^{j}\left( \eta _{j}\right) =e^{-\eta _{j}/2}\eta _{j}^{\nu _{j}}\Phi
\left( a_{j},c_{j};\eta _{j}\right) \,,  \notag \\
& y_{2}^{j}\left( \eta _{j}\right) =e^{\eta _{j}/2}\eta _{j}^{-\nu _{j}}\Phi
\left( 1-a_{j},2-c_{j};-\eta _{j}\right) \,,  \label{i.3.3}
\end{align}%
with arbitrary constants $b_{1}^{j}$ and $b_{2}^{j}$. The Wronskian of the
functions $y$ is%
\begin{equation}
y_{1}^{j}\left( \eta _{j}\right) \frac{d}{d\eta _{j}}y_{2}^{j}\left( \eta
_{j}\right) -y_{2}^{j}\left( \eta _{j}\right) \frac{d}{d\eta _{j}}%
y_{1}^{j}\left( \eta _{j}\right) =\frac{1-c_{j}}{\eta _{j}}\,.  \label{i.3.4}
\end{equation}

As can be seen from Eq.~(\ref{ns4.0}), the peak electric field is switched
on at the infinitely remote past $t\rightarrow -\infty $ and switched{\Huge %
\ }off at the infinitely remote future $t\rightarrow +\infty $. At these
regions, the exact solutions represent free particles and the appropriate
superpositions from Eq.~(\ref{i.3.3}) obey the asymptotic conditions (\ref%
{t4.1a}), where $p_{0}\left( -\infty \right) =\omega _{1}$ denotes energy of
initial particles at $t\rightarrow -\infty $, $p_{0}\left( +\infty \right)
=\omega _{2}$ denotes energy of final particles at $t\rightarrow +\infty $
and$\;_{\zeta }\mathcal{N}$ and $\;^{\zeta }\mathcal{N}$ are given by Eq.~(%
\ref{t4.3.1}).

Using the initial conditions (\ref{t4.1a}), we fix the constants $b_{1}^{j}$
and $b_{2}^{j},$ and then we find the \textrm{in-} and \textrm{out}-electron
and positron states in the intervals $\mathrm{I}$ and $\mathrm{II}$:%
\begin{eqnarray}
\ _{+}\varphi _{n}\left( t\right) &=&\;_{+}\mathcal{N}\exp \left( i\pi \nu
_{1}/2\right) y_{2}^{1}\left( \eta _{1}\right) \,,\,\ _{-}\varphi _{n}\left(
t\right) =\;_{-}\mathcal{N}\exp \left( -i\pi \nu _{1}/2\right)
y_{1}^{1}\left( \eta _{1}\right) \,,\ \ t\in \mathrm{I}\,;  \notag \\
\ ^{+}\varphi _{n}\left( t\right) &=&\;^{+}\mathcal{N}\exp \left( -i\pi \nu
_{2}/2\right) y_{1}^{2}\left( \eta _{2}\right) \,,\,\ ^{-}\varphi _{n}\left(
t\right) =\;^{-}\mathcal{N}\exp \left( i\pi \nu _{2}/2\right)
y_{2}^{2}\left( \eta _{2}\right) \,,\ \ t\in \mathrm{II}\,.  \label{i.4.1}
\end{eqnarray}

Taking into account the structure of exact solutions given by Eqs.~(\ref%
{i.3.3}) and (\ref{i.4.1}), we represent the functions$\ \ _{-}\varphi
_{n}\left( t\right) $ and $\ ^{+}\varphi _{n}\left( t\right) $ in the form%
\begin{eqnarray}
\ \ ^{+}\varphi _{n}\left( t\right) &=&\left\{
\begin{array}{l}
g\left( _{+}|^{+}\right) \ _{+}\varphi _{n}\left( t\right) +\kappa g\left(
_{-}|^{+}\right) \ _{-}\varphi _{n}\left( t\right) \,,\ \ t\in \mathrm{I} \\
\;^{+}\mathcal{N}\exp \left( -i\pi \nu _{2}/2\right) y_{1}^{2}\left( \eta
_{2}\right) \,,\ \ \ \ \ \ \ \ \ \ \,t\in \mathrm{II}%
\end{array}%
\right. \,,  \label{i.6.1} \\
\ _{-}\varphi _{n}\left( t\right) &=&\left\{
\begin{array}{l}
\;_{-}\mathcal{N}\exp \left( -i\pi \nu _{1}/2\right) y_{1}^{1}\left( \eta
_{1}\right) \,,\ \ \ \ \ \ \ \ \ \ \ \ \,t\in \mathrm{I} \\
g\left( ^{+}|_{-}\right) \ ^{+}\varphi _{n}\left( t\right) +\kappa g\left(
^{-}|_{-}\right) \ ^{-}\varphi _{n}\left( t\right) \,,\ \ t\in \mathrm{II}%
\end{array}%
\right. \,,  \label{i.6.2}
\end{eqnarray}%
already for any $t$. Here coefficients $g$\ are defined by Eq.~(\ref{t4.5}).
The constant $\kappa $ is defined by Eq.~(\ref{vacprob}). Its introduction
allows us to describe by one equation the case of scalar particles as well,
which is discussed in detail below. The functions$\ _{-}\varphi _{n}\left(
t\right) $ and $\ ^{+}\varphi _{n}\left( t\right) $ and their derivatives
satisfy the following continuity conditions:
\begin{equation}
\left. \ _{-}^{+}\varphi _{n}(t)\right\vert _{t=-0}=\left. \ _{-}^{+}\varphi
_{n}(t)\right\vert _{t=+0}\,,\ \ \left. \partial _{t}\ _{-}^{+}\varphi
_{n}(t)\right\vert _{t=-0}=\left. \partial _{t}\ _{-}^{+}\varphi
_{n}(t)\right\vert _{t=+0}\,.  \label{i.7}
\end{equation}

Using Eqs.~(\ref{i.7}) and (\ref{i.3.4}), one can find coefficients $g\left(
_{\zeta }|^{\zeta ^{\prime }}\right) $ and $g\left( ^{\zeta }|_{\zeta
^{\prime }}\right) $ from Eqs.~(\ref{i.6.1}) and (\ref{i.6.2}),%
\begin{eqnarray}
&&g\left( _{-}|^{+}\right) =\mathcal{C}\Delta \,,\ \ \mathcal{C}=-\frac{1}{2}%
\sqrt{\frac{q_{1}^{-}}{\omega _{1}q_{2}^{+}\omega _{2}}}\exp \left[ \frac{%
i\pi }{2}\left( \nu _{1}-\nu _{2}\right) \right] \,,  \notag \\
&&\Delta =\left. \left[ k_{1}h_{1}y_{1}^{2}\left( \eta _{2}\right) \frac{d}{%
d\eta _{1}}y_{2}^{1}\left( \eta _{1}\right) +k_{2}h_{2}y_{2}^{1}\left( \eta
_{1}\right) \frac{d}{d\eta _{2}}y_{1}^{2}\left( \eta _{2}\right) \right]
\right\vert _{t=0}\,;  \label{gp} \\
&&g\left( ^{+}|_{-}\right) =\mathcal{C}^{\prime }\Delta ^{\prime }\,,\ \
\mathcal{C}^{\prime }=-\frac{1}{2}\sqrt{\frac{q_{2}^{+}}{\omega
_{1}q_{1}^{-}\omega _{2}}}\exp \left[ \frac{i\pi }{2}\left( \nu _{2}-\nu
_{1}\right) \right] \,,  \notag \\
&&\Delta ^{\prime }=\left\{ k_{2}h_{2}y_{1}^{1}\left( \eta _{1}\right) \frac{%
d}{d\eta _{2}}y_{2}^{2}\left( \eta _{2}\right) +k_{1}h_{1}y_{2}^{2}\left(
\eta _{2}\right) \frac{d}{d\eta _{1}}y_{1}^{1}\left( \eta _{1}\right)
\right\} _{t=0}\,,  \label{nd5}
\end{eqnarray}%
respectively \cite{AdoGavGit16}. Comparing Eqs.~(\ref{gp}) and (\ref{nd5})
one can verify that the symmetry under a simultaneous change $%
k_{1}\leftrightarrows k_{2}$ and $\pi _{1}\leftrightarrows -\pi _{2}$ holds
true,%
\begin{equation}
g\left( ^{+}|_{-}\right) \leftrightarrows g\left( _{-}|^{+}\right) \,.
\label{nd6}
\end{equation}

A transition to solutions of the Klein-Gordon equation can be performed by
setting $\chi =0$ and $\kappa =-1$\ in Eqs.~(\ref{i.6.1}) and (\ref{i.6.2}),
and by using{\Huge \ }the normalization constants (\ref{KG8}). Thus, in the
case of scalar particles, the coefficient $g\left( _{-}|^{+}\right) $ reads%
\begin{equation}
g\left( _{-}|^{+}\right) =\mathcal{C}_{\mathrm{sc}}\left. \Delta \right\vert
_{\chi =0}\,,\ \ \mathcal{C}_{\mathrm{sc}}=\left( 4\omega _{1}\omega
_{2}\right) ^{-1/2}\exp \left[ i\pi \left( \nu _{1}-\nu _{2}\right) /2\right]
\,,  \label{gpKG}
\end{equation}%
where $\Delta $ is given by Eq.~(\ref{gp}). In this case, we have the
antisymmetry%
\begin{equation}
g\left( ^{+}|_{-}\right) \leftrightarrows -g\left( _{-}|^{+}\right) \,.
\label{nd6b}
\end{equation}%
under the simultaneous change $k_{1}\leftrightarrows k_{2}$ and $\pi
_{1}\leftrightarrows -\pi _{2}$.

Using $g\left( _{-}|^{+}\right) $ given by Eq.~(\ref{gp}), we find that in
the Fermi case, differential mean number of created particles is

\begin{equation}
N_{n}^{\mathrm{cr}}=\left\vert \mathcal{C}\Delta \right\vert ^{2}\,.
\label{4.0}
\end{equation}

In the Bose case, using $g\left( _{-}|^{+}\right) $ given by Eq.~(\ref{gpKG}%
), we find%
\begin{equation}
N_{n}^{\mathrm{cr}}=\left\vert \mathcal{C}_{\mathrm{sc}}\left. \Delta
\right\vert _{\chi =0}\right\vert ^{2}\,.  \label{4b}
\end{equation}%
It is clear that mean numbers $N_{n}^{\mathrm{cr}}$ depend on modulus
squared of the transversal momentum, $\mathbf{p}_{\perp }^{2}$. It follows
from Eqs.~(\ref{nd6}) and (\ref{nd6b}) that the numbers $N_{n}^{\mathrm{cr}}$
are invariant under the simultaneous change $k_{1}\leftrightarrows k_{2}$
and $\pi _{1}\leftrightarrows -\pi _{2}$ for fermions and bosons,
respectively. Then if $k_{1}=k_{2}$, the numbers $N_{n}^{\mathrm{cr}}$
appear to be even functions of the longitudinal momentum $p_{x}$.

\subsection{Slowly varying field\label{Ss6.2}}

The inverse parameters $k_{1}^{-1}$, $k_{2}^{-1}$ represent scales of time
duration of the electric field in the increasing and decreasing time
intervals \textrm{I} and \textsf{II}. In particular, slowly varying fields
correspond to small values of $k_{1}$ and $k_{2}$, satisfying the conditions%
\begin{equation}
\min \left( h_{1},h_{2}\right) \gg \max \left( 1,m^{2}/eE\right) \,,
\label{4.1}
\end{equation}%
where $h_{1}$ and $h_{2}$ are defined by Eq. (\ref{i.0}). In this case, we
have a two-parameter regularization for a constant electric field
(additional to the above presented one-parameter regularizations by the
Sauter-like electric field and the $T$-constant electric field).

Let us analyze how the differential numbers $N_{n}^{\mathrm{cr}}$ depend on
the quantities $p_{x}$ and $\pi _{\perp }$. A semiclassical consideration
show that $N_{n}^{\mathrm{cr}}$ are exponentially small for very large $\pi
_{\perp }\gtrsim \min \left( eEk_{1}^{-1},eEk_{2}^{-1}\right) $. Then the
range of fixed $\pi _{\perp }$ is of interest and in the following we assume
that condition (\ref{p-fin}) holds true, where in the case under
consideration any given number $K_{\bot }$ satisfies the inequality%
\begin{equation}
\min \left( h_{1},h_{2}\right) \gg K_{\bot }^{2}\gg \max \left(
1,m^{2}/eE\right) \,.  \label{4.3}
\end{equation}

By virtue of the symmetry properties of the numbers $N_{n}^{\mathrm{cr}}$
discussed above, one can only consider either{\Huge \ }positive or negative $%
p_{x}$. Let us, for example, consider the interval $-\infty <p_{x}\leq 0$.
In this case $\pi _{2}$ is negative, its modulus is large, $-\pi _{2}\geq
eE/k_{2}$, while $\pi _{1}$ varies from positive to negative values, $%
-\infty <\pi _{1}\leq eE/k_{1}$. The case of negative $\pi _{1}$ with large
modulus, $-2\pi _{1}/k_{1}>K_{1}$, where $K_{1}$ is any given large number, $%
K_{1}\gg K_{\bot }$, is quite simple. In this case, using the appropriate
asymptotic expressions of the confluent hypergeometric function one can see
that $N_{n}^{\mathrm{cr}}$ are negligibly small. To see this, Eq. (\ref{A10}%
) (see Appendix \ref{Ap}) is useful in the range $h_{1}\gtrsim -2\pi
_{1}/k_{1}>K_{1}$, while an expression for large $c_{2}$ with a fixed $a_{2}$
and $h_{2}$ and an expression for large $c_{1}$ with fixed $a_{1}-c_{1}$ and
$h_{1}$, given in \cite{BatE53}, are useful in the range $-2\pi
_{1}/k_{1}\gg h_{1}$.

We expect a significant contribution for the numbers $N_{n}^{\mathrm{cr}}$\
in the range%
\begin{equation}
h_{1}\geq 2\pi _{1}/k_{1}>-K_{1}.  \label{4.4}
\end{equation}%
This range can be divided in four subranges%
\begin{eqnarray}
\mathrm{(a)} &&\;h_{1}\geq 2\pi _{1}/k_{1}>h_{1}\left[ 1-\left( \sqrt{h_{1}}%
g_{2}\right) ^{-1}\right] ,  \notag \\
\mathrm{(b)} &&\;h_{1}\left[ 1-\left( \sqrt{h_{1}}g_{2}\right) ^{-1}\right]
>2\pi _{1}/k_{1}>h_{1}\left( 1-\varepsilon \right) ,  \notag \\
\mathrm{(c)} &&\;h_{1}\left( 1-\varepsilon \right) >2\pi
_{1}/k_{1}>h_{1}/g_{1},  \notag \\
\mathrm{(d)} &&\;h_{1}/g_{1}>2\pi _{1}/k_{1}>-K_{1},  \label{4.5}
\end{eqnarray}%
where $g_{1}$, $g_{2}$, and $\varepsilon $ are any given numbers satisfying
the conditions%
\begin{equation*}
g_{1}\gg 1,\ g_{2}\gg 1,\ \left( \sqrt{h_{1}}g_{2}\right) ^{-1}\ll
\varepsilon \ll 1.
\end{equation*}%
We note that%
\begin{equation*}
\tau _{1}=-ih_{1}/\left( 2-c_{1}\right) \approx \frac{h_{1}k_{1}}{%
2\left\vert \pi _{1}\right\vert }
\end{equation*}%
in the subranges (a), (b), and (c) and%
\begin{equation*}
\tau _{2}=ih_{2}/c_{2}\approx \frac{h_{2}k_{2}}{2\left\vert \pi
_{2}\right\vert }
\end{equation*}%
in the whole range (\ref{4.4}). In the subranges, $\tau _{2}$ satisfies the
inequalities$:$%
\begin{eqnarray}
\mathrm{(a)} &&\;1\leq \tau _{2}^{-1}<\left[ 1+\left( \sqrt{h_{2}}%
g_{2}\right) ^{-1}\right] ,  \notag \\
\mathrm{(b)} &&\;\left[ 1+\left( \sqrt{h_{2}}g_{2}\right) ^{-1}\right] <\tau
_{2}^{-1}<\left( 1+\varepsilon k_{2}/k_{1}\right) ,  \notag \\
\mathrm{(c)} &&\;\left( 1+\varepsilon k_{2}/k_{1}\right) <\tau _{2}^{-1}<%
\left[ 1+k_{2}/k_{1}\left( 1-1/g_{1}\right) \right] ,  \notag \\
\mathrm{(d)} &&\;\left[ 1+k_{2}/k_{1}\left( 1-1/g_{1}\right) \right] <\tau
_{2}^{-1}\lesssim \left( 1+k_{2}/k_{1}\right) .  \label{4.6}
\end{eqnarray}%
We see that $\tau _{1}-1\rightarrow 0$ and $\tau _{2}-1\rightarrow 0$ in the
range (a), while $\left\vert \tau _{1}-1\right\vert \sim 1$ in the range
(c), and $\left\vert \tau _{2}-1\right\vert \sim 1$ in the ranges (c) and
(d). In the range (b) these quantities vary from their values in the ranges
(a) and (c).

We choose $\chi =1$ for convenience in the Fermi case. In the range (a) we
can use the asymptotic expression for the confluent hypergeometric function
given by Eq.~(\ref{A1}) in Appendix \ref{Ap}. Using Eqs.~(\ref{A8}) and (\ref%
{A9}) (see Appendix \ref{Ap}), we can find that the differential means
numbers for fermions and bosons in the leading approximation have the same
form%
\begin{equation}
N_{n}^{\mathrm{cr}}=e^{-\pi \lambda }\left[ 1+O\left( \left\vert \mathcal{Z}%
_{1}\right\vert \right) \right] ,\ \ \max \left\vert \mathcal{Z}%
_{1}\right\vert \lesssim g_{2}^{-1}.  \label{4.7}
\end{equation}%
In the range (c), the confluent hypergeometric function $\Phi \left(
a_{2},c_{2};ih_{2}\right) $ is approximated by Eq.~(\ref{A10a}) and the
function $\Phi \left( 1-a_{1},2-c_{1};-ih_{1}\right) $ is approximated by
Eq.~(\ref{A10}) given in the Appendix \ref{Ap}. Then we find that%
\begin{eqnarray}
&&N_{n}^{\mathrm{cr}}=e^{-\pi \lambda }\left[ 1+O\left( \left\vert \mathcal{Z%
}_{1}\right\vert \right) ^{-1}+O\left( \left\vert \mathcal{Z}_{2}\right\vert
\right) ^{-1}\right] ,  \label{4.8} \\
&&\max \left\vert \mathcal{Z}_{1}\right\vert ^{-1}\lesssim \sqrt{g_{1}/h_{1}}%
,\ \ \max \left\vert \mathcal{Z}_{2}\right\vert ^{-1}\lesssim g_{2}^{-1}.
\notag
\end{eqnarray}%
Using asymptotic expression~(\ref{A1}) and taking into account Eqs.~(\ref%
{4.7}) and (\ref{4.8}), we obtain that in the range (b) the following
estimate holds $N_{n}^{\mathrm{cr}}\sim e^{-\pi \lambda }$. In the range
(d), the confluent hypergeometric function $\Phi \left(
a_{2},c_{2};ih_{2}\right) $ is approximated by Eq.~(\ref{A10a}) and the
function $\Phi \left( 1-a_{1},2-c_{1};-ih_{1}\right) $ is approximated by
Eq.~(\ref{A11}) given in Appendix \ref{Ap}. Then, in this range, we obtain
the following leading-order approximation for the differential mean numbers%
\begin{equation}
N_{n}^{\mathrm{cr}}\approx \frac{\exp \left[ -\frac{\pi }{k_{1}}\left(
\omega _{1}-\pi _{1}\right) \right] }{\sinh \left( 2\pi \omega
_{1}/k_{1}\right) }\times \left\{
\begin{array}{l}
\sinh \left[ \pi \left( \omega _{1}+\pi _{1}\right) /k_{1}\right] \ \mathrm{%
\ in\ Fermi\ case} \\
\cosh \left[ \pi \left( \omega _{1}+\pi _{1}\right) /k_{1}\right] \ \mathrm{%
in\ Bose\ case}%
\end{array}%
\right. \,.  \label{4.9a}
\end{equation}

It is clear that when $\pi _{1}\gg \pi _{\bot }$ the expressions given by
Eqs.~(\ref{4.9a}) take the form (\ref{4.8}), $N_{n}^{\mathrm{cr}}\rightarrow
e^{-\pi \lambda }$. Consequently, the result (\ref{4.9a}) is valid in the
whole range (\ref{4.4}). Assuming $m/k_{1}\gg 1$, we see that $N_{n}^{%
\mathrm{cr}}$ given by Eqs.~(\ref{4.9a}) are negligible in the range $\pi
_{1}\lesssim \pi _{\bot }$. Then one can see that substantial value of $%
N_{n}^{\mathrm{cr}}$ are formed in the range $\pi _{\bot }<\pi _{1}\leqslant
eE/k_{1}$ and are given the same formula%
\begin{equation}
N_{n}^{\mathrm{cr}}\approx \exp \left[ -\frac{2\pi }{k_{1}}\left( \omega
_{1}-\pi _{1}\right) \right] .  \label{4.10}
\end{equation}
both for bosons and fermions.

Considering positive $p_{x}>0$, we can take into account that numbers $%
N_{n}^{\mathrm{cr}}$ are invariant under the simultaneous exchange $%
k_{1}\leftrightarrows k_{2}$ and $\pi _{1}\leftrightarrows -\pi _{2}$. In
this case $\pi _{1}$ is positive and large, $\pi _{1}>eE/k_{1}$, while $\pi
_{2}$ varies from negative to positive values, $-eE/k_{2}<\pi _{2}<\infty $.
We find that a substantial contribution to $N_{n}^{\mathrm{cr}}$ are formed
in the range
\begin{equation}
-h_{2}<2\pi _{2}/k_{2}<K_{2},  \label{4.11}
\end{equation}%
where $K_{2}$ is any given large number, $K_{2}\gg K_{\bot }$. In this
range, similarly to the case of negative $p_{x}$, we obtain the following
leading-order approximation for the differential mean numbers%
\begin{equation}
N_{n}^{\mathrm{cr}}\approx \frac{\exp \left[ -\frac{\pi }{k_{2}}\left(
\omega _{2}+\pi _{2}\right) \right] }{\sinh \left( 2\pi \omega
_{2}/k_{2}\right) }\times \left\{
\begin{array}{c}
\sinh \left[ \pi \left( \omega _{2}-\pi _{2}\right) /k_{2}\right] \ \mathrm{%
in\ Fermi\ case} \\
\cosh \left( \pi \left( \omega _{2}-\pi _{2}\right) /k_{2}\right) \ \mathrm{%
in\ Bose\ case}%
\end{array}%
\right. \,.  \label{4.12a}
\end{equation}

Assuming $m/k_{2}\gg 1$, we see that substantial value of $N_{n}^{\mathrm{cr}%
}$ are formed in the range $-eE/k_{2}<\pi _{2}<-\pi _{\bot }$ and are given
the same formula%
\begin{equation}
N_{n}^{\mathrm{cr}}\approx \exp \left[ -\frac{2\pi }{k_{2}}\left( \omega
_{2}+\pi _{2}\right) \right]  \label{4.13}
\end{equation}%
both for bosons and fermions.

Consequently, for any given $\lambda $ satisfying Eqs.~ (\ref{p-fin}) and (%
\ref{4.3}), the quantities $N_{n}^{\mathrm{cr}}$ are quasiconstant over the
wide range of longitudinal momentum $p_{x}$. When $h_{1},h_{2}\rightarrow
\infty $, differential mean numbers coincide with (\ref{asy4}) in a constant
uniform electric field.

The above analysis\ shows that dominant contributions for mean numbers of
created particles by a slowly varying field are formed in ranges of large
kinetic momenta and have there asymptotic forms (\ref{4.10}) for $p_{x}<0$
and (\ref{4.13}) for $p_{x}>0$. In this case, the range $\Omega $\ in Eq.~(%
\ref{asy6}) is realized as $\pi _{\bot }<\pi _{1}\leqslant eE/k_{1}$\ for $%
p_{x}<0$\ and as $-eE/k_{2}<\pi _{2}<-\pi _{\bot }$\ for $p_{x}>0${\large . }%
Therefore, we can represent integral (\ref{asy6b}) as follows%
\begin{eqnarray}
&&\tilde{n}^{\mathrm{cr}}=\frac{J_{(d)}}{(2\pi )^{d-1}}\int_{\sqrt{\lambda }%
<K_{\bot }}d\mathbf{p}_{\bot }I_{\mathbf{p}_{\bot }},\ \ I_{\mathbf{p}_{\bot
}}=I_{\mathbf{p}_{\bot }}^{\left( 1\right) }+I_{\mathbf{p}_{\bot }}^{\left(
2\right) },  \notag \\
&&I_{\mathbf{p}_{\bot }}^{\left( 1\right) }=\int_{-\infty }^{0}dp_{x}N_{n}^{%
\mathrm{cr}}\approx \int_{\pi _{\perp }}^{eE/k_{1}}d\pi _{1}\exp \left[ -%
\frac{2\pi }{k_{1}}\left( \omega _{1}-\pi _{1}\right) \right] \,,  \notag \\
&&I_{\mathbf{p}_{\bot }}^{\left( 2\right) }=\int_{0}^{\infty }dp_{x}N_{n}^{%
\mathrm{cr}}\approx \int_{\pi _{\perp }}^{eE/k_{2}}d\left\vert \pi
_{2}\right\vert \exp \left[ -\frac{2\pi }{k_{2}}\left( \omega
_{2}-\left\vert \pi _{2}\right\vert \right) \right] \,.  \label{tot2}
\end{eqnarray}%
Using the variable changes%
\begin{equation*}
s=\frac{2}{k_{1}\lambda }\left( \omega _{1}-\pi _{1}\right) \,\ \mathrm{in\ }%
I_{\mathbf{p}_{\bot }}^{\left( 1\right) },\,\ s=\frac{2}{k_{2}\lambda }%
\left( \omega _{2}-\left\vert \pi _{2}\right\vert \right) \,\,\ \mathrm{in\ }%
I_{\mathbf{p}_{\bot }}^{\left( 2\right) },
\end{equation*}%
and neglecting exponentially small contributions, we respectively represent
the quantities $I_{\mathbf{p}_{\bot }}^{\left( 1\right) }$ and $I_{\mathbf{p}%
_{\bot }}^{\left( 2\right) }$ as%
\begin{equation}
I_{\mathbf{p}_{\bot }}^{\left( 1\right) }\approx \int_{1}^{\infty }\frac{ds}{%
s}\omega _{1}e^{-\pi \lambda s}\,,\ \ I_{\mathbf{p}_{\bot }}^{\left(
2\right) }\approx \int_{1}^{\infty }\frac{ds}{s}\omega _{2}e^{-\pi \lambda
s}\,.  \label{4.15}
\end{equation}%
The leading contributions for both integrals (\ref{4.15}) are from a range
near $s\rightarrow 1$, where $\omega _{1}$ and $\omega _{2}$ are
approximately given by,%
\begin{equation*}
\omega _{1}\approx \frac{eE}{sk_{1}}\,,\ \ \omega _{2}\approx \frac{eE}{%
sk_{2}}\,.
\end{equation*}%
Consequently the leading term in $I_{\mathbf{p}_{\bot }}$ takes the
following form:%
\begin{equation}
I_{\mathbf{p}_{\bot }}\approx eE\left( \frac{1}{k_{1}}+\frac{1}{k_{2}}%
\right) \int_{1}^{\infty }\frac{ds}{s^{2}}e^{-\pi \lambda s}=eE\left( \frac{1%
}{k_{1}}+\frac{1}{k_{2}}\right) e^{-\pi \lambda }G\left( 1,\pi \lambda
\right) ,  \label{4.17}
\end{equation}%
where%
\begin{equation}
G\left( \alpha ,x\right) =\int_{1}^{\infty }\frac{ds}{s^{\alpha +1}}%
e^{-x\left( s-1\right) }=e^{x}x^{\alpha }\Gamma \left( -\alpha ,x\right) ,
\label{4.19a}
\end{equation}%
and $\Gamma \left( -\alpha ,x\right) $ is the incomplete gamma function.

Neglecting an exponentially small contribution, one can extend the
integration limit over $\mathbf{p}_{\bot }$ in Eq.~(\ref{tot2}) from $\sqrt{%
\lambda }<K_{\bot }$ to $\sqrt{\lambda }<\infty .$ This allows us to
calculate the integral over $\mathbf{p}_{\bot }$ as Gaussian one. Thus, we
find%
\begin{equation}
\tilde{n}^{\mathrm{cr}}=r^{\mathrm{cr}}\left( \frac{1}{k_{1}}+\frac{1}{k_{2}}%
\right) G\left( \frac{d}{2},\pi \frac{m^{2}}{eE}\right) ,  \label{4.19}
\end{equation}%
where $r^{\mathrm{cr}}$ is given by Eq.~(\ref{t18.2}) \cite{AdoGavGit16}. We
see that the number density $\tilde{n}^{\mathrm{cr}}$, given by Eq.~(\ref%
{4.19}),\ is proportional to the total increment of the longitudinal kinetic
momentum,{\large \ }$\Delta U_{\mathrm{P}}=e\left\vert A_{x}\left( +\infty
\right) -A_{x}\left( -\infty \right) \right\vert =eE_{0}\left(
k_{1}^{-1}+k_{2}^{-1}\right) ${\large .} Note that if the electric field $E$
is weak, $m^{2}/eE\gg $ $1$, one obtains%
\begin{equation}
G\left( \frac{d}{2},\pi \frac{m^{2}}{eE}\right) \approx \frac{eE}{\pi m^{2}}.
\label{4.26}
\end{equation}%
If the electric field $E$ is strong enough, $m^{2}/eE\ll 1$, the leading
term of $G$-function, which is given by Eq. (\ref{4.19a}), is%
\begin{equation}
G\left( \frac{d}{2},\pi \frac{m^{2}}{eE}\right) \approx \frac{2}{d}.
\label{4.27}
\end{equation}

Using the above considerations we perform the summation (integration) in
Eq.~(\ref{vacprob}) and obtain the vacuum-to-vacuum probability $P_{v}$,%
\begin{eqnarray}
&&P_{v}=\exp \left( -\mu ^{\mathrm{P}}V_{\left( d-1\right) }\tilde{n}^{%
\mathrm{cr}}\right) ,\;\;\mu ^{\mathrm{P}}=\sum_{l=0}^{\infty }\frac{%
(-1)^{(1-\kappa )l/2}\epsilon _{l+1}^{\mathrm{P}}}{(l+1)^{d/2}}\exp \left(
-l\pi \frac{m^{2}}{eE}\right) \;,  \notag \\
&&\epsilon _{l}^{\mathrm{P}}=G\left( \frac{d}{2},l\pi \frac{m^{2}}{eE}%
\right) \left[ G\left( \frac{d}{2},\pi \frac{m^{2}}{eE}\right) \right] ^{-1},
\label{4.20}
\end{eqnarray}%
where $\tilde{n}^{\mathrm{cr}}$ is given by Eq.~(\ref{4.19}).

These results allow us to establish an immediate comparison with the
one-parameter regularizations of the constant field, namely the $T$-constant
and Sauter-like electric fields. The number densities of created particles
in such fields, given by Eqs.~(\ref{asy10}), (\ref{t18.2}), and (\ref{4.19}%
), are proportional to the corresponding increments of longitudinal kinetic
momenta. This fact allows one to compare pair creation effects in such
fields. Thus, for a given magnitude of the electric field $E$ one can
compare the pair creation effects in fields with equal increment of the
longitudinal kinetic momentum, or one can determine such increments of the
longitudinal kinetic momenta,\emph{\ }for which particle creation effects
are the same. Equating the number densities $\tilde{n}^{\mathrm{cr}}$ for
Sauter-like field and for the peak field to the density $\tilde{n}^{\mathrm{%
cr}}$ for the $T$-constant field, we find an effective duration time $%
T_{eff} $ in both cases,%
\begin{eqnarray}
T_{eff} &=&T_{\mathrm{S}}\delta \;\mathrm{for\;Sauter}\text{\textrm{-}}%
\mathrm{like\ field},  \notag \\
T_{eff} &=&\left( k_{1}^{-1}+k_{2}^{-1}\right) G\left( \frac{d}{2},\pi \frac{%
m^{2}}{eE}\right) \;\mathrm{for\;the\ peak\ field}.  \label{4.24}
\end{eqnarray}%
By the definition $T_{eff}=T$ for the $T$-constant field. One can say that
the Sauter-like and the peak electric fields with the same $T_{eff}=T$ are
equivalent to the $T$-constant field in pair production.

\subsection{Short pulse field\label{Ss6.3}}

Choosing parameters the peak field in a certain way, one can obtain electric
fields that exist only for a short time in a vicinity of the time instant $%
t=0$. The latter fields switch on and (or) switch off{\Huge \ }%
\textquotedblleft abruptly\textquotedblright\ near the time instant $t=0$.
This situation can be realized for large parameters $k_{1}$, $%
k_{2}\rightarrow \infty $ with a fixed ratio $k_{1}/k_{2}$. The
corresponding asymptotic potentials, $U\left( +\infty \right) =eEk_{2}^{-1}$
and $U\left( -\infty \right) =-eEk_{1}^{-1}$ define finite increments of the
longitudinal kinetic momenta $\Delta U_{1}$ and $\Delta U_{2}$ for
increasing and decreasing parts, respectively,%
\begin{equation}
\Delta U_{1}=U\left( 0\right) -U\left( -\infty \right)
=eEk_{1}^{-1},\;\;\Delta U_{2}=U\left( +\infty \right) -U\left( 0\right)
=eEk_{2}^{-1}.  \label{5.1}
\end{equation}%
Effectively we have a very short pulse of the electric field. At the same
time this configuration imitates well enough a $t$-electric rectangular
potential step (it is an analog of the Klein step, which is an $x$-electric
rectangular step; see Ref.~\cite{GavGit15}) and coincides with it as $k_{1}$%
, $k_{2}\rightarrow \infty $. Thus, these field configurations can be
considered as regularizations of a rectangular step. We assume that
sufficiently large $k_{1}$ and $k_{2}$ for any given $\pi _{\perp }$ and $%
\pi _{1,2}=p_{x}-U\left( \mp \infty \right) $ satisfy the following
inequalities:
\begin{equation}
\Delta U_{1}/k_{1}\ll 1,\;\;\Delta U_{2}/k_{2}\ll 1,\;\;\max \left( \omega
_{1}/k_{1},\omega _{2}/k_{2}\right) \ll 1.  \label{5.2}
\end{equation}%
In this case the confluent hypergeometric function can be approximated by
the first two terms in Eq. (\ref{chf}), which are $\Phi \left( a,c;\eta
\right) $, $c_{j}\approx 1$, and $\ \ a_{j}\approx \left( 1+\chi \right) /2$%
. Then we obtain \cite{AdoGavGit16}%
\begin{equation}
N_{n}^{\mathrm{cr}}=\left\{
\begin{array}{l}
\frac{\left( \omega _{1}+\pi _{1}\right) \left( \Delta U_{2}+\Delta
U_{1}+\omega _{2}-\omega _{1}\right) ^{2}}{4\omega _{1}\omega _{2}\left(
\omega _{2}-\pi _{2}\right) }\ \mathrm{in\ Fermi\ case} \\
\frac{\left( \omega _{2}-\omega _{1}\right) ^{2}}{4\omega _{1}\omega _{2}}\
\ \ \mathrm{in\ Bose\ case}%
\end{array}%
\right. .  \label{5.3}
\end{equation}

In contrast to the Fermi case, where $N_{n}^{\mathrm{cr}}\leq 1$, in the
Bose case, the differential numbers $N_{n}^{\mathrm{cr}}$ are unbounded in
two ranges of the longitudinal kinetic momenta, in the range where $\omega
_{1}/\omega _{2}\rightarrow \infty $ and in the range where $\omega
_{2}/\omega _{1}\rightarrow \infty $. In these ranges they have the form%
\begin{equation}
N_{n}^{\mathrm{cr}}\approx \frac{1}{4}\max \left\{ \omega _{1}/\omega
_{2},\omega _{2}/\omega _{1}\right\} .  \label{5.5}
\end{equation}%
We can treat this fact as an indication that the back reaction is big. If
so, the concept of the external field in the scalar QED is limited by the
condition $\min \left( \omega _{1}/k_{1},\omega _{2}/k_{2}\right) \gtrsim 1$
for the fields under consideration. We do not see similar problem in the
spinor QED.

If $k_{1}=k_{2}$ (in this case $\Delta U_{2}=\Delta U_{1}=\Delta U/2$), we
can compare the above results with the regularization of rectangular steps
by the Sauter-like potential, given by Eqs.~(\ref{ta10.2}) and (\ref{ta10.4}%
). We see that both regularizations are in agreement, for fermions under the
condition $\left\vert \omega _{2}-\omega _{1}\right\vert \ll \Delta U$ , and
for bosons under condition (\ref{ta10.3b}).

\subsection{Exponentially decaying field\label{Ss6.4}}

In the examples, considered above, increasing and decreasing phases of the
fields are almost symmetric. Here we consider an essentially asymmetric
configuration of the peak field, when, for example, the field switches
abruptly on at $t=0$, that is, $k_{1}$ is sufficiently large, while the
parameter $k_{2}>0$ is arbitrary, including, for example, the case of smooth
switching off process. Note, that due to the invariance of the mean numbers $%
N_{n}^{\mathrm{cr}}$ under the simultaneous change $k_{1}\leftrightarrows
k_{2}$ and $\pi _{1}\leftrightarrows -\pi _{2}$, one can easily transform
this situation to the case with a large $k_{2}$ and arbitrary $k_{1}>0$.

Let us assume that a sufficiently large $k_{1}$ satisfies the inequalities
\begin{equation}
\Delta U_{1}/k_{1}\ll 1,\;\;\omega _{1}/k_{1}\ll 1.  \label{5.6}
\end{equation}%
Then Eqs.~(\ref{4.0}) and (\ref{4b}) can be reduced to the following form%
\begin{equation}
\left\vert \Delta \right\vert ^{2}\approx \left\vert \Delta _{\mathrm{ap}%
}\right\vert ^{2}=e^{i\pi \nu _{2}}\left\vert \left[ \chi \Delta
U_{1}+\omega _{2}-\omega _{1}+k_{2}h_{2}\left( -\frac{1}{2}+\frac{d}{d\eta
_{2}}\right) \right] \Phi \left( a_{2},c_{2};\eta _{2}\right) \right\vert
_{t=0}^{2}\,.  \label{5.7}
\end{equation}%
Under the condition%
\begin{equation}
\Delta U_{1}/\omega _{1}\ll 1,  \label{5.8}
\end{equation}%
one can disregard the term $\chi \Delta U_{1}$ in Eq.~(\ref{5.7}) and set
approximately $\pi _{1}\approx p_{x}$. Thus, $\omega _{1}\approx \sqrt{%
p_{x}^{2}+\pi _{\perp }^{2}}$. In this approximation, leading terms do not
contain $\Delta U_{1}$, so that we obtain
\begin{equation}
N_{n}^{\mathrm{cr}}\approx \left\{
\begin{array}{l}
\left\vert \mathcal{C}\Delta _{\mathrm{ap}}\right\vert ^{2}\ \mathrm{for\
fermions} \\
\left\vert \mathcal{C}_{\mathrm{sc}}\left. \Delta _{\mathrm{ap}}\right\vert
_{\chi =0}\right\vert ^{2}\ \mathrm{for\ bosons}%
\end{array}%
\right. \,.  \label{5.9}
\end{equation}%
In fact, differential mean numbers obtained in these approximations are the
same as in the so-called exponentially decaying electric field, given by the
potential (\ref{com1}). Under condition (\ref{5.8}), the results presented
by Eqs.~(\ref{5.9}) for arbitrary $k_{2}>0$ are in agreement with ones
obtained in Ref. \cite{AdoGavGit14}.

Let us consider the most asymmetric case when Eqs.~(\ref{5.9}) hold and when
the increment of the longitudinal kinetic momentum due to exponentially
decaying electric field is sufficiently large ($k_{2}$ are sufficiently
small),%
\begin{equation}
h_{2}=2\Delta U_{2}/k_{2}\gg \max \left( 1,m^{2}/eE\right) \,.  \label{5.11}
\end{equation}%
In this case only the range of $\pi _{\perp }$ (\ref{p-fin}) is essential,
in which $K_{\bot }$ is any given number satisfying the condition%
\begin{equation}
h_{2}\gg K_{\bot }^{2}\gg \max \left( 1,m^{2}/eE\right) \,.  \label{5.12}
\end{equation}

It should be noted that the distribution $N_{n}^{\mathrm{cr}}$, given by
Eqs.~(\ref{5.9}) for this most asymmetric case,\emph{\ }coincides with the
one obtained in our work \cite{AdoGavGit14}. However, the detailed study of
this distribution was not performed there. The main contribution from this
distribution to the total number $N^{\mathrm{cr}}$ (\ref{asy5}) was
estimated in our recent work \cite{AdoGavGit16}. Here we consider a detailed
dependence of distribution ~(\ref{5.9}) on the physical parameters $p_{x}$
and $\pi _{\perp }$ and the corresponding consequences to the global
quantities $N^{\mathrm{cr}}$ and $P_{v}$ . We choose $\chi =-1$ for
convenience in the Fermi case.

In the case of large negative momenta $p_{x}$, $p_{x}<0$ and $-2\pi
_{2}/k_{2}>g_{1}h_{2}$ (where $g_{1}$ is any given number, $g_{1}$ $\gg 1$),
using an expression for the confluent hypergeometric function with large $%
c_{2}$ and fixed $a_{2}$ and $h_{2}$, given in \cite{BatE53}, one can verify
that the mean numbers $N_{n}^{\mathrm{cr}}$ are negligibly small both for
fermions and bosons. The same holds true for very large positive $p_{x}$,
such that $2\pi _{2}/k_{2}>K_{2}$, where $K_{2}$ is any given large number, $%
K_{2}\gg K_{\bot }$. We see that the mean numbers $N_{n}^{\mathrm{cr}}$
differ from zero\ only in the range $-g_{1}h_{2}<2\pi _{2}/k_{2}<K_{2}$ .
This range can be divided in the following subranges:%
\begin{eqnarray}
\mathrm{(a)} &&\;\left( 1+\varepsilon \right) h_{2}\leq -2\pi
_{2}/k_{2}<g_{1}h_{2},  \notag \\
\mathrm{(b)} &&\;h_{2}\left[ 1+\left( \sqrt{h_{2}}g_{2}\right) ^{-1}\right]
\leq -2\pi _{2}/k_{2}<\left( 1+\varepsilon \right) h_{2},  \notag \\
\mathrm{(c)} &&\;h_{2}\left[ 1-\left( \sqrt{h_{2}}g_{2}\right) ^{-1}\right]
\leq -2\pi _{2}/k_{2}<h_{2}\left[ 1+\left( \sqrt{h_{2}}g_{2}\right) ^{-1}%
\right] ,  \notag \\
\mathrm{(d)} &&\;\left( 1-\varepsilon \right) h_{2}\leq -2\pi
_{2}/k_{2}<h_{2}\left[ 1-\left( \sqrt{h_{2}}g_{2}\right) ^{-1}\right] ,
\notag \\
\mathrm{(e)} &&\;h_{2}/g_{1}<-2\pi _{2}/k_{2}<\left( 1-\varepsilon \right)
h_{2},  \notag \\
\mathrm{(f)} &&\;-K_{2}<-2\pi _{2}/k_{2}<h_{2}/g_{1},  \label{5.14}
\end{eqnarray}%
where $g_{2}$ and $\varepsilon $ are any given numbers satisfying the
conditions $g_{2}$ $\gg 1$ and $\varepsilon \ll 1$. We assume that $%
\varepsilon \sqrt{h_{2}}\gg 1$. Note that in the ranges (\ref{5.14}) $\tau
_{2}=ih_{2}/c_{2}\approx \frac{h_{2}k_{2}}{2\left\vert \pi _{2}\right\vert }%
\ .$ Then in the ranges from (a) to (e), $\tau _{2}$ varies from $1/g_{1}$
to $g_{1}$.

In the range (a), the confluent hypergeometric function $\Phi \left(
a_{2},c_{2};ih_{2}\right) $ is approximated by Eq.~(\ref{A10a}) given in
Appendix \ref{Ap}. In this range the differential mean numbers in the
leading-order approximation are%
\begin{equation}
N_{n}^{\mathrm{cr}}\approx \frac{\omega _{1}-\left\vert p_{x}\right\vert }{%
2\omega _{1}}\left[ 1+O\left( \left\vert \mathcal{Z}_{2}\right\vert
^{-1}\right) \right] \ \times \left\{
\begin{array}{l}
1\ \mathrm{for\ fermions} \\
\frac{\omega _{1}-\left\vert p_{x}\right\vert }{\left\vert p_{x}\right\vert }%
\ \mathrm{for\ bosons}%
\end{array}%
\right. \,,  \label{6.1}
\end{equation}%
where $\max \left\vert \mathcal{Z}_{2}\right\vert ^{-1}\sim \left(
\varepsilon \sqrt{h_{2}}\right) ^{-1}\lesssim \sqrt{g_{1}/h_{2}}$ . We see
from Eq.~(\ref{6.1}) that $N_{n}^{\mathrm{cr}}\rightarrow 0$ if $\left\vert
p_{x}\right\vert \gg \pi _{\bot }$. Note that $\varepsilon
eE/k_{2}<\left\vert p_{x}\right\vert <\left( g_{1}-1\right) eE/k_{2}$. \
Taking into account the inequality (\ref{p-fin}), we see that the numbers (%
\ref{6.1}) are negligibly small if $\varepsilon \sqrt{h_{2}}\gg K_{\bot }$.

In the range (c), $\tau _{2}-1\rightarrow 0$ and, using Eqs.~(\ref{A5}), (%
\ref{A2}), and (\ref{A4}) given in Appendix \ref{Ap} we find that%
\begin{equation}
N_{n}^{\mathrm{cr}}=\frac{\omega _{1}+p_{x}}{2\omega _{1}}e^{-\pi \lambda /2}%
\left[ 1+O\left( \left\vert \mathcal{Z}_{2}\right\vert \right) ^{-1}\right]
\times \left\{
\begin{array}{l}
\cosh \left( \frac{\pi \lambda }{4}\right) \ \mathrm{for\ fermions} \\
\frac{\pi \left( \omega _{1}+p_{x}\right) }{\sqrt{8eE}\left\vert \Gamma
\left( \frac{3+i\lambda }{4}\right) \right\vert ^{2}}\ \mathrm{for\ bosons}%
\end{array}%
\right. ,  \label{6.2}
\end{equation}%
where $\max \left\vert \mathcal{Z}_{2}\right\vert ^{-1}\lesssim g_{2}^{-1}$.
Note that $N_{n}^{\mathrm{cr}}$ given by Eq.~(\ref{6.2}) are finite and
restricted, $N_{n}^{\mathrm{cr}}\leq 1$ for fermions and $N_{n}^{\mathrm{cr}%
}\lesssim 1/g_{2}$ for bosons. In the range (b) the distributions $N_{n}^{%
\mathrm{cr}}$ vary between their values in the ranges (a) and (c).

In the range (e), parameters $\eta _{2}$ and $c_{2}$ are large with $a_{2}$
fixed and $\tau _{2}>1$. In this case, using the asymptotic expression of
the confluent hypergeometric function given by Eq.~(\ref{A10}) in Appendix %
\ref{Ap}, we find that%
\begin{equation}
N_{n}^{\mathrm{cr}}=\exp \left[ -\frac{2\pi }{k_{2}}\left( \omega _{2}+\pi
_{2}\right) \right] \left[ 1+O\left( \left\vert \mathcal{Z}_{2}\right\vert
^{-1}\right) \right] ,  \label{5.15}
\end{equation}%
where $\mathcal{Z}_{2}$ is given by Eq.~(\ref{A5}) in the Appendix \ref{Ap},
both for fermions and bosons, . We note that\textrm{\ }modulus $\left\vert
\mathcal{Z}_{2}\right\vert ^{-1}$ varies from $\left\vert \mathcal{Z}%
_{2}\right\vert ^{-1}\sim \left( \varepsilon \sqrt{h_{2}}\right) ^{-1}$ to $%
\left\vert \mathcal{Z}_{2}\right\vert ^{-1}\sim \left[ \left( g_{1}-1\right)
\sqrt{h_{2}}\right] ^{-1}$. Approximately, expression (\ref{5.15}) can be
written as%
\begin{equation}
N_{n}^{\mathrm{cr}}\approx \exp \left( -\frac{\pi \pi _{\perp }^{2}}{%
k_{2}\left\vert \pi _{2}\right\vert }\right) .  \label{5.16}
\end{equation}%
Note that $eE/g_{1}<k_{2}\left\vert \pi _{2}\right\vert <\left(
1-\varepsilon \right) eE$ in the range (e) and the distribution $N_{n}^{%
\mathrm{cr}}$ given by Eq.~(\ref{5.16}) has the following limiting form:%
\begin{equation}
N_{n}^{\mathrm{cr}}\rightarrow e^{-\pi \lambda }\ \ \mathrm{as\ \ }%
k_{2}\left\vert \pi _{2}\right\vert \rightarrow \left( 1-\varepsilon \right)
eE\ .  \label{5.16b}
\end{equation}%
Thus, the distribution ~(\ref{asy4}) is reproduced in the case of an
exponentially decaying electric field in the wide range of a large increment
of the longitudinal kinetic momentum, $-\pi _{2}\sim eE/k_{2}$. Taking into
account the condition $\varepsilon \sqrt{h_{2}}\gg 1,$ we see that $p_{x}/%
\sqrt{eE}\gg 1$ in this range. Thus, condition (\ref{5.8}) holds if%
\begin{equation}
\Delta U_{1}/\sqrt{eE}\lesssim 1.  \label{6.3}
\end{equation}%
Under this condition, the form of the distribution $N_{n}^{\mathrm{cr}}$
does not depend on the details of the switching on before the time instant $%
t=0$. In the range (d), the distributions $N_{n}^{\mathrm{cr}}$ vary from
their values in the ranges (c) and (e) for fermions and bosons.

In the range (f), we can use an asymptotic expression of the confluent
hypergeometric function for large $h_{2}$ at fixed $a_{2}$ and $c_{2}$ given
by Eq.~(\ref{A11}) in Appendix \ref{Ap} to get the following result:
\begin{equation}
N_{n}^{\mathrm{cr}}\approx \frac{\exp \left[ -\frac{\pi }{k_{2}}\left(
\omega _{2}+\pi _{2}\right) \right] }{\sinh \left( 2\pi \omega
_{2}/k_{2}\right) }\times \left\{
\begin{array}{c}
\sinh \left[ \pi \left( \omega _{2}-\pi _{2}\right) /k_{2}\right] \ \mathrm{%
for\ fermions} \\
\cosh \left( \pi \left( \omega _{2}-\pi _{2}\right) /k_{2}\right) \ \mathrm{%
for\ bosons}%
\end{array}%
\right. \,  \label{5.17}
\end{equation}%
in the leading-order approximation. The same distribution takes place in a
slowly varying field for $p_{x}>0$, see Eq.~(\ref{4.12a}). In the range (f)
the form of $N_{n}^{\mathrm{cr}}$ given by Eq.~(\ref{5.17}) does not depend
on details of switching on at $t=0$ if condition (\ref{6.3}) holds true. For
$m/k_{2}\gg 1$, distribution~(\ref{5.17}) is approximated by Eq. (\ref{5.15}%
).

Note that WKB approximation holds true under the condition $\left( \omega
_{2}+\pi _{2}\right) /k_{2}\gg 1$ for $N_{n}^{\mathrm{cr}}$ given by Eq.~(%
\ref{5.15}). In the range (e) $N_{n}^{\mathrm{cr}}$ given by (\ref{5.15})
coincides exactly with an estimation, obtained previously in \cite%
{Spokoinyia,Spokoinyib} in the framework of the semiclassical consideration.
We stress that in our consideration (which does not use the WKB) the the
approximation~(\ref{5.15}) holds for any value of $\left( \omega _{2}+\pi
_{2}\right) /k_{2}$ and exact results given by Eqs. (\ref{6.1}), (\ref{6.2}%
), and (\ref{5.17}) are quite different from the corresponding semiclassical
ones.

Now, we can estimate the number density{\large \ }$n^{\mathrm{cr}}$ of pairs
created by an exponentially decaying electric field, defined by Eq.~(\ref%
{asy5}). To this end, we represent the leading terms of integral (\ref{asy5}%
) as a sum of two contributions, one due to the ranges (e) and\textrm{\ }(f)
and another one due to the ranges \ (b), (c), and (d):
\begin{eqnarray}
&&n^{\mathrm{cr}}\approx \frac{J_{(d)}}{(2\pi )^{d-1}}\int_{\sqrt{\lambda }%
<K_{\bot }}d\mathbf{p}_{\bot }I_{\mathbf{p}_{\bot }},\ \ I_{\mathbf{p}_{\bot
}}=I_{\mathbf{p}_{\bot }}^{\left( 1\right) }+I_{\mathbf{p}_{\bot }}^{\left(
2\right) },  \notag \\
&&I_{\mathbf{p}_{\bot }}^{\left( 1\right) }=\int_{\pi _{2}\in \mathrm{%
(b)\cup (c)\cup (d)}}d\pi _{2}N_{n}^{\mathrm{cr}}\,,\ \ I_{\mathbf{p}_{\bot
}}^{\left( 2\right) }=\int_{\pi _{2}\in \mathrm{(e)\cup (f)}}d\pi _{2}N_{n}^{%
\mathrm{cr}}\,.  \label{5.18}
\end{eqnarray}%
Note that the mean numbers $N_{n}^{\mathrm{cr}}$ given by Eq.~(\ref{6.1}) in
the total range (a) and the mean numbers given by Eqs.~(\ref{5.17}) in the
range $-\pi _{2}\lesssim \pi _{\bot }$ are negligibly small. The main
contribution to the number density (\ref{5.18}) is due to the wide range (e)$%
\mathrm{\cup }$(f) of a large increment of the longitudinal kinetic momentum
$\pi _{2}$ with a relatively small transversal momentum $\left\vert \mathbf{p%
}_{\perp }\right\vert $. The contribution to this quantity from the
relatively narrow momentum ranges \ (b), (c), and (d)\ is finite and the
corresponding integral $I_{\mathbf{p}_{\bot }}^{\left( 1\right) }$ is of the
order $\sqrt{eE}/g_{2}$ . The integral $I_{\mathbf{p}_{\bot }}^{\left(
2\right) }$ can be taken from Eq.~(\ref{tot2}). Using the results of Sec. %
\ref{Ss6.2}, we can find the leading term in $I_{\mathbf{p}_{\bot }}^{\left(
2\right) }$,%
\begin{equation}
I_{\mathbf{p}_{\bot }}^{\left( 2\right) }\approx \frac{eE}{k_{2}}%
\int_{1}^{\infty }\frac{ds}{s^{2}}e^{-\pi \lambda s}=\frac{eE}{k_{2}}e^{-\pi
\lambda }G\left( 1,\pi \lambda \right) ,  \label{5.19}
\end{equation}%
where $G\left( \alpha ,x\right) $ is given by Eq.~(\ref{4.19a}). The
integral $I_{\mathbf{p}_{\bot }}^{\left( 1\right) }$ in Eq.~(\ref{5.18}) is
much less than the integral $I_{\mathbf{p}_{\bot }}^{\left( 2\right) }$ (\ref%
{5.19}), which represents the dominant contribution, $I_{\mathbf{p}_{\bot
}}\approx I_{\mathbf{p}_{\bot }}^{\left( 2\right) }$. In this case, the
range $\Omega $\ in Eq.~(\ref{asy6b}) is realized by the{\large \ }condition
$\sqrt{\lambda }<K_{\bot }$ and{\large \ \ }$\pi _{2}\in \mathrm{(e)\cup (f)}
${\large . }Then, calculating the Gaussian integral, we find%
\begin{equation}
\tilde{n}^{\mathrm{cr}}=\frac{r^{\mathrm{cr}}}{k_{2}}G\left( \frac{d}{2},\pi
\frac{m^{2}}{eE}\right) ,  \label{5.20}
\end{equation}%
where $r^{\mathrm{cr}}$ is given by Eq.~(\ref{t18.2}). We see that $\tilde{n}%
^{\mathrm{cr}}$ given by Eq.~(\ref{5.20}) is the $k_{2}$-dependent part of
the number density of pairs created in the slowly varying peak field~(\ref%
{4.19}).

Calculating the vacuum-to-vacuum probability, we obtain%
\begin{equation}
P_{v}=\exp \left( -\mu ^{\mathrm{P}}V_{\left( d-1\right) }\tilde{n}^{\mathrm{%
cr}}\right) ,  \label{5.21}
\end{equation}%
where $\tilde{n}^{\mathrm{cr}}$ is given by Eq.~(\ref{5.20}) and $\mu ^{%
\mathrm{P}}$ is given by Eq.~(\ref{4.20}).

As it was mentioned above, in the ranges of dominant contribution (e) and
(f) under condition (\ref{6.3}), the form of $N_{n}^{\mathrm{cr}}$ does not
depend on the details of the switching on at $t=0$. Therefore, calculations
in an exponentially decaying field are quite representative for a large
class of decaying electric fields switching  on abruptly.

\section{Universal behavior of the vacuum mean values in slowly varying
electric fields\label{S7}}

\subsection{Total density of created pairs\label{Ss7.1}}

As was recently discovered in our work \cite{GavGit17}, an information
derived from considerations of exactly solvable cases allows one to make
some general conclusions about quantum effects in slowly varying strong
fields for which no closed form solutions of the Dirac equation are known.
Below, we briefly represent such conclusions about an universal behavior of
vacuum mean values in slowly varying strong electric fields.

We note that in all these cases the quantity\ $N_{n}^{\mathrm{cr}}$ is
quasiconstant over the wide range of the longitudinal momentum $p_{x}$\ for
any given\emph{\ }$\lambda ,$ namely\emph{\ }$N_{n}^{\mathrm{cr}}\sim
e^{-\pi \lambda }$\emph{. }Pair creation effects in such fields are
proportional to large increments of the longitudinal kinetic momentum, $%
\Delta U=e\left\vert A_{x}\left( +\infty \right) -A_{x}\left( -\infty
\right) \right\vert $. Defining the slowly varying regime in general terms,
one can observe an universal character of vacuum effects caused by strong
electric field.\

We call $E(t)$ a slowly varying electric field on a time interval $\Delta t$
if the following condition holds true:%
\begin{equation}
\left\vert \frac{\overline{\dot{E}(t)}\Delta t}{\overline{E(t)}}\right\vert
\ll 1,\ \ \Delta t/\Delta t_{\mathrm{st}}^{\mathrm{m}}\gg 1,  \label{svf1}
\end{equation}%
where $\overline{E(t)}$ and $\overline{\dot{E}(t)}$ are mean values of $E(t)$
and $\dot{E}(t)$ on the time interval $\Delta t$, respectively, and $\Delta
t $ is significantly larger than the time scale $\Delta t_{\mathrm{st}}^{%
\mathrm{m}}$ which is
\begin{equation}
\;\Delta t_{\mathrm{st}}^{\mathrm{m}}=\Delta t_{\mathrm{st}}\max \left\{
1,m^{2}/e\overline{E(t)}\right\} ,\,\;\,\Delta t_{\mathrm{st}}=\left[ e%
\overline{E(t)}\right] ^{-1/2}.  \label{svf2}
\end{equation}

We are primarily interested in strong electric fields, $m^{2}/e\overline{E(t)%
}\lesssim 1$. In this case, inequality (\ref{svf2}) is simplified to the
form $\Delta t/\Delta t_{\mathrm{st}}\gg 1$, in which the mass $m$ is
absent. In such cases, the potential of the corresponding electric steps
hardly differs from the potential of a constant electric field,
\begin{equation}
U\left( t\right) =-eA_{x}\left( t\right) \approx U_{c}\left( t\right) =e%
\overline{E(t)}t+U_{0},  \label{4.39}
\end{equation}%
on the time interval $\Delta t$, where $U_{0}$ is a given constant. This
behavior is inherent to the fields of exact solvable cases presented above.

If the electric field is not very strong, mean numbers $N_{n}^{\mathrm{cr}}$
of created pairs (or distributions) at the final time instant are
exponentially small, $N_{n}^{\mathrm{cr}}\ll 1$. In this case the
probability of the vacuum to remain a vacuum and probabilities of particle
scattering and pair creation have simple representations in terms of these
numbers,%
\begin{equation}
\left\vert w_{n}\left( +-|0\right) \right\vert ^{2}\approx N_{n}^{\mathrm{cr}%
},\;\left\vert w_{n}\left( -|-\right) \right\vert ^{2}\approx \left(
1+N_{n}^{\mathrm{cr}}\right) ,\;P_{v}\approx 1-\sum_{n}N_{n}^{\mathrm{cr}}.
\label{a3.1}
\end{equation}%
The latter relations are often used in semiclassical calculations\emph{\ }to
find $N_{n}^{\mathrm{cr}}$\ and the total number of created pair, $N^{%
\mathrm{cr}}=\sum_{n}N_{n}^{\mathrm{cr}}$, from the representation of $P_{v}$%
\ given by Schwinger's effective action.

However, when the electric field cannot be considered as a weak one (e.g. in
some situations in astrophysics and condensed matter), the mean numbers $%
N_{n}^{\mathrm{cr}}$ can achieve their limited values $N_{n}^{\mathrm{cr}%
}\rightarrow 1$ already at finite time instants $t$ and the sum $N^{\mathrm{%
cr}}$ cannot be considered as a small quantity. Moreover, for slowly varying
strong electric fields this sum is proportional to the large parameter $%
T_{eff}/\Delta t_{\mathrm{st}}$. In such a case relations (\ref{a3.1}) are
not correct anymore.{\large \ }However, as shown next, for arbitrary slowly
varying strong electric field one can derive in the leading-term
approximation an universal form for the total density of created pairs.

Let us define the range $D\left( t\right) $ as follows:
\begin{equation}
D\left( t\right) :\left\langle P_{x}\left( t\right) \right\rangle
<0,\;\;\left\vert \left\langle P_{x}\left( t\right) \right\rangle
\right\vert \gg \pi _{\perp }.  \label{4.31a}
\end{equation}%
In this range the longitudinal kinetic momentum $\left\langle P_{x}\left(
t\right) \right\rangle =p_{x}-U\left( t\right) $ is negative and big enough.
If $p_{x}$ components of the particle momentum belongs to the range $D\left(
t\right) $, then the particle energy is in main determined by an increment
of the longitudinal kinetic momentum\emph{, }$U\left( t\right) -U\left( t_{%
\mathrm{in}}\right) $,\emph{\ }during the time interval\emph{\ }$t-t_{%
\mathrm{in}}$ and $\left\langle P_{x}\left( t\right) \right\rangle =$ $%
\left\langle P_{x}\left( t_{\mathrm{in}}\right) \right\rangle -\left[
U\left( t\right) -U\left( t_{\mathrm{in}}\right) \right] $. Note that $%
D\left( t\right) \subset D\left( t^{\prime }\right) $ if $t<t^{\prime }$.
The leading term of the total number density of created pairs, $\tilde{n}^{%
\mathrm{cr}}$, is formed over the range $D\left( t_{\mathrm{out}}\right) $,
that is, the range $D\left( t_{\mathrm{out}}\right) $ is chosen as a
realization of the subrange $\Omega $\ in Eq.~(\ref{asy6b}).

In the case when the electric field does not switch abruptly on and off,
that is, the field slowly weakens at $t\rightarrow \pm \infty $ and one of
the time instants $t_{\mathrm{in}}$ and $t_{\mathrm{out}}$ , or both are
infinite $t_{\mathrm{in}}\rightarrow -\infty $ and $t_{\mathrm{out}%
}\rightarrow \infty $, one can ignore exponentially small contributions to $%
\tilde{n}^{\mathrm{cr}}$ from the time intervals $\left( t_{\mathrm{in}},t_{%
\mathrm{in}}^{eff}\right] $ and $\left( t_{\mathrm{out}}^{eff},t_{\mathrm{out%
}}\right) $, where electric fields are much less than the maximum field $E$,
$E\left( t_{\mathrm{in}}^{eff}\right) ,E\left( t_{\mathrm{out}}^{eff}\right)
\ll E$. Thus, in the general case it is enough to consider a finite interval
$\left( t_{\mathrm{in}}^{eff},t_{\mathrm{out}}^{eff}\right] $. Denoting $%
t_{1}=t_{\mathrm{in}}^{eff}$ and $t_{M+1}=t_{\mathrm{out}}^{eff}$, we divide
this interval into $M$ intervals $\Delta t_{i}=t_{i+1}-t_{i}>0$, $i=1,...,M$%
, $\sum_{i=1}^{M}\Delta t_{i}=t_{\mathrm{out}}^{eff}-t_{\mathrm{in}}^{eff}$.
We suppose that Eqs.~(\ref{svf1}) and (\ref{svf2}) hold true for all the
intervals, respectively. That allows us to treat the electric field as
approximately constant within each interval, $\overline{E(t)}\approx
\overline{E}(t_{i})$, for $t\in \left( t_{i},t_{i+1}\right] $. Note that
inside of each interval $\Delta t_{i}$ abrupt changes of the electric field $%
E(t)$ whose duration is much less than $\Delta t_{i}$, cannot change
significantly the total value of $\tilde{n}^{\mathrm{cr}}$, since $N_{n}^{%
\mathrm{cr}}\leq 1$ for fermions. Using Eq.~(\ref{t18.2}) for the case of $T$%
-constant field, we can represent $\tilde{n}^{\mathrm{cr}}$ as the following
sum
\begin{eqnarray}
&&\tilde{n}^{\mathrm{cr}}=\sum_{i=1}^{M}\Delta \tilde{n}_{i}^{\mathrm{cr}%
},\;\Delta \tilde{n}_{i}^{\mathrm{cr}}\approx \frac{J_{(d)}}{(2\pi )^{d-1}}%
\int_{e\overline{E}(t_{i})t_{i}}^{e\overline{E}(t_{i})\left( t_{i}+\Delta
t_{i}\right) }dp_{x}\int_{\sqrt{\lambda _{i}}<K_{\bot }}d\mathbf{p}_{\bot
}N_{n}^{\left( i\right) }\,,  \notag \\
&&N_{n}^{\left( i\right) }=e^{-\pi \lambda _{i}},\ \ \lambda _{i}=\frac{\pi
_{\bot }^{2}}{e\overline{E}(t_{i})}\,,\   \label{uni1}
\end{eqnarray}%
where $K_{\bot }$ is any given number satisfying the condition$\;\sqrt{e%
\overline{E}(t_{i})}\Delta t_{i}\gg K_{\bot }^{2}\gg \max \left\{ 1,m^{2}/e%
\overline{E}(t_{i})\right\} $. Taking into account Eq.~(\ref{4.31a}), we
represent the variable $p_{x}$ as follows
\begin{equation}
p_{x}=U\left( t\right) ,\!\!\;\;U\left( t\right) =\int_{t_{\mathrm{in}%
}}^{t}dt^{\prime }eE\left( t^{\prime }\right) +U\left( t_{\mathrm{in}%
}\right) .  \label{uni2b}
\end{equation}%
Then neglecting small contributions to the integral (\ref{uni1}), we find
the following universal form for the total density of created pairs in the
leading-term approximation for a slowly varying, but otherwise arbitrary
strong electric field%
\begin{equation}
\tilde{n}^{\mathrm{cr}}\approx \frac{J_{(d)}}{(2\pi )^{d-1}}\int_{t_{\mathrm{%
in}}}^{t_{\mathrm{out}}}dteE\left( t\right) \int d\mathbf{p}_{\bot }N_{n}^{%
\mathrm{uni}},\ \ N_{n}^{\mathrm{uni}}=\exp \left[ -\pi \frac{\pi _{\bot
}^{2}}{eE\left( t\right) }\right] .  \label{uni2}
\end{equation}%
Note that $N_{n}^{\mathrm{uni}}$ is written in an universal form which can
be used to calculate any total characteristics of the pair creation effect.\
After the integration over $\mathbf{p}_{\bot }$, we finally obtain%
\begin{equation}
\tilde{n}^{\mathrm{cr}}=\frac{J_{(d)}}{(2\pi )^{d-1}}\int_{t_{\mathrm{in}%
}}^{t_{\mathrm{out}}}dt\left[ eE\left( t\right) \right] ^{d/2}\exp \left\{
-\pi \frac{m^{2}}{eE\left( t\right) }\right\} .  \label{uni3}
\end{equation}

These universal forms can be derived for bosons as well, if to restrict
forms of external electric fields. Namely, by fields that have no abrupt
variations of $E(t)$ that can produce significant grow of $N_{n}^{\mathrm{cr}%
}$ on a finite time interval. In fact, in this case we have to include in
the range $D\left( t\right) $ the only subranges where $N_{n}^{\mathrm{cr}%
}\leq 1$. In this case the universal forms for bosons are the same (\ref%
{uni2}) and (\ref{uni3}) assuming that $J_{(d)}$ is the number of the boson
spin degrees of freedom, in particular, $J_{(d)}=1$ for scalar particles and
$J_{(4)}=3$ for vector particles.

Using the identity $-\kappa \ln \left( 1-\kappa N_{n}^{\mathrm{uni}}\right)
=N_{n}^{\mathrm{uni}}+\left( -1\right) ^{\left( 1-\kappa \right) /2}\left(
N_{n}^{\mathrm{uni}}\right) ^{2}\ldots $, in the same manner one can derive
an universal form of the vacuum-to-vacuum transition probability $P_{v}$
defined for fermions ($\kappa =+1$) and bosons ($\kappa =-1$) by Eq.~(\ref%
{vacprob}). Performing the integration over $\mathbf{p}_{\bot }$, we obtain
that
\begin{equation}
P_{v}\approx \exp \left\{ -\frac{V_{\left( d-1\right) }J_{(d)}}{(2\pi )^{d-1}%
}\sum_{l=0}^{\infty }\int_{t_{\mathrm{in}}}^{t_{\mathrm{out}}}dt\left(
-1\right) ^{\left( 1-\kappa \right) l/2}\frac{\left[ eE\left( t\right) %
\right] ^{d/2}}{\left( l+1\right) ^{d/2}}\exp \left[ -\pi \frac{\left(
l+1\right) m^{2}}{eE\left( t\right) }\right] \right\} .  \label{uni5}
\end{equation}

Using Eqs.~(\ref{uni3}) and (\ref{uni5}),{\large \ }one obtains the same
expressions{\large \ }(\ref{asy10}), (\ref{t18.2}), and (\ref{4.19}) for the
total densities and expressions (\ref{asy11}), (\ref{t19}), and (\ref{4.20})
for the vacuum-to-vacuum transition probabilities that were found in the
corresponding exactly solvable cases. Thus, we have an independent
confirmation of the universal forms obtained above. These representations do
not require knowledge of corresponding solutions of the relativistic wave
equations.

The representation (\ref{uni5}) coincides with the leading term
approximation of derivative expansion results from field-theoretic
calculations obtained in Refs.~\cite{DunHal98,GusSh99a,GusSh99b} for $d=3$
and $d=4$, respectively. In this approximation the probability $P_{v}$ was
derived from a formal expansion in increasing numbers of derivatives of the
background field strength for Schwinger's effective action:%
\begin{equation}
S=S^{\left( 0\right) }[F_{\mu \nu }]+S^{\left( 2\right) }[F_{\mu \nu
},\partial _{\mu }F_{\nu \rho }]+...  \label{uni7b}
\end{equation}%
where $S^{\left( 0\right) }$ involves no derivatives of the background field
strength $F_{\mu \nu }$ (that is, $S^{\left( 0\right) }$ is a locally
constant field approximation for $S$), while the first correction $S^{\left(
2\right) }$ involves two derivatives of the field strength, and so on, see
Ref.~\cite{Dunn04} for a review. It was found that%
\begin{equation}
P_{v}=\exp \left( -2\mathrm{Im}S^{\left( 0\right) }\right) .  \label{uni7}
\end{equation}%
In the derivative expansion the fields are assumed to vary very slowly and
satisfy the condition (\ref{svf1}). A very convenient formalism for doing
such an expansion is the worldline formalism, see \cite{Schub01} for the
review, in which the effective action is written as a quantum mechanical
path integral.

However, for a general background field, it is extremely difficult to
estimate and compare the magnitude of various terms in the derivative
expansion. Only under the assumption $m^{2}/eE>1,$ one can demonstrate that
the derivative expansion is completely consistent with the semiclassical WKB
analysis of the imaginary part of the effective action \cite{DunnH99}. It is
shown only for a constant electric field that Eq.~(\ref{uni7}) is given
exactly by the semiclassical WKB limit when the leading order of
fluctuations is taken into account \cite{GordonS15}.

It should be stressed that unlike to the authors of Refs.~\cite%
{DunHal98,GusSh99a,GusSh99b}, Eq.~(\ref{uni5}) is derived in the framework
of the general exact formulation of strong-field QED \cite{FGS,GavGT06},
where $P_{v}$ are defined by Eq.~(\ref{vacprob}). Therefore, the obtained
result holds true for any strong field under consideration.{\large \ }In
particular, it is proven that Eq.~(\ref{uni7}) is given exactly by the
semiclassical WKB limit for arbitrary slowly varying electric field.

\subsection{Time evolution of vacuum instability\label{Ss7.2}}

In this section details of the time evolution of vacuum instability effects
are of interest. In particular, the study of the time evolution of the mean
electric current, energy, and momentum\emph{\ }provides us with new
characteristics of the effect, related, in particular, to the back reaction.
Due to the translational invariance of the spatially uniform external field,
all the corresponding mean values are proportional to the space volume.
Therefore, it is enough to calculate the vacuum mean values of the current
density vector $\langle j^{\mu }(t)\rangle $ and of the EMT $\langle T_{\mu
\nu }(t)\rangle $, defined by Eq.~(\ref{int1}). Note that these densities
depend on the initial vacuum, on the evolution of the electric field from
the initial time instant\ up to the current time instant $t$, but they do
not depend on the further history of the system and definition of
particle-antiparticle at the time $t$.

Let us consider the time dependence of the current density vector $\langle
j^{\mu }(t)\rangle $\ and of the EMT $\langle T_{\mu \nu }(t)\rangle $,
given by Eqs.~(\ref{A1.4}). Due to the uniform character of the
distributions $N_{n}^{\mathrm{cr}}$, the only diagonal matrix elements of
EMT differ from zero and the only longitudinal current components are not
zero if $d\neq 3$ . In $d=3$ dimensions, a non-zero current component $%
\langle j^{2}(t)\rangle $ can exist, this fact is related to the so-called
Chern-Simons term in the effective action, see details in Ref.~\cite%
{GavGitY12}. However, if there are both fermion species in a model, as it
takes place, for example, in the Dirac model of the graphene, then $\langle
j^{2}(t)\rangle =0$.

It follows from Eqs.~(\ref{3.25}) and (\ref{A1.4}) that the nonzero terms $%
\mathrm{Re}\langle j^{\mu }(t)\rangle ^{p}\,$ and $\mathrm{Re}\langle T_{\mu
\nu }(t)\rangle ^{p}$ appear due to the vacuum instability. These terms are
growing with time due to an increase of the number of states that are
occupied by created pairs. In any system of Fermi particles the mean value $%
\langle j^{2}(t)\rangle $ is finite.

As a consequence of Eq.~(\ref{4.31a}){\large , }we have
\begin{equation}
i\partial _{t}\;^{\pm }\varphi _{n}\left( t\right) \approx \pm \left\vert
\left\langle P_{x}\left( t\right) \right\rangle \right\vert \;^{\pm }\varphi
_{n}\left( t\right) ,  \label{4.31b}
\end{equation}%
which means that at the time $t$ we deal with an ultrarelativistic particle\
and its kinetic momentum $\left\langle P_{x}\left( t\right) \right\rangle $
can be considered as a large parameter. Considering time dependence of means
$\mathrm{Re}\,\langle j^{1}(t)\rangle ^{p}$ and $\mathrm{Re}\,\langle T_{\mu
\mu }(t)\rangle ^{p}$, we suppose that the time difference $t-t_{\mathrm{in}%
} $ is big enough to satisfy Eq.~(\ref{4.31b}). Using exact relation Eq.~(%
\ref{t4.4}) to express solutions $_{\pm }\psi _{n}$ via $^{\pm }\psi _{n}$,
and neglecting strongly oscillating terms, we find that leading contribution
to the function $S^{p}(x,x^{\prime })$ (defined by Eq.~(\ref{3.25})) at $%
t\sim t^{\prime }$\ can be represented by the following expression
\begin{equation}
S^{p}(x,x^{\prime })\approx -i\sum_{n}N_{n}^{\mathrm{cr}}\left[ ^{+}{\psi }%
_{n}(x)^{+}{\bar{\psi}}_{n}(x^{\prime })-\,^{-}{\psi }_{n}(x)^{-}{\bar{\psi}}%
_{n}(x^{\prime })\right] \,.  \label{4.30}
\end{equation}%
It is clear that for any large enough difference{\large \ }$t-t_{\mathrm{in}%
} ${\large \ }the integral over momentum $p$\ in the right hand side of Eq. (%
\ref{4.30}) can be approximated by an integral over the range{\large \ }$%
D\left( t_{\mathrm{out}}\right) ${\large \ }that gives the dominant
contribution to the mean number of created particles with respect to the
total increment of the longitudinal kinetic momentum. Moreover, taking into
account Eqs.~(\ref{4.31a}) and (\ref{uni2b}), we see that{\large \ }$D\left(
t\right) \subset D\left( t^{\prime }\right) \subset D\left( t_{\mathrm{out}%
}\right) ${\large \ }if{\large \ }$t<t^{\prime }<t_{\mathrm{out}}${\large \ }%
and for a given difference{\large \ }$t-t_{\mathrm{in}}$ the dominant
contribution to the right hand side of Eq. (\ref{4.30}) is from a subrange%
{\large \ }$D\left( t\right) \subset D\left( t_{\mathrm{out}}\right) $%
{\large .}

We recall that, according to Eq.~(\ref{t4.10}), one can choose the
corresponding \textrm{in}- and \textrm{out}- Dirac solutions either with $%
\chi =+1$ or with $\chi =-1$. Using this possibility, we choose $\chi =+1$
for $^{+}{\psi }_{n}(x)$ and $\chi =-1$ for $^{-}{\psi }_{n}(x).$ With such
a choice, taking into account that $\mathbf{p\in }D\left( t\right) $, we
simplify essentially the matrix structure of the representation (\ref{4.30}%
). Thus, after a summation over spin polarizations $\sigma $, we obtain the
following result:%
\begin{equation}
S^{p}(x,x^{\prime })\approx (\gamma P+m)\Delta ^{p}(x,x^{\prime }),
\label{4.32}
\end{equation}%
where the function $\Delta ^{p}(x,x^{\prime })$ reads%
\begin{eqnarray*}
&&\Delta ^{p}(x,x^{\prime })=-i\sum_{\mathbf{p\in }D\left( t\right) }N_{n}^{%
\mathrm{cr}}\left\vert \left\langle P_{x}\left( t\right) \right\rangle
\right\vert \exp \left[ i\mathbf{p}\left( \mathbf{r-r}^{\prime }\right) %
\right] \\
&&\times \left\{ \left( 1+\gamma ^{0}\gamma ^{1}\right) \left. \left[
\;^{+}\varphi _{n}\left( t\right) \;^{+}\varphi _{n}^{\ast }\left( t^{\prime
}\right) \right] \right\vert _{\chi =+1}+\left( 1-\gamma ^{0}\gamma
^{1}\right) \left. \left[ \;^{-}\varphi _{n}\left( t\right) \;^{-}\varphi
_{n}^{\ast }\left( t^{\prime }\right) \right] \right\vert _{\chi
=-1}\right\} .
\end{eqnarray*}

Using Eq. (\ref{4.32}) in~Eq. (\ref{A1.4}), we find{\large \ }the following
representations for the vacuum means of current density and EMT components:
\begin{align}
& \langle j^{1}(t)\rangle ^{p}\approx 2e\frac{V_{\left( d-1\right) }J_{(d)}}{%
(2\pi )^{d-1}}\int_{\mathbf{p\in }D\left( t\right) }d\mathbf{p}N_{n}^{%
\mathrm{cr}}\rho \left( t\right) \left\vert \left\langle P_{x}\left(
t\right) \right\rangle \right\vert ;  \notag \\
& \langle T_{00}(t)\rangle ^{p}\approx \langle T_{11}(t)\rangle
^{p}\,\approx \frac{V_{\left( d-1\right) }J_{(d)}}{(2\pi )^{d-1}}\int_{%
\mathbf{p\in }D\left( t\right) }d\mathbf{p}N_{n}^{\mathrm{cr}}\rho \left(
t\right) \left\langle P_{x}\left( t\right) \right\rangle ^{2},  \notag \\
& \langle T_{ll}(t)\rangle ^{p}\,\approx \frac{V_{\left( d-1\right) }J_{(d)}%
}{(2\pi )^{d-1}}\int_{\mathbf{p\in }D\left( t\right) }d\mathbf{p}N_{n}^{%
\mathrm{cr}}\rho \left( t\right) p_{l}^{2},\;l=2,...,D,  \notag \\
& \rho \left( t\right) =2\left\vert \left\langle P_{x}\left( t\right)
\right\rangle \right\vert \left\{ \left. \left\vert \;^{+}\varphi _{n}\left(
t\right) \right\vert ^{2}\right\vert _{\chi =+1}+\left. \left\vert
\;^{-}\varphi _{n}\left( t\right) \right\vert ^{2}\right\vert _{\chi
=-1}\right\} .  \label{4.33}
\end{align}

In what follows we show that some universal behavior of the densities{\large %
\ }$\langle j^{1}(t)\rangle ^{p}${\large \ }and{\large \ }$\langle T_{\mu
\mu }(t)\rangle ^{p}${\large \ }can be derived from general forms~(\ref{4.33}%
) for any large difference{\large . }We begin the demonstration of this fact
with the case of a finite interval of time when the electric field potential
can be approximated by a potential of a constant electric field~(\ref{4.39}%
). At the same time, we assume that $\left\langle P_{x}\left( t\right)
\right\rangle $ satisfies condition (\ref{4.31a}) at the time $t$. It is
convenient to compare the cases of $T$-constant and exponentially decaying
fields, which both are abruptly switching on but their ways of switching off
may be different.

In the case of exponentially decaying field, the functions $^{\pm }\varphi
_{n}\left( t\right) $ in Eq.~(\ref{4.33}) are given by the second line in
Eq.~(\ref{i.4.1}) and approximation (\ref{4.39}) holds if $k_{2}t\ll 1$.
Then $\left\vert \left\langle P_{x}\left( t\right) \right\rangle \right\vert
\ll \left\vert \pi _{2}\right\vert $. To obtain functions $\;^{\pm }\varphi
_{n}\left( t\right) $ in such an approximation we use the asymptotic
representation (\ref{A10a}). Thus, we obtain%
\begin{equation}
\rho \left( t\right) =\left[ V_{\left( d-1\right) }\left\vert \left\langle
P_{x}\left( t\right) \right\rangle \right\vert \right] ^{-1}.  \label{4.40}
\end{equation}%
In the range {\large \ }$D\left( t\right) $, the distribution $N_{n}^{%
\mathrm{cr}}$ is approximately given by Eq.~(\ref{asy4}). Finally we obtain%
\begin{eqnarray}
&\langle j^{1}(t)\rangle ^{p}\approx &2er^{\mathrm{cr}}\Delta t,  \notag \\
&\langle T_{00}(t)\rangle ^{p}\approx &\langle T_{11}(t)\rangle ^{p}\approx
eEr^{\mathrm{cr}}\Delta t^{2},  \notag \\
&\langle T_{ll}(t)\rangle ^{p}\approx &\pi ^{-1}r^{\mathrm{cr}}\ln \left(
\sqrt{eE}\Delta t\right) \;\mathrm{if}\;l=2,...,D,  \label{4.41}
\end{eqnarray}%
where $\Delta t=t-t_{\mathrm{in}}$ is the duration time of a constant field.
In this case $t_{\mathrm{in}}=0$.

The field potential of the $T$-constant field (\ref{t7}) has the form (\ref%
{4.39}) in the intermediate region{\Huge \ }$\mathrm{II}$. For{\Large \ }%
sufficiently large times $t<t_{\mathrm{out}}$,\ when the longitudinal
kinetic momentum belongs to the range {\large \ }$D\left( t\right) $, the
distribution $N_{n}^{\mathrm{cr}}$ is approximately given by Eq.~(\ref{asy4}%
). In this case, exact expressions for the functions $\;^{+}\varphi
_{n}\left( t\right) $, given by Eq.~(\ref{t8}), and similar expressions for
the functions $\;^{-}\varphi _{n}\left( t\right) $ can be approximated as
the following WPCFs:%
\begin{eqnarray}
&&\ ^{+}\varphi _{n}\left( t\right) \approx V_{\left( d-1\right)
}^{-1/2}CD_{-1-\rho }[(1+i)\xi ],\ \;^{-}\varphi _{n}\left( t\right) \approx
V_{\left( d-1\right) }^{-1/2}CD_{\rho }[(1-i)\xi ],  \notag \\
&&\xi =\left( eEt-p_{x}\right) \left( eE\right) ^{-1/2},\ \ \ C=\left(
2eE\right) ^{-1/2}\exp \left( -\pi \lambda /8\right) \,.  \label{4.42}
\end{eqnarray}%
Then we find from Eq.~(\ref{4.33}) that the densities $\langle
j^{1}(t)\rangle ^{p}$ and $\langle T_{\mu \mu }(t)\rangle ^{p}$ have the
same form (\ref{4.41}) with $t_{\mathrm{in}}=-T/2$.

Note that the above results are obtained by using functions $^{\pm }\varphi
_{n}\left( t\right) $, which have in and out-asymptotics at $t_{\mathrm{out}%
} $.{\large \ }Nevertheless, these results show also that densities (\ref%
{4.41}) are not affected by evolution of the functions $^{\pm }\varphi
_{n}\left( t\right) $\ from $t$ to $t_{\mathrm{out}}$ in the range $p\in
D\left( t\right) $, assuming that the corresponding electric field exists
during a macroscopically large time period $\Delta t$, satisfying Eq.~(\ref%
{svf1}). This fact is closely related with a characteristic property of the
kernel of integrals (\ref{4.33}){\large , }which will be derived from an
universal form of the total density of created pairs given by Eq.~(\ref{uni2}%
).{\large \ }Let $t_{\mathrm{out}}^{\prime }<t_{\mathrm{out}}$\ is another
possible final time instant, then%
\begin{equation}
\tilde{n}^{\mathrm{cr}}\left( t_{\mathrm{out}}^{\prime }\right) \approx
\frac{J_{(d)}}{(2\pi )^{d-1}}\int_{t_{\mathrm{in}}}^{t_{\mathrm{out}%
}^{\prime }}dt\left[ eE\left( t\right) \right] ^{d/2}\exp \left\{ -\pi \frac{%
m^{2}}{eE\left( t\right) }\right\}  \label{uni8}
\end{equation}%
Eq.~(\ref{uni8}) corresponds to the assumption that in the range $p\in \
D\left( t_{\mathrm{out}}^{\prime }\right) \subset D\left( t_{\mathrm{out}%
}\right) $ the electric field is switched on at $t_{\mathrm{in}}$\ and
switched off at $t_{\mathrm{out}}^{\prime }$. Then instead of functions\ \ $%
^{\zeta }\psi _{n}\left( x\right) $\ satisfying the eigenvalue problem (\ref%
{t4b}), we have to use solutions of the following eigenvalue problem%
\begin{equation*}
H\left( t\right) \ ^{\zeta }\psi _{n}^{\left( t_{\mathrm{out}}^{\prime
}\right) }\left( x\right) =\ ^{\zeta }\varepsilon _{n}\ ^{\zeta }\psi
_{n}^{\left( t_{\mathrm{out}}^{\prime }\right) }\left( x\right) \,,\ \ t\in %
\left[ t_{\mathrm{out}}^{\prime },+\infty \right) \,,\ ^{\zeta }\varepsilon
_{n}=\zeta p_{0}\left( t_{\mathrm{out}}^{\prime }\right) \,.
\end{equation*}%
Using the representation%
\begin{equation*}
\ ^{\zeta }\psi _{n}^{\left( t_{\mathrm{out}}^{\prime }\right) }\left(
x\right) =\left[ i\partial _{t}+H\left( t\right) \right] \gamma ^{0}\exp
\left( i\mathbf{pr}\right) \ ^{\zeta }\varphi _{n}^{\left( t_{\mathrm{out}%
}^{\prime }\right) }\left( t\right) v_{\chi ,\sigma }
\end{equation*}%
we obtain%
\begin{eqnarray}
&&\ ^{\zeta }\varphi _{n}^{\left( t_{\mathrm{out}}^{\prime }\right) }\left(
t\right) =\ ^{\zeta }N^{\left( t_{\mathrm{out}}^{\prime }\right) }\exp \left[
-i\zeta p_{0}\left( t_{\mathrm{out}}^{\prime }\right) \left( t-t_{\mathrm{out%
}}^{\prime }\right) \right] \,,\ \ t\in \left[ t_{\mathrm{out}}^{\prime
},+\infty \right) \,,  \notag \\
&&\ ^{\zeta }N^{\left( t_{\mathrm{out}}^{\prime }\right) }=\left(
2p_{0}\left( t_{\mathrm{out}}^{\prime }\right) \left\{ p_{0}\left( t_{%
\mathrm{out}}^{\prime }\right) -\chi \zeta \left[ p_{x}-U\left( t_{\mathrm{%
out}}^{\prime }\right) \right] \right\} V_{\left( d-1\right) }\right)
^{-1/2}\ .  \label{uni9}
\end{eqnarray}%
Thus, leading contribution to the function $S^{p}(x,x^{\prime })$ (defined
by Eq.~(\ref{3.25})) at $t^{\prime }\sim $ $t<t_{\mathrm{out}}^{\prime }$
can be expressed via $\ ^{\zeta }\psi _{n}^{\left( t_{\mathrm{out}}^{\prime
}\right) }\left( x\right) $ as follows%
\begin{equation}
S^{p}(x,x^{\prime })\approx -i\sum_{\sigma ,\mathbf{p\in }D\left( t\right)
}N_{n}^{\mathrm{cr}}\left[ \ ^{+}{\psi }_{n}^{\left( t_{\mathrm{out}%
}^{\prime }\right) }(x)\ ^{+}{\bar{\psi}}_{n}^{\left( t_{\mathrm{out}%
}^{\prime }\right) }(x^{\prime })-\,\ ^{-}{\psi }_{n}^{\left( t_{\mathrm{out}%
}^{\prime }\right) }(x)\ ^{-}{\bar{\psi}}_{n}^{\left( t_{\mathrm{out}%
}^{\prime }\right) }(x^{\prime })\right] \,.  \label{uni10}
\end{equation}%
Then $\rho \left( t\right) $\ in Eq.~(\ref{4.33}) can be represented as%
\begin{equation*}
\rho \left( t\right) =2\left\vert \left\langle P_{x}\left( t\right)
\right\rangle \right\vert \left\{ \left. \left\vert \ ^{+}\varphi
_{n}^{\left( t_{\mathrm{out}}^{\prime }\right) }\left( t\right) \right\vert
^{2}\right\vert _{\chi =+1}+\left. \left\vert \ ^{-}\varphi _{n}^{\left( t_{%
\mathrm{out}}^{\prime }\right) }\left( t\right) \right\vert ^{2}\right\vert
_{\chi =-1}\right\} .
\end{equation*}%
Taken into account Eq.~(\ref{uni9}), we can see that Eq.~(\ref{4.40}) holds
for any large time difference $t-t_{\mathrm{in}}$. Using the universal form
of the differential numbers of created pairs, \ $N_{n}^{\mathrm{cr}}\approx
N_{n}^{\mathrm{uni}}$, given by Eq.~(\ref{uni2}), changing the variable
according to Eq.~(\ref{uni2b}), and performing the integration over $p_{\bot
}$, we find from Eq.~(\ref{4.33}) that the vacuum mean values of current and
EMT components have the following universal behavior for any large
difference $t-t_{\mathrm{in}}$:%
\begin{eqnarray}
&\langle j^{1}(t)\rangle ^{p}\approx &2e\tilde{n}^{\mathrm{cr}}\left(
t\right) ,  \notag \\
&\langle T_{00}(t)\rangle ^{p}\approx &\langle T_{11}(t)\rangle ^{p}\approx
\frac{J_{(d)}}{(2\pi )^{d-1}}\int_{t_{\mathrm{in}}}^{t}dt^{\prime }\left[
U\left( t\right) -U\left( t^{\prime }\right) \right] \left[ eE\left(
t^{\prime }\right) \right] ^{d/2}\exp \left[ -\frac{\pi m^{2}}{eE\left(
t^{\prime }\right) }\right] ,  \notag \\
&\langle T_{ll}(t)\rangle ^{p}\approx &\frac{J_{(d)}}{(2\pi )^{d}}\int_{t_{%
\mathrm{in}}}^{t}\frac{dt^{\prime }\left[ eE\left( t^{\prime }\right) \right]
^{d/2+1}}{\left[ U\left( t\right) -U\left( t^{\prime }\right) \right] }\exp %
\left[ -\frac{\pi m^{2}}{eE\left( t^{\prime }\right) }\right] ,\;l=2,...,D\ .
\label{uni11}
\end{eqnarray}%
{\large \ }Here $\tilde{n}^{\mathrm{cr}}\left( t\right) $\ is given by Eq.~(%
\ref{uni8}).

\emph{\ }For $t>t_{\mathrm{out}}$, the pair production stops, vacuum
polarization effects disappear, and quantities ~(\ref{uni11}) for $t>t_{%
\mathrm{out}}$ maintain their values at $t=t_{\mathrm{out}}$,%
\begin{equation}
\left. \langle j^{1}(t)\rangle \right\vert _{t>t_{\mathrm{out}}}\approx
\langle j^{1}(t_{\mathrm{out}})\rangle ^{p},\;\left. \langle T_{\mu \mu
}(t)\rangle \right\vert _{t>t_{\mathrm{out}}}\approx \langle T_{\mu \mu }(t_{%
\mathrm{out}})\rangle ^{p}.  \label{uni12}
\end{equation}%
Note that $\tilde{n}^{\mathrm{cr}}\left( t_{\mathrm{out}}\right) =\tilde{n}^{%
\mathrm{cr}}$\ is the number density of created real pairs, given by Eq.~(%
\ref{uni3}), that is, it is the number density of electrons and positrons
detectable at any time instant after switching of an electric field off. The
quantities $\langle j^{1}(t_{\mathrm{out}})\rangle ^{p}$\ and $\langle
T_{\mu \mu }(t_{\mathrm{out}})\rangle ^{p}$\ are the mean current density
and the EMT components of real pairs created from the vacuum.{\large \ }The
energy density $\langle T_{00}(t_{\mathrm{out}})\rangle $ is equal to the
pressure $\langle T_{11}(t_{\mathrm{out}})\rangle $ along the direction of
the electric field at the time instant $t_{\mathrm{out}}$. This equality is
a natural equation of state for noninteracting particles accelerated by an
electric field to relativistic velocities.

In particular, for fields admitting exactly solvable cases, we find from
Eqs.~(\ref{uni11}) and (\ref{uni12}):

(a) For $T$-constant field
\begin{eqnarray}
&&\langle T_{00}(t_{\mathrm{out}})\rangle ^{p}\approx \langle T_{11}(t_{%
\mathrm{out}})\rangle ^{p}\approx eEr^{\mathrm{cr}}\left( t_{\mathrm{out}%
}-t_{\mathrm{in}}\right) ^{2},  \notag \\
&&\langle T_{ll}(t_{\mathrm{out}})\rangle ^{p}\approx \pi ^{-1}r^{\mathrm{cr}%
}\ln \left[ \sqrt{eE}\left( t_{\mathrm{out}}-t_{\mathrm{in}}\right) \right]
,\;l=2,...,D.  \label{4.37}
\end{eqnarray}

(b) For the peak field:%
\begin{eqnarray}
&&\langle T_{00}(t_{\mathrm{out}})\rangle ^{p}\approx \langle T_{11}(t_{%
\mathrm{out}})\rangle ^{p}\approx eEr^{\mathrm{cr}}\left[
k_{2}^{-1}+k_{1}^{-1}\right]  \notag \\
&&\times \left\{ \left[ k_{2}^{-1}-k_{1}^{-1}\right] G\left( \frac{d}{2}+1,%
\frac{\pi m^{2}}{eE}\right) +k_{1}^{-1}G\left( \frac{d}{2},\frac{\pi m^{2}}{%
eE}\right) \right\} ,  \notag \\
&&\langle T_{ll}(t_{\mathrm{out}})\rangle ^{p}\approx \frac{r^{\mathrm{cr}}}{%
2\pi }\left[ G\left( \frac{d}{2}-1,\frac{\pi m^{2}}{eE}\right) +\frac{k_{2}}{%
k_{1}}G\left( \frac{d}{2},\frac{\pi m^{2}}{eE}\right) \right] ,\;l=2,...,D.
\label{4.38}
\end{eqnarray}%
Densities (\ref{4.38}) correspond to the case of an exponentially decaying
field as $k_{1}^{-1}\rightarrow 0$.

(c) For\ Sauter-like\ field:%
\begin{eqnarray}
&&\langle T_{00}(t_{\mathrm{out}})\rangle ^{p}\approx \langle T_{11}(t_{%
\mathrm{out}})\rangle ^{p}\approx eEr^{\mathrm{cr}}T_{\mathrm{S}}^{2}\left[
\delta -G\left( \frac{d}{2},\frac{\pi m^{2}}{eE}\right) \right] ,  \notag \\
&&\langle T_{ll}(t_{\mathrm{out}})\rangle ^{p}\approx \frac{r^{\mathrm{cr}}}{%
2\pi }\left[ \sqrt{\pi }\Psi \left( \frac{1}{2},2-\frac{d}{2};\frac{\pi m^{2}%
}{eE}\right) +G\left( \frac{d}{2}-1,\frac{\pi m^{2}}{eE}\right) \right] .
\label{4.38b}
\end{eqnarray}

Note that using the differential mean numbers of created pairs given by
Eqs.~(\ref{asy2}), (\ref{asy4}), (\ref{4.10}), and (\ref{4.13}) for the
exactly solvable cases [without the use of the universal form given by Eq.~(%
\ref{uni2})], we obtain from Eq.~(\ref{4.33})~literally expressions{\large \
}(\ref{4.37}) (earlier obtained in Refs.~\cite{GG06-08a,GG06-08b,GavGitY12}%
), (\ref{4.38}), and (\ref{4.38b}). It is an independent confirmation of
universal form (\ref{uni11}).{\large \ }

The obtained results show that the scale $\Delta t_{\mathrm{st}}^{\mathrm{m}%
} $ plays the role of the stabilization time for the densities $\langle
j^{1}(t)\rangle ^{p}$ and $\langle T_{\mu \mu }(t)\rangle ^{p}$. The
characteristic parameter $m^{2}/eE$ can be represented as the ratio of two
characteristic lengths: $c^{3}m^{2}/\hslash eE=\left( c\Delta t_{\mathrm{st}%
}/\Lambda _{C}\right) ^{2}$, where $\Lambda _{C}=\hbar /mc$ is the Compton
wavelength. In strong electric fields, $\left( c\Delta t_{\mathrm{st}%
}/\Lambda _{C}\right) ^{2}\lesssim 1$, inequality (\ref{svf2}) is simplified
to the form $\Delta t/\Delta t_{\mathrm{st}}\gg 1$, in which the Compton
wavelength is absent. We see that the scale $\Delta t_{\mathrm{st}}$ plays
the role of the stabilization time for a strong electric field. This means
that $\Delta t_{\mathrm{st}}$ is a characteristic time scale which allows us
to distinguish fields that have microscopic or macroscopic time change, it
plays similar role as the Compton wavelength plays in the case of a weak
field. Therefore, calculations in a $T$-constant field are quite
representative for a large class of slowly varying electric fields.

Under natural assumptions, the parameter $e\overline{E(t)}\Delta t^{2}$ is
limited. Considering problems of high-energy physics in $d=4,$ it is usually
assumed that just from the beginning there exists an uniform classical
electric field with a given energy density. This field can be modelled by the%
{\large \ }$T$-constant{\large \ }field. The system of particles interacting
with this field is closed, that is, the total energy of the system is
conserved. Under such an assumption, the pair creation is a transient
process and, for example, the applicability of the constant field
approximation is limited by the smallness of the back reaction, which
implies the following restriction from above:%
\begin{equation}
\left( \Delta t/\Delta t_{\mathrm{st}}\right) ^{2}\ll \frac{\pi ^{2}}{%
J\alpha }\,\exp \left( \pi \frac{c^{3}m^{2}}{\hslash eE}\right) ,
\label{4.48}
\end{equation}%
on time $\Delta t$ for a given electric field strength $E$. Here $\alpha $
is the fine structure constant and $J$ is the number of the spin degrees of
freedom, see \cite{GavG08}. Thus, there is a {range of} the parameters $E$
and $\Delta t$ where the approximation of the constant external field is
consistent.

It is well known that at certain conditions (the so-called charge neutrality
point) electronic excitations in graphene monolayer behave as relativistic
Dirac massless fermions in $2+1$ dimensions, with the Fermi velocity $%
v_{F}\simeq 10^{6}$ \textrm{m/s} playing the role of the speed of light, see
details in recent reviews \cite{dassarma,VafVish14}. Then in the range of
the applicability of the Dirac model to the graphene physics, any electric
field is strong. There appears a timescale specific to graphene (and to
similar nanostructures with the Dirac fermions), $\Delta t_{\mathrm{st}}^{%
\mathrm{g}}=\left( eEv_{F}/\hbar \right) ^{-1/2}\,,$ which plays the role of
the stabilization time in the case under consideration. The $T$-constant
field is suitable for imitating a slowly varying field under condition $%
\Delta t/\Delta t_{\mathrm{st}}^{\mathrm{g}}\gg 1$. The transport in
graphene can be considered as ballistic then the ballistic flight time $%
T_{bal}$ to be the effective time duration, $\Delta t=T_{bal}$. The external
constant electric field can be considered as a good approximation of the
effective mean field as long as the field produced by the induced current of
created particles, $\langle j^{1}(t)\rangle ^{p}$ given by Eq.~(\ref{4.41}),
is negligible compared to the applied field. {This implies the consistency
restriction }$\Delta t\ll \Delta t_{\mathrm{br}}=\Delta t_{\mathrm{st}}^{%
\mathrm{g}}/4\alpha $ \cite{GavGitY12}. Thus, in this case there is a {window%
} in voltages, $7\times 10^{-4}\,\mathrm{V}\ll V\ll 8\,\mathrm{V}$, where
the model with constant external field is consistent. These voltages are in
the range typically used in experiments with the graphene.

\subsection{Vacuum polarization\label{Ss7.3}}

In which follows we use the example of the $T$-constant field to consider the%
{\large \ }contributions{\large \ }$\mathrm{Re}\langle j^{\mu }(t)\rangle
^{c}${\large \ }and{\large \ }$\mathrm{Re}\langle T_{\mu \nu }(t)\rangle
^{c} ${\large \ }to the mean values of the current density $\langle j^{\mu
}(t)\rangle $\ and the EMT $\langle T_{\mu \nu }(t)\rangle $, given by Eqs.~(%
\ref{A1.4}). Note that the mean current density $\,\langle j^{\mu
}(t)\rangle $ and the physical part of the mean value $\langle T_{\mu \nu
}(t)\rangle $ are zero for any $t<t_{\mathrm{in}}$. For $t>t_{\mathrm{in}}$,
we are interested in these mean values only for a large time periods $\Delta
t=t-t_{\mathrm{in}}$ satisfying Eq.~ (\ref{svf1}). In this case, the
longitudinal kinetic momentum belongs to the range (\ref{4.31a}) and
distributions $N_{n}^{\mathrm{cr}}$ are approximated by Eq.~(\ref{asy4}).
Using approximation (\ref{4.42}), the functions $\;_{-}\varphi _{n}\left(
t\right) $, given by Eq.~(\ref{t8}), {\large \ }and similar functions $%
\;_{+}\varphi _{n}\left( t\right) $ can be taken in the following form%
\begin{equation}
\ _{-}\varphi _{n}\left( t\right) =V_{\left( d-1\right) }^{-1/2}CD_{-1-\rho
}[-(1+i)\xi ],\;_{+}\varphi _{n}\left( t\right) =V_{\left( d-1\right)
}^{-1/2}CD_{\rho }[-(1-i)\xi ].  \label{4.44}
\end{equation}%
In the same approximation, the causal propagator $S^{c}(x,x^{\prime })$ (\ref%
{3.22}) can be calculated using solutions $\ ^{\pm }\psi _{n}(x)$ and $\
_{\pm }\psi _{n}(x)$ with scalar functions given by Eqs.~(\ref{4.42}) and (%
\ref{4.44}) in the range (\ref{4.31a}). It can be shown that the main
contributions to $\mathrm{Re}\,\langle j^{\mu }(t)\rangle ^{c}$, $\langle
j^{2}(t)\rangle $ and $\mathrm{Re}\,\langle T_{\mu \mu }(t)\rangle ^{c}$ are
formed in the range (\ref{4.31a}) for a large time period $\Delta t$. It is
important that these contributions are independent of the interval $\Delta t$%
, that is, the densities $\mathrm{Re}\,\langle j^{\mu }(t)\rangle ^{c}$, $%
\langle j^{2}(t)\rangle $, and $\mathrm{Re}\,\langle T_{\mu \mu }(t)\rangle
^{c}$ are local quantities describing only vacuum polarization\ effects.
Then we integrate in Eq.~(\ref{3.22}) over all the momenta. Thus, we see
that in the case under consideration, the propagator $S^{c}(x,x^{\prime })$
can be approximated by the propagator in a constant uniform electric field.

The propagator $S^{c}(x,x^{\prime })$ in a constant uniform electric field
can be represented as the Fock--Schwinger proper time integral:%
\begin{equation}
S^{c}(x,x^{\prime })=(\gamma P+m)\Delta ^{c}(x,x^{\prime }),\;\;\Delta
^{c}(x,x^{\prime })=\int_{0}^{\infty }f(x,x^{\prime },s)ds\,,  \label{4.45}
\end{equation}%
see \cite{Nikis70a} and \cite{GGG98a,GGG98b}, where the Fock--Schwinger
kernel $f(x,x^{\prime },s)$ reads%
\begin{align}
& f(x,x^{\prime },s)=\exp \left( i\frac{e}{2}\sigma ^{\mu \nu }F_{\mu \nu
}s\right) f^{(0)}(x,x^{\prime },s)\,,\ \ f^{(0)}(x,x^{\prime },s)=-\frac{%
eEs^{-d/2+1}}{(4\pi i)^{d/2}\sinh (eEs)}  \notag \\
& \times \exp \left[ -i\left( e\Lambda +m^{2}s\right) +\frac{1}{4i}%
(x-x^{\prime })eF\coth (eFs)(x-x^{\prime })\right] .  \notag
\end{align}%
Here $\coth (eFs)$ is the matrix with the components $[\coth (eFs)]^{\mu
}{}_{\nu }$, $F_{\mu \nu }=E\left( \delta _{\mu }^{0}\delta _{\nu
}^{1}-\delta _{\mu }^{1}\delta _{\nu }^{0}\right) $, and $\Lambda
=(t+t^{\prime })(x_{1}-x_{1}^{\prime })E/2$, see \cite{Schwinger51,F37}.\

It is easy to see that $\langle j^{1}\left( t\right) \rangle ^{c}=0$, as
should be expected due to the translational symmetry. If $d=3$ there is a
transverse vacuum-polarization current,
\begin{equation}
\langle j^{2}(t)\rangle =\pm \frac{e^{2}}{4\pi ^{3/2}}\gamma \left( \frac{1}{%
2},\frac{\pi m^{2}}{eE}\right) E,  \label{emt2d}
\end{equation}%
for each $\pm $ fermion species, \cite{GavGitY12}, where $\gamma \left(
1/2,x\right) $ is the incomplete gamma function. Note that the transverse
current of created particles is absent, $\langle j^{2}(t)\rangle =0$ if $%
t>t_{\mathrm{out}}$. The factor in the front of $E$ in Eq. (\ref{emt2d}) can
be considered as a nonequilibrium Hall conductivity for large duration of
the electric field. In the presence of both $\pm $ species in a model, $%
\langle j^{2}(t)\rangle =0$ for any $t$.

Using Eq.~(\ref{4.45}), we obtain components of the EMT for the $T$-constant
field in the following form
\begin{align}
& \mathrm{Re}\langle T_{00}(t)\rangle ^{c}=-\mathrm{Re}\langle
T_{11}(t)\rangle ^{c}=E_{0}\frac{\partial \mathrm{Re}\mathcal{L}\left[ E%
\right] }{\partial E}-\mathrm{Re}\mathcal{L}\left[ E\right] \,,  \notag \\
& \mathrm{Re}\langle T_{ll}(t)\rangle ^{c}=\mathrm{Re}\mathcal{L}\left[ E%
\right] \,,\;l=2,...,D,  \label{4.46a}
\end{align}%
where
\begin{equation}
\mathcal{L}\left[ E\right] =\frac{1}{2}\int_{0}^{\infty }\frac{ds}{s}\mathrm{%
tr}f(x,x,s),\ \mathrm{tr}f(x,x,s)=2^{[d/2]}\cosh (eEs)f^{(0)}(x,x,s).
\label{ELaa}
\end{equation}

The quantity $\mathcal{L}\left[ E\right] $ can be identified with a
non-renormalized one-loop effective Euler-Heisenberg Lagrangian of the Dirac
field in an uniform constant electric field $E$. Note that components $%
\mathrm{Re}\langle T_{\mu \nu }(t)\rangle ^{c}$ do not depend of the time
duration $\Delta t$\ of the $T$-constant field if $\Delta t$ is sufficiently
large.

This result can be generalized to the case of arbitrary slowly varying
electric field{\large . }To this end we divide as before the finite interval
$\left( t_{\mathrm{in}}^{eff},t_{\mathrm{out}}^{eff}\right] $\ into $M$\
intervals{\large \ }$\Delta t_{i}=t_{i+1}-t_{i}>0${\large , }such that Eq.~(%
\ref{svf1}) holds true for each of them. That allows us to treat the
electric field as approximately constant within each{\large \ }interval,%
{\large \ }$\overline{E(t)}\approx \overline{E}(t_{i})${\large \ }for{\large %
\ }$t\in \left( t_{i},t_{i+1}\right] ${\large . }In each such an interval,
we obtain expressions similar to the ones{\large \ }(\ref{4.46a}) and (\ref%
{ELaa}), where the constant electric field $E$\ has to be substituted by%
{\large \ }$\overline{E}(t_{i})$. Then{\large \ }components of the EMT for
arbitrary slowly varying strong electric field $E\left( t\right) $\ in the
leading-term approximation can be represented as
\begin{align}
& \mathrm{Re}\langle T_{00}(t)\rangle ^{c}=-\mathrm{Re}\langle
T_{11}(t)\rangle ^{c}=E\left( t\right) \frac{\partial \mathrm{Re}\mathcal{L}%
\left[ E\left( t\right) \right] }{\partial E\left( t\right) }-\mathrm{Re}%
\mathcal{L}\left[ E\left( t\right) \right] \,,  \notag \\
& \mathrm{Re}\langle T_{ll}(t)\rangle ^{c}=\mathrm{Re}\mathcal{L}\left[
E\left( t\right) \right] \,,\;l=2,...,D,  \label{4.46b}
\end{align}%
where
\begin{eqnarray}
&&\mathcal{L}\left[ E\left( t\right) \right] =\frac{1}{2}\int_{0}^{\infty }%
\frac{ds}{s}\mathrm{tr}\tilde{f}(x,x,s),\ \mathrm{tr}\tilde{f}%
(x,x,s)=2^{[d/2]}\cosh \left[ eE\left( t\right) s\right] \tilde{f}%
^{(0)}(x,x,s),  \notag \\
&&\tilde{f}^{(0)}(x,x,s)=-\frac{eE\left( t\right) s^{-d/2+1}\exp \left(
-im^{2}s\right) }{(4\pi i)^{d/2}\sinh \left[ eE\left( t\right) s\right] }.
\label{4.46c}
\end{eqnarray}%
Note that{\large \ }$\mathcal{L}\left[ E\left( t\right) \right] ${\large \ }%
evolves in time due to the time dependence of the field $E\left( t\right) $%
{\large .}

The quantity $\mathcal{L}\left[ E\left( t\right) \right] $ describes the
vacuum polarization. The quantities (\ref{4.46b}) are divergent due to the
real part of the effective Lagrangian (\ref{4.46c}), which is ill defined.
This real part must be regularized and renormalized. In low dimensions, $%
d\leq 4$, $\mathrm{Re}\mathcal{L}\left[ E\left( t\right) \right] $ can be
regularized in the proper-time representation and renormalized by the
Schwinger renormalizations of the charge and the electromagnetic field \cite%
{Schwinger51}. In particular, for $d=4$, the renormalized effective
Lagrangian $\mathcal{L}_{ren}\left[ E\left( t\right) \right] $ is
\begin{equation}
\mathcal{L}_{ren}\left[ E\left( t\right) \right] =\int_{0}^{\infty }\frac{%
ds\exp \left( -im^{2}s\right) }{8\pi ^{2}s}\left\{ \frac{eE\left( t\right)
\coth \left[ eE\left( t\right) s\right] }{s}-\frac{1}{s^{2}}-\frac{\left[
eE\left( t\right) s\right] ^{2}}{3}\right\} .  \label{HEL1}
\end{equation}%
In higher dimensions, $d>4$, a different approach is required. One can give
a precise meaning and calculate the one-loop effective action using
zeta-function regularization, see details in Ref.~\cite{GavGitY12}. Making
the same renormalization for $\langle T_{\mu \mu }(t)\rangle ^{c}$, we can
see that for the renormalized EMT the following relations hold true
\begin{align}
& \mathrm{Re}\,\langle T_{00}(t)\rangle _{ren}^{c}=-\mathrm{Re}\langle
T_{11}(t)\rangle _{ren}^{c}=E\left( t\right) \frac{\partial \mathrm{Re}%
\mathcal{L}_{ren}\left[ E\left( t\right) \right] }{\partial E\left( t\right)
}-\mathrm{Re}\mathcal{L}_{ren}\left[ E\left( t\right) \right] ,  \notag \\
& \mathrm{Re}\,\langle T_{ll}(t)\rangle _{ren}^{c}=\mathrm{Re}\mathcal{L}%
_{ren}\left[ E\left( t\right) \right] \,,\quad \ l=2,3,\dots ,D.
\label{L-ren}
\end{align}%
In the strong-field case ( $m^{2}/eE\left( t\right) \ll 1$), the leading
contributions to the renormalized EMT are
\begin{equation}
\mathrm{Re}\,\langle T_{\mu \mu }(t)\rangle _{ren}^{c}\sim \left\{
\begin{array}{l}
\left[ eE\left( t\right) \right] ^{d/2}\,,\quad d\neq 4k\, \\
\left[ eE\left( t\right) \right] ^{d/2}\ln \left[ eE\left( t\right) /M^{2}%
\right] ,\quad d=4k%
\end{array}%
\right. .  \label{emt-lead}
\end{equation}

The final form of the vacuum mean components of the EMT are
\begin{equation}
\langle T_{\mu \mu }(t)\rangle _{ren}=\mathrm{Re}\,\langle T_{\mu \mu
}(t)\rangle _{ren}^{c}+\mathrm{Re}\,\langle T_{\mu \mu }(t)\rangle ^{p},
\label{emt-ren}
\end{equation}%
where the {components }$\mathrm{Re}\,\langle T_{\mu \mu }(t)\rangle
_{ren}^{c}${\ and $\mathrm{Re}\,\langle T_{\mu \mu }(t)\rangle ^{p}$ are
given by }Eqs.~(\ref{L-ren}) and (\ref{uni11}), respectively.{\ }For $%
t<t_{in}$ and $t>t_{out}$ the electric field is absent such that $\mathrm{Re}%
\langle T_{\mu \mu }(t)\rangle _{ren}^{c}=0$.

On the right hand side of Eq.~(\ref{emt-ren}), the term $\mathrm{Re}%
\,\langle T_{\mu \mu }(t)\rangle ^{p}$ represents contributions due to the
vacuum instability, whereas the term $\mathrm{Re}\,\langle T_{\mu \mu
}(t)\rangle _{ren}^{c}$ represents\textrm{\ }vacuum polarization effects.
For weak electric fields, $m^{2}/eE\gg 1,$ contributions due to the vacuum
instability are exponentially small, so that the vacuum polarization effects
play the principal role. For strong electric fields, $m^{2}/eE\ll 1$, the
energy density of the vacuum polarization $Re\,\langle T_{00}(t)\rangle
_{ren}^{c}$\ is negligible compared to the energy density due to the vacuum
instability $\langle T_{00}\left( t\right) \rangle ^{p}$,%
\begin{equation}
\langle T_{\mu \mu }(t)\rangle _{ren}\approx \mathrm{Re}\,\langle T_{\mu \mu
}(t)\rangle ^{p}.  \label{emt-cr}
\end{equation}%
The latter density is formed on the whole time interval $t-t_{\mathrm{in}},$%
{\large \ }however, dominant contributions are due to time intervals $\Delta
t_{i}$ with $m^{2}/e\overline{E}(t_{i})<1$\ and the large dimensionless
parameters $\sqrt{e\overline{E}(t_{i})}\Delta t_{i}$.

We note that effective Lagrangian (\ref{4.46c}) and its renormalized form $%
\mathcal{L}_{ren}\left[ E\left( t\right) \right] $\ coincide with leading
term approximation of derivative expansion results from field-theoretic
calculations obtained in Refs.~\cite{DunHal98,GusSh99a,GusSh99b} for{\large %
\ }$d=3${\large \ }and{\large \ }$d=4${\large . }In this approximation the $%
S^{\left( 0\right) }$\ term of the Schwinger's effective action, given by
the expansion (\ref{uni7b}), has the form
\begin{equation}
S^{\left( 0\right) }[F_{\mu \nu }]=\int dx\mathcal{L}_{ren}\left[ E\left(
t\right) \right] .  \label{derexp}
\end{equation}

It should be stressed that unlike to the authors of Refs.~~\cite%
{DunHal98,GusSh99a,GusSh99b}, expression (\ref{4.46c}) and its renormalized
form were derived in the framework of the general exact formulation of
strong-field QED \cite{FGS,GavGT06}, using QED definition of the mean value
of the EMT, given by Eq.~(\ref{A1.4}).{\large \ }Therefore $\mathcal{L}_{ren}%
\left[ E\left( t\right) \right] $ is obtained independently from the
derivative expansion approach and the obtained result holds true for any
strong field under consideration.{\large \ }Moreover, it is proven that in
this case not only the imaginary part of $S^{\left( 0\right) }$ but its real
part as well is given exactly by the semiclassical WKB limit. It is clearly
demonstrated that the imaginary part of the effective action,{\large \ }$%
\mathrm{Im}S^{\left( 0\right) }$, is related to the vacuum-to-vacuum
transition probability $P_{v}$\ and can be represented as an integral of $%
\mathcal{L}_{ren}\left[ E\left( t\right) \right] $ over the total field
history, whereas the kernel of the real part of this effective action,%
{\large \ }$\mathrm{Re}\mathcal{L}_{ren}\left[ E\left( t\right) \right] $,
is related to the local EMT which defines the vacuum polarization. Obtained
results justify the derivative expansion as an asymptotic expansion that can
be useful to find the corrections for mean values of the EMT components. We
also note that some authors have argued that the locally constant field
approximation, which amounts to limiting oneself to the leading contribution
of the\ derivative expansion of the effective action, allows for reliable
results for electromagnetic fields of arbitrary strength; cf., e.g., \cite%
{GalN83,GiesK17}.

\section{Concluding remarks}

We have presented in detail consistent QED calculations of zero order
quantum effects in external electromagnetic field that correspond to the
most important three{\Large \ } exactly solvable cases of $t$-electric
potential steps, namely, the Sauter-like electric field, the $T$-constant
electric field, and the exponentially growing and decaying electric fields.
In all the cases, we present some new important details, unpublished so far.
Nontrivial details underlying{\Large \ } the derivation of number density of
pairs created from vacuum due to the strong Sauter-like case are presented.
Next-to-leading term approximation for the differential mean number of pair
created, $N_{n}^{\mathrm{cr}}$, due to $T$-constant electric field of long
duration is obtained. A detailed study of differential mean numbers of pair
created in the most asymmetric case of the exponentially growing and
decaying electric fields is presented. On the base of exactly solvable
cases, we consider in detail distributions $N_{n}^{\mathrm{cr}}$~ as
functions on the particle momenta and establish the ranges of dominant
contributions for mean numbers of created particles due to a slowly varying
field. This allows us to to gain new insights on the universal behavior of
the vacuum mean values in slowly varying intense electric fields. Comparing
results for three exactly solvable cases, one can see the appearance of a
large parameter, which is an increment of the longitudinal kinetic momentum,
and which corresponds to a large number of quantum states, in which
particles can be created. One can define the slowly varying regime in
general terms. Using the cases of the $T$-constant and exponentially growing
and decaying electric fields, we find universal forms of the vacuum mean
values of current, EMT components and the total density of created pairs in
the leading-term approximation for any large duration of an electric field.
One can find a close relation of obtained universal forms with a leading
term approximation of derivative expansion results in field-theoretic
calculations. In fact, it is the possibility to adopt a locally constant
field approximation which makes an effect universal. These results allow one
to formulate semiclassical approximations, that are not restricted by
smallness of differential mean numbers of created pairs, and could be
helpful for the development of numerical methods in strong-field QED.

\subparagraph{\protect\large Acknowledgement}

The reported study of T.C.A., S.P.G., and D.M.G. was supported by a grant
from the Russian Science Foundation, Research Project No. 15-12-10009.

\appendix

\section{Some asymptotic expansions \label{Ap}}

The asymptotic expression of the confluent hypergeometric function for large
$\eta $ and $c$ with fixed $a$ and $\tau =\eta /c\sim 1$ is given by
Eq.~(13.8.4) in \cite{DLMF} as%
\begin{eqnarray}
&&\Phi \left( a,c;\eta \right) \simeq c^{a/2}e^{\mathcal{Z}^{2}/4}F\left(
a,c;\tau \right) ,\ \ \mathcal{Z=-}\left( \tau -1\right) \mathcal{W}\left(
\tau \right) \sqrt{c},  \notag \\
&&F\left( a,c;\tau \right) =\tau \mathcal{W}^{1-a}D_{-a}\left( \mathcal{Z}%
\right) +\mathcal{R}D_{1-a}\left( \mathcal{Z}\right) ,  \notag \\
&&\mathcal{R}=\left( \mathcal{W}^{a}-\tau \mathcal{W}^{1-a}\right) /\mathcal{%
Z},\ \ \mathcal{W}\left( \tau \right) =\left[ 2\left( \tau -1-\ln \tau
\right) /\left( \tau -1\right) ^{2}\right] ^{1/2}  \label{A1}
\end{eqnarray}%
where $D_{-a}\left( \mathcal{Z}\right) $ is the Weber parabolic cylinder
function (WPCF) \cite{BatE53}. Using Eq.~(\ref{A1}) we present the functions
$y_{1}^{2}\left( \eta _{2}\right) $, $y_{2}^{1}\left( \eta _{1}\right) $ and
their derivatives at $t=0$ as%
\begin{eqnarray}
&&\left. y_{2}^{1}\left( \eta _{1}\right) \right\vert _{t=0}\simeq
e^{ih_{1}/2}\left( ih_{1}\right) ^{-\nu _{1}}\left( 2-c_{1}\right) ^{\left(
1-a_{1}\right) /2}e^{\mathcal{Z}_{1}^{2}/4}F\left( 1-a_{1},2-c_{1};\tau
_{1}\right) ,  \notag \\
&&\mathcal{Z}_{1}\mathcal{=}\mathcal{-}\left( \tau _{1}-1\right) \mathcal{W}%
\left( \tau _{1}\right) \sqrt{2-c_{1}},\ \ \tau _{1}=-ih_{1}/\left(
2-c_{1}\right) ,  \notag \\
&&\left. \frac{\partial y_{2}^{1}\left( \eta _{1}\right) }{\partial \eta _{1}%
}\right\vert _{t=0}\simeq e^{ih_{1}/2}\left( ih_{1}\right) ^{-\nu
_{1}}\left( 2-c_{1}\right) ^{\left( 1-a_{1}\right) /2}e^{\mathcal{Z}%
_{1}^{2}/4}\left[ -\frac{1}{2ih_{1}}-\frac{1}{2-c_{1}}\frac{\partial }{%
\partial \tau _{1}}\right] F\left( 1-a_{1},2-c_{1};\tau _{1}\right) ;  \notag
\\
&&\left. y_{1}^{2}\left( \eta _{2}\right) \right\vert _{t=0}\simeq
e^{-ih_{2}/2}\left( ih_{2}\right) ^{\nu _{2}}c_{2}^{a_{2}/2}e^{\mathcal{Z}%
_{2}^{2}/4}F\left( a_{2},c_{2};\tau _{2}\right) ,  \notag \\
&&\mathcal{Z}_{2}\mathcal{=}\mathcal{-}\left( \tau _{2}-1\right) \mathcal{W}%
\left( \tau _{2}\right) \sqrt{c_{2}},\ \ \tau _{2}=ih_{2}/c_{2},  \notag \\
&&\left. \frac{\partial y_{1}^{2}\left( \eta _{2}\right) }{\partial \eta _{2}%
}\right\vert _{t=0}\simeq e^{-ih_{2}/2}\left( ih_{2}\right) ^{\nu
_{2}}c_{2}^{a_{2}/2}e^{\mathcal{Z}_{2}^{2}/4}\left[ -\frac{1}{2ih_{2}}+\frac{%
1}{c_{2}}\frac{\partial }{\partial \tau _{2}}\right] F\left(
a_{2},c_{2};\tau _{2}\right) .  \label{A5}
\end{eqnarray}

Assuming $\tau -1\rightarrow 0$, one has%
\begin{eqnarray*}
&&\mathcal{W}^{1-a}\approx 1+\frac{a-1}{3}\left( \tau -1\right) ,\ \
\mathcal{R}\approx \frac{2\left( a+1\right) }{3\sqrt{c}},\ \ \mathcal{%
Z\approx -}\left( \tau -1\right) \sqrt{c}, \\
&&\frac{\partial F\left( a,c;\tau \right) }{\partial \tau }\approx \frac{2+a%
}{3}D_{-a}\left( \mathcal{Z}\right) +\frac{\partial D_{-a}\left( \mathcal{Z}%
\right) }{\partial \tau }+\mathcal{R}\frac{\partial D_{1-a}\left( \mathcal{Z}%
\right) }{\partial \tau }.
\end{eqnarray*}%
Expanding WPCFs near $\mathcal{Z}=0$, in the leading approximation at $%
\mathcal{Z}\rightarrow 0$ one obtains that%
\begin{eqnarray}
&&\frac{\partial F\left( a,c;\tau \right) }{\partial \tau }\approx -\sqrt{%
\eta }D_{-a}^{\prime }\left( 0\right) +O\left( \eta \right) ,  \notag \\
&&F\left( a,c;\tau \right) \approx D_{-a}\left( 0\right) +O\left(
c^{-1/2}\right) ,  \label{A2}
\end{eqnarray}%
and%
\begin{equation}
D_{-a}\left( 0\right) =\frac{2^{-a/2}\sqrt{\pi }}{\Gamma \left( \frac{a+1}{2}%
\right) },\ \ D_{-a}^{\prime }\left( 0\right) =\frac{2^{\left( 1-a\right) /2}%
\sqrt{\pi }}{\Gamma \left( \frac{a}{2}\right) },  \label{A4}
\end{equation}%
where $\Gamma (z)$ is the Euler gamma function. We find under condition (\ref%
{4.1}) that%
\begin{eqnarray}
&&\omega _{1,2}\approx \left\vert \pi _{1,2}\right\vert \left( 1+\lambda
/h_{1,2}\right) ,\ \ a_{1,2}\approx \left( 1+\chi \right) /2+i\lambda /2,
\notag \\
&&2-c_{1}\approx 1-i\left( \lambda +\frac{2\pi _{1}}{k_{1}}\right) ,\ \
c_{2}\approx 1+i\left( \lambda -\frac{2\pi _{2}}{k_{2}}\right) ,  \notag \\
&&\tau _{1}-1\approx -\frac{1}{h_{1}}\left( i+\lambda +\frac{2p_{x}}{k_{1}}%
\right) ,\ \ \tau _{2}-1\approx \frac{1}{h_{2}}\left( i-\lambda +\frac{2p_{x}%
}{k_{2}}\right) .  \label{A6}
\end{eqnarray}%
Using Eqs.~(\ref{A5}), (\ref{A2}), and (\ref{A6}) we represent Eq.~(\ref{4.0}%
) in the form%
\begin{eqnarray}
N_{n}^{\mathrm{cr}} &=&e^{-\pi \lambda /2}\left[ \left\vert \delta
_{0}\right\vert ^{2}+O\left( h_{1}^{-1/2}\right) +O\left(
h_{2}^{-1/2}\right) \right] ,  \notag \\
\delta _{0} &=&e^{i\pi /4}D_{-a_{2}}\left( 0\right) D_{a_{1}-1}^{\prime
}\left( 0\right) -e^{-i\pi /4}D_{-a_{2}}^{\prime }\left( 0\right)
D_{a_{1}-1}\left( 0\right) .  \label{A8}
\end{eqnarray}%
Assuming $\chi =1$ for fermions and $\chi =0$ for bosons, and using the
relations of the Euler gamma function we find that%
\begin{equation}
\delta _{0}=\exp \left( i\pi /2+i\pi \chi /4\right) e^{-\pi \lambda /4}.
\label{A9}
\end{equation}

Assuming $\left\vert \tau -1\right\vert \sim 1$, one can use the asymptotic
expansions of WPCFs in Eq.~(\ref{A1}), e.g., see \cite{BatE53,DLMF}. Note
that $\arg \left( \mathcal{Z}\right) \approx \frac{1}{2}\arg \left( c\right)
$ if $1-\tau >0$. Then one finds that%
\begin{equation}
\Phi \left( a,c;\eta \right) =\left( 1-\tau \right) ^{-a}\left[ 1+O\left(
\left\vert \mathcal{Z}\right\vert ^{-1}\right) \right] \ \ \mathrm{if}\
\;1-\tau >0.  \label{A10a}
\end{equation}%
In the case of $1-\tau <0$, one has
\begin{equation*}
\arg \left( \mathcal{Z}\right) \approx \left\{
\begin{array}{c}
\frac{1}{2}\arg \left( c\right) +\pi \ \ \mathrm{if}\ \ \arg \left( c\right)
<0 \\
\frac{1}{2}\arg \left( c\right) -\pi \ \ \mathrm{if}\ \ \arg \left( c\right)
>0%
\end{array}%
\right. .
\end{equation*}%
Then one obtains finally that%
\begin{equation}
\Phi \left( a,c;\eta \right) =\left\{
\begin{array}{l}
\left( \tau -1\right) ^{-a}e^{-i\pi a}\left[ 1+O\left( \left\vert \mathcal{Z}%
\right\vert ^{-1}\right) \right] \ \ \mathrm{if}\ \;\arg \left( c\right) <0
\\
\left( \tau -1\right) ^{-a}e^{i\pi a}\left[ 1+O\left( \left\vert \mathcal{Z}%
\right\vert ^{-1}\right) \right] \ \ \mathrm{if}\ \;\arg \left( c\right) >0%
\end{array}%
\right. .  \label{A10}
\end{equation}

The asymptotic expression of the confluent hypergeometric function $\Phi
\left( a,c;\pm ih\right) $ for large real $h$ with fixed $a$ and $c$ is
given by Eq.~(6.13.1(2)) in \cite{BatE53} as%
\begin{equation}
\Phi \left( a,c;\pm ih\right) =\frac{\Gamma \left( c\right) }{\Gamma \left(
c-a\right) }e^{\pm i\pi a/2}h^{-a}+O\left( \left\vert h\right\vert
^{-a-1}\right) +\frac{\Gamma \left( c\right) }{\Gamma \left( a\right) }%
e^{\pm ih}\left( e^{\pm i\pi /2}h\right) ^{a-c}+O\left( \left\vert
h\right\vert ^{a-c-1}\right) .  \label{A11}
\end{equation}

\end{document}